\documentclass[a4paper, 11pt, oneside]{article}

\usepackage{jheppub}

\newcommand\fverb{\setbox\fverbbox=\hbox\bgroup\verb}
\newcommand\fverbdo{\egroup\medskip\noindent%
            \fbox{\unhbox\fverbbox}\ }
\newcommand\fverbit{\egroup\item[\fbox{\unhbox\fverbbox}]}
\newbox\fverbbox

\title{Comparative Study of $B_{c}\to D_{s}^{\ast}\ell^{+}\ell^{-}$ Decays in Standard Model and Supersymmetric Models}
\author[1]{Aqeel Ahmed,}
\author[1,2]{Ishtiaq Ahmed,}
\author[1,2]{M. Ali Paracha,}
\author[1,2]{Muhammad Junaid,}
\author[1]{Abdur Rehman,}
\author[2]{and M. Jamil Aslam}
\affiliation[1]{National Centre for Physics,\\
Quaid-i-Azam University Campus, Islamabad 45320, Pakistan }
\affiliation[2]{Department of Physics,\\
Quaid-i-Azam University, Islamabad 45320, Pakistan}
\emailAdd{aqeel@ncp.edu.pk}
\emailAdd{ishtiaq@ncp.edu.pk}
\emailAdd{ali@ncp.edu.pk}
\emailAdd{mjunaid@ncp.edu.pk}
\emailAdd{rehman@ncp.edu.pk}
\emailAdd{jamil@ncp.edu.pk}
\date{\today}

\abstract{A comparative study of the exclusive rare
$B_{c}\rightarrow D_{s}^{\ast}\ell^{+}\ell^{-}$ ($\ell=\mu , \tau$)
decays has been made in the minimal supersymmetric models (MSSM) and
the SUSY SO(10) GUT models. In this context, various physical
observables such as branching ratios $(\mathcal{BR})$,
forward-backward asymmetries $(\mathcal{A}_{FB})$, lepton
polarization asymmetries $(P_{L,N,T})$ and helicity fractions
($f_{L,T}$) of $D_{s}^{\ast}$ meson by using the the QCD sum rules
form factors have been investigated. It is found that the SUSY
effects are characteristically prominent to that of the SM values
for these observables. For instance, in SUSY I and SUSY II, the
forward-backward asymmetry does not cross zero which is mainly due
to the same sign of the $C_{7}^{eff}$ and $C_{9}^{eff}$ Wilson
coefficients. Similarly in SUSY SO(10) GUT models due to the complex
nature of the new Wilson coefficients -- corresponding to the new
operators arising due to the contribution of neutral Higgs bosons
(NHBs) -- the above mentioned observables are sizably affected.
Therefore the analysis of said observables in charmed semileptonic
$B$ meson decays can put some stringent constraints on the parameter
space of SUSY variants and can serve as a windowpane to look beyond
the SM.}


\begin{document}
\maketitle

\section{Introduction}\label{intro}

The standard model (SM) of particle physics is considered to be one
of the most successful theory of the twentieth century but we can
not accept it as the ultimate theory of nature since there are many
open questions beyond the scope of the SM to be addressed. These
questions include: gauge and fermion mass hierarchy,
matter-antimatter asymmetry, number of generations of quarks and
leptons, the nature of the dark matter and the unification of
fundamental forces. Therefore, we need to search new physics (NP)
beyond the SM that may help us to answer these open problems of the
SM. One of the plausible extension of the SM is supersymmetry (SUSY)
\cite{Chung} which is a promising candidate of NP and can shed light
on some of the issues mentioned above \cite{JE}. For example, SUSY
solves the hierarchy problem of the SM by the cancellation of the
quadratic divergences in the radiative loop corrections since in
SUSY fermions and bosons contribute with opposite sign
\cite{EW,EW1}. It is also considered that the lightest
supersymmetric particles (LSP) are the dark matter candidates
because they are stable and interact very weakly with the ordinary
matter. It is also an important ingredient in the superstring theory
which is a suitable candidate for the unification of all the known
forces including gravity.

There are two ways to search for the SUSY, one is to discover the
SUSY particles (sparticles) at high energy colliders directly
\cite{lhclamp} and the other is to search for its effects through
indirect methods. Since no sparticle has been seen so far, they must
have higher masses than their SM partners, implying that the SUSY is
a broken symmetry. The other option is to investigate the SUSY
through the indirect searches where SUSY can show its footprints. In
this regard, the processes involving the flavor changing neutral
currents (FCNC) are the elegant way to probe the indirect searches
of the SUSY. So the rare decays, involving FCNC, induced by $b\to
s(d)$ transitions provide a windowpane to look for the physics
beyond the SM. This lies in the fact that FCNC processes are not
allowed at the tree level in the SM but can occur at the loop level
through Glashow-Iliopoulos-Maiani (GIM) mechanism \cite{1} thereby
making them quite sensitive to the indirect searches of the SUSY
\cite{2,2a}. The sparticles can contribute to FCNC transitions
through the quantum loop due to R-parity conservation, hence, making
these processes a handy tool for studying the possible effects of
sparticles in the rare decays \cite{aali,RNH,RNH1}. This gives us a
good motivation to study the rare decays both theoretically and
experimentally to look beyond the SM especially in the LHC era
\cite{LHC,LHC1}.

Since FCNC processes are only possible at the loop level and have
small branching ratios as compared to the tree level transitions but
they have implications for the results obtained in the ongoing
experiments after the observations of the rare radiative
$b\rightarrow s\gamma $ decay at CLEO \cite{3}. Since then there
have been many studies on the rare semileptonic, radiative and
leptonic decays of $B_{u,d,s}$ mesons induced by the FCNC
transitions of $b\rightarrow s(d)$ \cite{4}. These studies will be
even more complete if one considers similar decays of the charmed
$B$ mesons $(B_{c})$. The charmed $B_{c}$ meson is a bound state of
two heavy quarks, the bottom $b$ and the charm $c$, and was first
observed in 1998 at the Tevatron in Fermilab \cite{8}. Because of
the two heavy quarks, the $B_{c}$ mesons are rich in phenomenology
compared to the other $B$ mesons. At Large Hadron Collider (LHC) the
expected number of events for the production of $B_{c}$ mesons are
about $10^{8}-10^{10}$ per year \cite{9, 9a, 9b, 9c, 9d,10} which is
a reasonable number to work on the phenomenology of the $B_{c}$
meson. In the literature, some of the possible radiative and
semileptonic exclusive decays of $B_{c}$ mesons such as
$B_{c}\rightarrow \left( \rho ,K^{\ast },D_{s}^{\ast },B_{u}^{\ast
}\right)\gamma,$ $B_{c}\rightarrow \ell\nu \gamma ,$
$B_{c}\rightarrow B_{u}^{\ast}\ell^{+}\ell^{-},$ $B_{c}\rightarrow
D_{1}^{0}\ell\nu ,$ $B_{c}\rightarrow D_{s0}^{\ast
}\ell^{+}\ell^{-}$ and $B_{c}\rightarrow D_{s,d}^{\ast
}\ell^{+}\ell^{-}$ have been studied using the frame work of
relativistic constituent quark model, QCD Sum Rules, the Light Cone
Sum Rules and the Ward identities \cite{11a,11b, 11,12, 12a, 12b,
12c, 12d, 12e,13, ali}.

Theoretically what makes the $B_{c}\rightarrow D_{s}^{\ast
}\ell^{+}\ell ^{-}$ more important compared to the other $B$ meson
decays, such as $B^{0}\rightarrow (K^{\ast },K_{1},\rho ,\pi
)\ell^{+}\ell^{-}$, is the fact that it can have two types of
contributions, one is through the penguin (loop level) and the other
is due to weak annihilation (WA). In the ordinary $B$ meson decays
the WA contributions are very small and can be ignored as compared
to the corresponding penguin contributions. However, for the $B_c$
meson these WA contributions are proportional to the CKM matrix
elements $V_{cb}V^{*}_{cs}$ and hence can not be ignored as compared
to penguin contributions \cite{ali}. In perspective to the new physics (NP) effects in the presence of weak annihilation (WA) contribution, it is shown \cite{QQ} that for $B^0\rightarrow K^{*}\mu^{+}\mu^{-}$ the helicity fractions are mildly effected under the implication of SM4 while in the case of $B_{c}\rightarrow D_{s}^{*}\mu^{+}\mu^{-}$ the SM4 effects are very optimistic \cite{QQQ}. Similarly for this observables when tauns are the final state the effects are comparable i.e. for $B_{c}\rightarrow D_{s}^{*}\mu^{+}\mu^{-}$ the shift in the maximum(minimum) values in the longitudinal(transverse) helicity fractions is 0.23 and in the case of $B^0\rightarrow K^{*}\tau^{+}\tau^{-}$ this shift is 0.20. With this motivation it is also interesting to study the SUSY effects for the lepton polarization asymmetries in this decay channel which is calculated in the present manuscript. The present study shows that the lepton polarization asymmetries for $B_{c}\rightarrow D_{s}^{*}\l^{+}\l^{-}$ more influenced due to the supersymmetric models in comparison of $B^0\rightarrow K^{*}\l^{+}\l^{-}$ \cite{QQQQ} and one can clearly see from the graphs that in the transverse lepton polarization for muons and tauns the NP effects are more prominent for $B_{c}\rightarrow D_{s}^{*}\l^{+}\l^{-}$. This feature arises due to the interference between the WA and the Supersymmetric contribution which is encapsulated in the $f_{7}$ and $f_{8}$ auxilary functions. In the study of the exclusive
$B$-meson decays the form factors which are the non-perturbative
quantities and are the scalar functions of the square of momentum
transfer are important ingredients. In literature the form factors
for $B_{c}\rightarrow D_{s}^{\ast }\ell^{+}\ell^{-}$ decay were
calculated using different approaches including the light front
constituent quark models, the relativistic quark models, the Ward
identities and  the QCD sum rules \cite{11,12, 12a, 12b, 12c, 12d,
12e, 13,ali,53,53a}. In this study we borrow the form factors
calculated by the QCD sum rules approach for our numerical
calculations from Azizi et al. \cite{53a}. However, in the context of hadronic uncertinities which enter through the relevent form factors, it needs to be stressed that various asymmetries such as forward-backward asymmetries, polarization asymmetries of the final state meson and leptons (which are calculated in the manuscript submitted) have almost negligible influence. Therefore, any deviation from the SM in these observables is the clear indication of NP.

It is important to emphasize here that the NP effects manifest
themselves in the rare $B$ decays in two different ways, one is due
to the new contribution to the Wilson coefficients and other is due
to the appearance of new operators in the effective Hamiltonian,
which are absent in the SM. The manifestations of the NP due to the
SUSY is unique in a sense that it modifies the Wilson coefficients
as well as it introduces the new operators in the effective
Hamiltonian. In the present study, the NP effects are analyzed by
studying the branching ratio ${\cal BR}$, the forward-backward
asymmetry ${\cal A}_{FB}$, the lepton polarization asymmetries and
the helicity fractions of $D_{s}^{\ast }$ meson for
$B_{c}\rightarrow D_{s}^{\ast }\ell^{+}\ell^{-}$ decays both in the
SM and in different MSSM scenarios with special emphasis on the
effects of neutral Higgs bosons (NHBs)
\cite{mssm,mssma,mssmb,mssmc,mssmd,mssme}. A sizeable deviations to
the SM results due to the SUSY effects are observed in the above
mentioned observables for many rare decays
\cite{j11,j14,jam,jama,azizi,azizia}.

It is worth mention that previously an anomaly has been observed by
Belle \cite{belle} in the lepton forward-backward asymmetry (${\cal
A}_{FB}$) in the exclusive decay $\bar{B}\to
\bar{K}^{\ast}\mu^+\mu^-$ \cite{L9,L10,L11,L12}. It was noticed that
in the large $q^2$ region $q^{2}\geq14$ GeV$^2$, the ${\cal
A}_{FB}(q^{2})$ measurements tend to be larger than that of the SM
predictions, although the behavior was same for the both cases. The
discrepancy was more severe at the low $q^2$ region where the SM
prediction is negative \cite{L13}, whereas the data favor positive
values and Belle claimed that this is the clear indication of NP
\cite{DL}. Contrary to this the LHCb recently announced the first
result on the lepton forward-backward asymmetry (${\cal A}_{FB}$) in
the exclusive decay $\bar{B}\to \bar{K}^{\ast}\mu^+\mu^-$ which is
close to the SM predictions. In particular, the position of the zero
crossing of the ${\cal A}_{FB}(q^{2})$ lie on the same position as
predicted by the SM \cite{aali, L14,L16}. This situation is really
exciting and tighten the screws on scenarios those have not
zero-crossing point of the ${\cal A}_{FB}(q^{2})$. In this context,
SUSY which previously become a prime candidate to explain the Belle
anomaly because in some of the scenarios of MSSM such as SUSY-I and
SUSY-II, the ${\cal A}_{FB}(q^{2})$ do not cross the zero position
\cite{j11,j14,jam,jama,azizi,azizia} now seems to be ruled out after
the LHCb data. However, before to say any final remarks about SUSY-I
nd SUSY-II, it needs to be required some confirmation through
complementary information as LHCb collaboration said that they will
continue to collect more data and try to see deviation in the
observable data if there is any NP exist. With this motivation we have
predicted the MSSM effects in the $B_{c}\to D_s^{\ast}\l^+\l^-$,
where $l=\mu , \tau $, decay not only for the ${\cal A}_{FB}(q^{2})$
but also for the other asymmetries which depends on the polarization
of the final state particles such as lepton polarization asymmetries
and the helicity fractions.

In this paper, we investigate the decay processes $B_{c}\rightarrow
D_{s}^{\ast }\ell^{+}\ell^{-}$ in the context of the SUSY SO(10) GUT
models \cite{gut}. Since the effects of the counterparts of usual
chromo-magnetic and electro-magnetic dipole moment operators as well
as semileptonic operators with opposite chirality are suppressed by
$m_{s}/m_{b}$ and as such become insensitive in the SM. However, in
the SUSY SO(10) GUTs their effects can be significantly large, since
complex flavor non-diagonal down-type squark mass matrix elements of
2nd and 3rd generations in the RR sector ($\delta_{23}^{dRR}$) can
be as large as 0.5 \cite{gut,so10}. Furthermore, $\delta_{23}^{dRR}$
can induce new operators, the counterparts of usual scalar operators
($Q_{1,2}$) in the SUSY models, due to the NHBs penguins with
gluino-down type squark propagated in the loop.  The values of the
relevant new Wilson coefficients for the MSSM and the SUSY SO(10)
GUT models are collected from Refs. \cite{j11,j14}. The NHBs could
contribute largely to the inclusive processes $B\to
X_{s}\ell^{+}\ell^{-}$, because part of the SUSY contributions is
proportional to the $\tan^{3}\beta$ \cite{nhb,nhba}. Subsequently,
the physical observables, such as the branching ratio ${\cal BR}$,
forward-backward asymmetries ${\cal A}_{FB}$, lepton polarization
asymmetries and helicity fractions of the final state meson, in the
large $\tan\beta$ region of parameter space in the SUSY models can
be quite different from that in the SM. Hence, the measurement of
these observables will give us some hints of the SUSY effects in
these decays.

The paper is structured as follows. In Sec. \ref{tf} we present the
theoretical framework for the decay $B_{c}\rightarrow D_{s}^{\ast
}\ell^{+}\ell^{-}$ necessary for the study of the different SUSY
variants. In Sec. \ref{po} we present the basic formulas for
physical observables such as decay rate, forward-backward
asymmetries ${\cal A}_{FB}$, lepton polarization asymmetries and
helicity fractions of $D_{s}^{\ast }$ meson. The numerical analysis
and discussion on these observables is given in Sec. \ref{num}.
Section \ref{con} gives the summary of the results of our study.

\section{Theoretical Framework}\label{tf}
In this section we give the effective Hamiltonian for
$B_{c}\rightarrow D_{s}^{\ast } \ell^{+} \ell^{-}$ decays. We can
split the contributions of the total decay amplitude into two, one
is the penguin (FCNC trasitions) and the second is weak annihilation
contribution.

\subsection{Penguin Amplitude}

At quark level, the semileptonic decay $B_{c}\rightarrow D_{s}^{\ast
}\ell^{+}\ell^{-}$ is governed by the transition $b\rightarrow
s\ell^{+}\ell^{-}$ for which the general effective Hamiltonian in
the SUSY SO(10) GUT model, can be written, after integrating out the
heavy degrees of freedom in the full theory, as \cite{j14}
\begin{align}
H_{eff}=-\frac{4G_{F}}{\sqrt{2}}V_{tb}V_{ts}^{\ast }&\bigg[\sum\limits_{i=1}^{2}C_{i}(\mu )O_{i}(\mu)+\sum\limits_{i=3}^{10}\left\{C_{i}(\mu )O_{i}(\mu)+C^{\prime}_{i}(\mu )O^{\prime}_{i}(\mu)\right\}\notag\\
&+\sum\limits_{i=1}^{8}\left\{C_{Qi}(\mu )Q_{i}(\mu)+C^{\prime}_{Qi}(\mu )Q^{\prime}_{i}(\mu)\right\} \bigg],  \label{1}
\end{align}%
where $O_{i}(\mu )$ $(i=1,2, \cdots,10)$ are the four quark operators and $%
C_{i}(\mu )$ are the corresponding Wilson coefficients at the energy scale $%
\mu $ \cite{mssm,mssma,mssmb,mssmc,mssmd,mssme} which is usually taken to be the $b$-quark mass $\left(
m_{b}\right) $. The theoretical uncertainties related to the renormalization
scale can be reduced when the next to leading logarithm corrections are
included. The new operators $Q_{i}(i=1, 2, \cdots, 8)$ come from the NHBs exchange diagrams,
whose manifest forms and corresponding Wilson coefficients can be found in \cite{j38,j38a,j38b,j38c,j38d}. The primed operators
are the counterparts of the unprimed operators, which can be obtained by flipping the chiralities in the corresponding
unprimed operators. It is worth mentioning that these primed operators will appear only in the SUSY SO(10) GUT model and are absent in the SM and the MSSM \cite{j11}.

The explicit forms of the operators responsible for the decay $%
B_{c}^{-}\rightarrow D_{s}^{\ast -}\ell ^{+}\ell ^{-}$, in the SM and the SUSY models, are
\begin{subequations}
\begin{eqnarray}
O_{7} &=&\frac{e^{2}}{16\pi ^{2}}m_{b}(\bar{s}\sigma _{\mu \nu }Rb)F^{\mu
\nu }  \label{2} \\
O^{\prime}_{7} &=&\frac{e^{2}}{16\pi ^{2}}m_{b}(\bar{s}\sigma _{\mu \nu }Lb)F^{\mu
\nu } \label{2p}\\
O_{9} &=&\frac{e^{2}}{16\pi ^{2}}\left( \bar{s}\gamma _{\mu }Lb\right) \bar{\ell
}\gamma ^{\mu }\ell  \label{3} \\
O^{\prime}_{9} &=&\frac{e^{2}}{16\pi ^{2}}\left( \bar{s}\gamma _{\mu
}Rb\right) \bar{\ell }\gamma ^{\mu }\ell\label{3p}\\
O_{10} &=&\frac{e^{2}}{16\pi ^{2}}\left( \bar{s}\gamma _{\mu }Lb\right) \bar{\ell}\gamma ^{\mu }\gamma ^{5}\ell  \label{4}\\
O^{\prime}_{10} &=&\frac{e^{2}}{16\pi ^{2}}\left( \bar{s}\gamma
_{\mu }Rb\right) \bar{\ell}\gamma ^{\mu }\gamma ^{5}\ell
\label{4p}\end{eqnarray}
\begin{eqnarray}
Q_{1} &=&\frac{e^{2}}{16\pi ^{2}}\left( \bar{s}Rb\right) \bar{\ell}\ell  \label{q1}\\
Q^{\prime}_{1} &=&\frac{e^{2}}{16\pi ^{2}}\left( \bar{s}Lb\right) \bar{\ell}\ell \label{q1p}\\
Q_{2} &=&\frac{e^{2}}{16\pi ^{2}}\left( \bar{s}Rb\right) \bar{\ell}\gamma ^{5}\ell  \label{q2}\\
Q^{\prime}_{2} &=&\frac{e^{2}}{16\pi ^{2}}\left( \bar{s}Lb\right) \bar{\ell}\gamma ^{5}\ell \label{q2p}
\end{eqnarray}%
\end{subequations}
with $L,R=\frac{1}{2}\left( 1\mp \gamma ^{5}\right)$.

Using the effective Hamiltonian given in Eq.(\ref{1}) the free quark
amplitude for $b\rightarrow s\ell ^{+}\ell ^{-}$ can be written as%
\begin{align}
\mathcal{M}^{\text{PENG}}(b\rightarrow s\ell^{+}\ell^{-}) &=-\frac{%
G_{F}\alpha }{\sqrt{2}\pi }V_{tb}V_{ts}^{\ast
}\bigg[C_{9}^{eff}\left( \mu \right) (\bar{s}\gamma _{\mu
}Lb)(\bar{\ell}\gamma ^{\mu }\ell)+C_{9}^{\prime eff}\left( \mu
\right) (\bar{s}\gamma _{\mu }R b)(\bar{\ell}\gamma ^{\mu }\ell)\notag\\
&+C_{10}(\bar{s}\gamma _{\mu }Lb)(\bar{\ell}\gamma ^{\mu }\gamma ^{5}\ell) +C^{\prime}_{10}(\bar{s}\gamma _{\mu }R b)(\bar{\ell}\gamma ^{\mu }\gamma ^{5}\ell) \notag \\
&-2C_{7}^{eff}\left( \mu \right) \frac{m_{b}}{q^{2}}(\bar{s}i\sigma
_{\mu \nu }q^{\nu }R b)\bar{\ell}\gamma ^{\mu }\ell-2C_{7}^{\prime
eff}\left( \mu \right) \frac{m_{b}}{q^{2}}(\bar{s}i\sigma _{\mu \nu
}q^{\nu
}Lb)\bar{\ell}\gamma ^{\mu }\ell\notag\\
&+C_{Q1}\left( \bar{s}Rb\right) \left(\bar{\ell}\ell
\right)+C^{\prime}_{Q1}\left( \bar{s}Lb\right) \left(\bar{\ell}\ell
\right) +C_{Q2}\left( \bar{s}Rb\right)
\left(\bar{\ell}\gamma^{5}\ell \right) +C^{\prime}_{Q2}\left(
\bar{s}Lb\right) \left(\bar{\ell}\gamma^{5}\ell \right)\bigg],
\label{5a}
\end{align}
where $q$ is the momentum transfer. Note that the operator $%
O_{10}$ given in Eq.(\ref{4}) can not be induced by the insertion of four
quark operators because of the absence of $Z$-boson in the effective theory.
Therefore, the Wilson coefficient $C_{10}$ does not renormalize under QCD
corrections and is independent of the energy scale $\mu .$ Additionally the
above quark level decay amplitude can get contributions from the matrix
element of four quark operators, $\sum_{i=1}^{6}\left\langle
\ell^{+}\ell^{-}s\left\vert O_{i}\right\vert b\right\rangle ,$ which are usually
absorbed into the effective Wilson coefficient $C_{9}^{eff}(\mu )$ and can
be written as \cite{25, 26, 27, 28, 29, 30, 31}
\begin{equation*}
C_{9}^{eff}(\mu )=C_{9}(\mu )+Y_{SD}(z,s^{\prime })+Y_{LD}(z,s^{\prime }).
\end{equation*}%
where $z=m_{c}/m_{b}$ and $s^{\prime }=q^{2}/m_{b}^{2}$. $Y_{SD}(z,s^{\prime
})$ describes the short distance contributions from four-quark operators far
away from the $c\bar{c}$ resonance regions, and this can be calculated
reliably in the perturbative theory. However the long distance contribution $%
Y_{LD}(z,s^{\prime })$ cannot be calculated by using the first principles of
QCD, so they are usually parametrized in the form of a phenomenological
Breit-Wigner formula making use of the vacuum saturation approximation and
quark hadron duality. The expressions for the short-distance and the long-distance contributions $Y_{SD}(z,s^{\prime })$ is given as
\begin{eqnarray}
Y_{SD}(z,s^{\prime }) &=&h(z,s^{\prime })\left[3C_{1}(\mu )+C_{2}(\mu
)+3C_{3}(\mu )+C_{4}(\mu )+3C_{5}(\mu )+C_{6}(\mu )\right]  \notag \\
&&-\frac{1}{2}h(1,s^{\prime })\left[4C_{3}(\mu )+4C_{4}(\mu )+3C_{5}(\mu
)+C_{6}(\mu )\right]  \notag \\
&&-\frac{1}{2}h(0,s^{\prime })\left[C_{3}(\mu )+3C_{4}(\mu )\right]+{\frac{2}{9}}\left[3C_{3}(\mu )+C_{4}(\mu )+3C_{5}(\mu )+C_{6}(\mu )\right],
\end{eqnarray}
\begin{eqnarray}
Y_{LD}(z,s^{\prime }) &=&\frac{3}{\alpha _{em}^{2}}(3C_{1}(\mu )+C_{2}(\mu
)+3C_{3}(\mu )+C_{4}(\mu )+3C_{5}(\mu )+C_{6}(\mu ))  \notag \\
&&\times\sum_{j=\psi ,\psi ^{\prime }}\omega _{j}(q^{2})k_{j}\frac{\pi \Gamma
(j\rightarrow l^{+}l^{-})M_{j}}{q^{2}-M_{j}^{2}+iM_{j}\Gamma _{j}^{tot}},
\label{LD}
\end{eqnarray}%
with
\begin{eqnarray}
h(z,s^{\prime }) &=&-{\frac{8}{9}}\mathrm{ln}z+{\frac{8}{27}}+{\frac{4}{9}}x-%
{\frac{2}{9}}(2+x)|1-x|^{1/2}\left\{
\begin{array}{l}
\ln \left| \frac{\sqrt{1-x}+1}{\sqrt{1-x}-1}\right| -i\pi \quad \mathrm{for}{%
{\ }x\equiv 4z^{2}/s^{\prime }<1} \\
2\arctan \frac{1}{\sqrt{x-1}}\qquad \mathrm{for}{{\ }x\equiv
4z^{2}/s^{\prime }>1}%
\end{array}%
\right. ,  \notag \\
h(0,s^{\prime }) &=&{\frac{8}{27}}-{\frac{8}{9}}\mathrm{ln}{\frac{m_{b}}{\mu
}}-{\frac{4}{9}}\mathrm{ln}s^{\prime }+{\frac{4}{9}}i\pi \,\,.
\end{eqnarray}%
Here $M_{j}(\Gamma _{j}^{tot})$ are the masses (widths) of the intermediate
resonant states and $\Gamma (j\rightarrow l^{+}l^{-})$ denote the partial
decay width for the transition of vector charmonium state to massless lepton
pair, which can be expressed in terms of the decay constant of charmonium
through the relation \cite{32}
\begin{equation*}
\Gamma (j\rightarrow \ell^{+}\ell^{-})=\pi \alpha _{em}^{2}{\frac{16}{27}}{\frac{%
f_{j}^{2}}{M_{j}}}.
\end{equation*}%
The phenomenological parameter $k_{j}$ in Eq.(\ref{LD}) is to account for
inadequacies of the factorization approximation, and it can be determined
from
\begin{equation*}
{\cal BR}(B_{c}\rightarrow D_{s}^{\ast} J/\psi \rightarrow  D_{s}^{\ast}
\ell^{+}\ell^{-})={\cal BR}(B_{c}\rightarrow D_{s}^{\ast} J/\psi )\cdot {\cal BR}(J/\psi\rightarrow \ell^{+}\ell^{-}).
\end{equation*}%
The function $\omega _{j}(q^{2})$ introduced in Eq.(\ref{LD}) is to
compensate the naive treatment of long distance contributions due to the
charm quark loop in the spirit of quark-hadron duality, which can
overestimate the genuine effect of the charm quark at small $q^{2}$
remarkably \footnote{%
For a more detailed discussion on long-distance and short-distance
contributions from the charm loop, one can refer to references \cite{aali, 32, b to s 2, b to s 3,charm loop 1, charm loop 2,charm
loop 3,yuming}.}. The quantity $\omega _{j}(q^{2})$ can be normalized to $\omega
_{j}(M_{\psi _{j}}^{2})=1$, but its exact form is unknown at present. Since
the dominant contribution of the resonances is in the vicinity of the
intermediate $\psi _{i}$ masses, we will simply use $\omega _{j}(q^{2})=1$
in our numerical calculations.

Moreover, the non factorizable effects from the charm quark loop brings further
corrections to the radiative transition $b\rightarrow s\gamma ,$ and these
can be absorbed into the effective Wilson coefficients $C_{7}^{eff}$ which
then takes the form \cite{yuming,32, 33, 34, 35, 36}
\begin{equation*}
C_{7}^{eff}(\mu )=C_{7}(\mu )+C_{b\rightarrow s\gamma }(\mu )
\end{equation*}%
with
\begin{eqnarray}
C_{b\rightarrow s\gamma }(\mu ) &=&i\alpha _{s}\left[ \frac{2}{9}\eta
^{14/23}(G_{1}(x_{t})-0.1687)-0.03C_{2}(\mu )\right]  \label{8} \\
G_{1}(x_{t}) &=&\frac{x_{t}\left( x_{t}^{2}-5x_{t}-2\right) }{8\left(
x_{t}-1\right) ^{3}}+\frac{3x_{t}^{2}\ln ^{2}x_{t}}{4\left( x_{t}-1\right)
^{4}}  \label{9}
\end{eqnarray}%
where $\eta =\alpha _{s}(m_{W})/\alpha _{s}(\mu ),$ \ $%
x_{t}=m_{t}^{2}/m_{W}^{2}$ and $C_{b\rightarrow s\gamma }$ is the absorptive
part for the $b\rightarrow sc\bar{c}\rightarrow s\gamma $ rescattering.

\subsection{Weak Annihilation Amplitude}

The charmed B-meson $(B_{c})$ is made up of two different heavy flavors, $b$%
-quark and $c$-quark, which brings WA contributions into the play. Using the
procedure developed in refs. \cite{Cheng, Du} for $B_{c}\rightarrow
D_{s}^{\ast }\gamma $, the WA amplitude for the decay $B_{c}\rightarrow
D_{s}^{\ast }\ell ^{+}\ell ^{-}$ can be written as
\begin{equation}
\mathcal{M}_{B_{c}\rightarrow D_{s}^{\ast }\ell ^{+}\ell ^{-}}^{\text{WA}}%
\mathcal{=}\frac{G_{F}\alpha }{2\sqrt{2}\pi }V_{cb}V_{cs}^{\ast }\left[
-i\epsilon _{\mu \nu \alpha \beta }\varepsilon ^{\ast \nu }p^{\alpha
}q^{\beta }{\cal K}_{1}^{ann}(q^{2})+\left( \varepsilon \cdot qp_{\mu }+p\cdot
q\varepsilon _{\mu }\right) {\cal K}_{2}^{ann}(q^{2})\right] \bar{\ell}\gamma ^{\mu }\ell
\label{1a}
\end{equation}%
where ${\cal K}_{1}^{ann}(q^{2})$ and ${\cal K}_{2}^{ann}(q^{2})$ are the weak annihilation form factors.

Before proceeding further we would like to mention that we use parametrizations for weak annihilation form factors ${\cal K}_{1}^{ann}(q^{2})$ and ${\cal K}_{2}^{ann}(q^{2})$, i.e.
\begin{equation}
{\cal K}_{1,2}^{ann}(q^{2}) =\frac{{\cal K}_{1,2}^{ann}(0)}{1+\alpha
\frac{q^{2}}{M_{B_{c}}^{2}}+\beta \frac{q^{4}}{M_{B_{c}}^{4}}}.
\label{1b}
\end{equation}%
The values of ${\cal K}_{1}^{ann}(0)$ and ${\cal K}_{2}^{ann}(0)$ are calculated by using QCD sum rules \cite{53,53a} and the
values of the parameter $\alpha $ and $\beta $ are given in Ref. \cite{12, 12a, 12b, 12c, 12d, 12e, 53,53a}, which are summarized in Table \ref{wavalues}.
\begin{table}[ht]
\centering
\begin{tabular}{cccc}
\hline\hline
 ${\cal K}^{ann}(q^{2})$ & $\hspace{2cm}{\cal K}_{0}^{ann}$ & $\hspace{2cm}\alpha$ & $\hspace{%
2cm}\beta$ \\ \hline
 ${\cal K}_{1}^{ann}\left( q^{2}\right) $ & $\hspace{2cm}0.23$ & $%
\hspace{2cm}-1.25$ & $\hspace{2cm}-0.097$ \\ \hline
 ${\cal K}_{2}^{ann}(q^{2})$ & $\hspace{2cm}0.25$ & $\hspace{2cm}%
-0.10$ & $\hspace{2cm}-0.097$ \\ \hline
\end{tabular}
\caption{$B_{c}\rightarrow D_{s}^{\ast }$ form factors corresponding
to WA in the QCD Sum Rules. ${\cal K}_{1,2}^{ann}(0)$ denote the
value of form factors at $q^{2}=0$ while $\alpha$ and $\beta$ are
the parameters in the parametrizations shown in Eq. (\ref {1b})
\cite{53a}.} \label{wavalues}
\end{table}

\subsection{Parameterizations of the Matrix Elements and Form Factors}\label{ff}

The exclusive $B_{c}\rightarrow D_{s}^{\ast }\ell^{+}\ell^{-}$ decay involves the
hadronic matrix elements which can be obtained by sandwiching the quark
level operators give in Eq. (\ref{5a}) between initial state $%
B_{c}$ meson and final state $D_{s}^{\ast }$ meson. These can be
parametrized in terms of form factors which are the scalar functions of the
square of the four momentum transfer($q^{2}=(p-k)^{2}).$ The non vanishing
matrix elements for the process $B_{c}\rightarrow D_{s}^{\ast }$ can be
parametrized in terms of the seven form factors as follows%
\begin{eqnarray}
\left\langle D_{s}^{\ast }(k,\varepsilon )\left\vert \bar{s}\gamma _{\mu
}b\right\vert B_{c}(p)\right\rangle &=&\frac{2A_{V}(q^{2})}{%
M_{B_{c}}+M_{D_{s}^{\ast}}}\epsilon _{\mu \nu \alpha \beta }\varepsilon
^{\ast \nu }p^{\alpha }k^{\beta }  \label{10} \\
\left\langle D_{s}^{\ast }(k,\varepsilon )\left\vert \bar{s}\gamma _{\mu
}\gamma _{5}b\right\vert B_{c}(p)\right\rangle &=&i\left(
M_{B_{c}^{-}}+M_{D_{s}^{\ast }}\right) \varepsilon ^{\ast \mu }A_{0}(q^{2})-i\frac{A_{+}\left( q^{2}\right) }{M_{B_{c}}+M_{D_{s}^{\ast }}}%
(\varepsilon ^{\ast }\cdot p)\left( p+k\right) ^{\mu }\notag\\&& -i\frac{A_{-}\left( q^{2}\right) }{M_{B_{c}}+M_{D_{s}^{\ast }}}%
(\varepsilon ^{\ast }\cdot p)q^{\mu }    \label{11}\\
\left\langle D_{s}^{\ast }(k,\varepsilon )\left\vert \bar{s}(1\pm\gamma_{5})b\right\vert B_{c}(p)\right\rangle &=&\mp i\frac{2M_{D_{s}^{\ast}}}{m_{b}+m_{s}}\left(\varepsilon ^{\ast}\cdot p\right) \tilde{A}_{0}(q^{2})\label{12}
\end{eqnarray}%
where $p$ is the momentum of the $B_{c}$, $\varepsilon $ and $k$ are the
polarization vector and momentum of the final state $D_{s}^{\ast }$ vector meson. Whereas $\tilde{A}_{0}(q^{2})$ can be parametrized as
\begin{equation}
\tilde{A}_{0}(q^{2})=-\frac{M_{B_{c}}+M_{D_{s}^{\ast }}}{2M_{D_{s}^{\ast }}}%
A_{0}(q^{2})+\frac{M_{B_{c}}-M_{D_{s}^{\ast }}}{2M_{D_{s}^{\ast }}}
A_{+}(q^{2})+\frac{q^{2}}{\left(M_{B_{c}}-M_{D_{s}^{\ast }}\right)M_{D_{s}^{\ast }}}
A_{-}(q^{2})  \label{12a}
\end{equation}%

In addition to the above form factors there are some penguin form factors,
which we can write as
\begin{eqnarray}
\left\langle D_{s}^{\ast }(k,\varepsilon )\left\vert \bar{s}\sigma _{\mu \nu
}q^{\nu }b\right\vert B_{c}(p)\right\rangle &=&2iT_{1}(q^{2})\epsilon _{\mu
\nu \alpha \beta }\varepsilon ^{\ast \nu }p^{\alpha }k^{\beta }  \label{13a}
\\
\left\langle D_{s}^{\ast }(k,\varepsilon )\left\vert \bar{s}\sigma _{\mu \nu
}q^{\nu }\gamma ^{5}b\right\vert B_{c}(p)\right\rangle &=&\left[ \left(
M_{Bc}^{2}-M_{D_{s}^{\ast }}^{2}\right) \varepsilon _{\mu }^{\ast
}-(\varepsilon ^{\ast }\cdot p)(p+k)_{\mu }\right] T_{2}(q^{2})  \notag \\
&&  \label{13b} \\
&&+(\varepsilon ^{\ast }\cdot p)\left[ q_{\mu }-\frac{q^{2}}{%
M_{Bc}^{2}-M_{D_{s}^{\ast }}^{2}}(p+k)_{\mu }\right] T_{3}(q^{2}).  \notag
\end{eqnarray}%
The form factors $A_{V}\left( q^{2}\right) ,~A_{0}\left( q^{2}\right) ,$ $%
~A_{+}\left( q^{2}\right) ,~A_{-}\left( q^{2}\right) ,~T_{1}\left(
q^{2}\right) ,~T_{2}\left( q^{2}\right) ,~T_{3}\left( q^{2}\right) $ are the
non-perturbative quantities and to calculate them one has to rely on some
non-perturbative approaches and in our numerical analysis we use the form
factors calculated by using QCD Sum Rules \cite{53a}. The dependence of these
form factors on square of the momentum transfer $(q^{2})$ can be written as%
\begin{equation}
F\left( q^{2}\right) =\frac{F\left( 0\right) }{1+a
\frac{q^{2}}{M_{B_{c}}^{2}}+b \frac{q^{4}}{M_{B_{c}}^{4}}}.
\label{ff-param}
\end{equation}%
where the values of the parameters $F\left( 0\right) $, $a$ and $b$ is given in Table \ref{formfactor}.

\begin{table}[tbh]
\centering
\begin{tabular}{cccc}
\hline\hline
 $F(q^{2})$ & $\hspace{2cm}F(0)$ & $\hspace{2cm}a$ & $\hspace{%
2cm}b$ \\ \hline
 $A_{V}\left( q^{2}\right) $ & $\hspace{2cm}0.54\pm 0.018$ & $%
\hspace{2cm}-1.28$ & $\hspace{2cm}-0.23$ \\ \hline
 $A_{0}(q^{2})$ & $\hspace{2cm}0.30\pm 0.017$ & $\hspace{2cm}%
-0.13$ & $\hspace{2cm}-0.18$ \\ \hline
 $A_{+}(q^{2})$ & $\hspace{2cm}0.36\pm 0.013$ & $\hspace{2cm}%
-0.67$ & $\hspace{2cm}-0.066$ \\ \hline
 $A_{-}(q^{2})$ & $\hspace{2cm}-0.57\pm 0.04$ & $\hspace{2cm}%
-1.11$ & $\hspace{2cm}-0.14$ \\ \hline
 $T_{1}(q^{2})$ & $\hspace{2cm}0.31\pm 0.017$ & $\hspace{2cm}%
-1.28$ & $\hspace{2cm}-0.23$ \\ \hline
 $T_{2}(q^{2})$ & $\hspace{2cm}0.33\pm 0.016$ & $\hspace{2cm}%
-0.10$ & $\hspace{2cm}-0.097$ \\ \hline
 $T_{3}(q^{2})$ & $\hspace{2cm}0.29\pm 0.034$ & $\hspace{2cm}%
-0.91$ & $\hspace{2cm}0.007$ \\ \hline\hline
\end{tabular}
\caption{$B_{c}\rightarrow D_{s}^{\ast }$ form factors corresponding to penguin
contributions in the QCD Sum Rules.
$F(0)$ denotes the value of form factors at $q^{2}=0$ while $a$ and $b$
are the parameters in the parametrizations shown in Eq. (\ref{ff-param}) \cite{53a}.}
\label{formfactor}
\end{table}

From Eq. (\ref{5a}) it is straightforward to write the penguin
amplitude
\begin{equation}
\mathcal{M}^{\text{PENG}}=-\frac{G_{F}\alpha }{2\sqrt{2}\pi }%
V_{tb}V_{ts}^{\ast }\left[ {\cal T}_{\mu }^{1}(\bar{\ell}\gamma ^{\mu }\ell)+{\cal T}_{\mu
}^{2}\left( \bar{\ell}\gamma ^{\mu }\gamma ^{5}\ell\right) +{\cal T}\left( \bar{\ell}\ell\right)\right]\label{59}
\end{equation}%
where%
\begin{eqnarray}
{\cal T}_{\mu }^{1} &=&f_{1}(q^{2})\epsilon _{\mu \nu \alpha \beta }\varepsilon
^{\ast \nu }p^{\alpha }k^{\beta }-if_{2}(q^{2})\varepsilon _{\mu }^{\ast
}+if_{3}(q^{2})(\varepsilon ^{\ast }\cdot p)P_{\mu }  \label{60} \\
{\cal T}_{\mu }^{2} &=&f_{4}(q^{2})\epsilon _{\mu \nu \alpha \beta }\varepsilon
^{\ast \nu }p^{\alpha }k^{\beta }-if_{5}(q^{2})\varepsilon _{\mu }^{\ast
}+if_{6}(q^{2})(\varepsilon ^{\ast }\cdot p)P_{\mu }+if_{7}(q^{2})(\varepsilon ^{\ast }\cdot p)P_{\mu }   \label{61}\\
{\cal T}&=&2if_{8}(q^{2})(\varepsilon ^{\ast }\cdot p)
\end{eqnarray}
with $P_{\mu}=p_{\mu}+k_{\mu}$.

The next task is to calculate the decay rate and the helicity fractions of $D_{s}^{\ast}$ meson in terms of these auxiliary functions which contains both long distance (form
factors) and short distance (Wilson coefficients) effects and these can be
written as
\begin{align}
f_{1}(q^{2}) =&4(C_{7}^{eff}+C_{7}^{\prime eff })\frac{m_{b}+m_{s}}{q^{2}}T_{1}(q^{2})+(C_{9}^{eff}+C_{9}^{\prime eff})\frac{2A_{V}(q^{2})}{M_{B_{c}}+M_{D_{s}^{\ast }}}  \label{621a}
\\
f_{2}(q^{2}) =&2(C_{7}^{eff}-C_{7}^{\prime eff})\frac{m_{b}-m_{s}}{q^{2}}T_{2}(q^{2})\left(
M_{B_{c}}^{2}-M_{D_{s}^{\ast }}^{2}\right) \notag\\ & +(C_{9}^{eff}-C_{9}^{\prime eff})A_{0}(q^{2})\left(
M_{B_{c}}+M_{D^{\ast }}\right)  \label{621b}\\
f_{3}(q^{2}) = &4(C_{7}^{eff}-C_{7}^{\prime eff})\frac{m_{b}-m_{s}}{q^{2}}\left( T_{2}(q^{2})+q^{2}%
\frac{T_{3}(q^{2})}{\left( M_{B_{c}}^{2}-M_{D_{s}^{\ast }}^{2}\right) }
\right) \notag\\ &+(C_{9}^{eff}-C_{9}^{\prime eff})\frac{A_{+}(q^{2})}{M_{B_{c}}+M_{D_{s}^{\ast }}}
\label{621c}\\
f_{4}(q^{2}) =&(C_{10}+C_{10}^{\prime})\frac{2A_{V}(q^{2})}{M_{B_{c}}+M_{D_{s}^{\ast }}}  \label{621d}
\\
f_{5}(q^{2}) =&2(C_{10}-C_{10}^{\prime})A_{0}(q^{2})\left(
M_{B_{c}}+M_{D_{s}^{\ast }}\right) \label{621e}\\
f_{6}(q^{2})
=&2(C_{10}-C_{10}^{\prime})\frac{A_{+}(q^{2})}{M_{B_{c}}+M_{D_{s}^{\ast
}}}
\label{621f} \\
f_{7}(q^{2})=&4(C_{10}-C_{10}^{\prime})\frac{A_{-}(q^{2})}{M_{B_{c}}+M_{D_{s}^{\ast }}}+(C_{Q2}-C_{Q2}^{\prime})\frac{M_{D_{s}^{\ast }}}{m(m_{b}+m_{s})} \tilde{A}_{0}(q^{2}) \label{621g}\\
f_{8}(q^{2}) =&-(C_{Q1}-C_{Q1}^{\prime})\frac{M_{D_{s}^{\ast }}}{(m_{b}+m_{s})} \tilde{A}_{0}(q^{2}) \label{621h}
\end{align}
Here the NHBs contribution are encoded in the auxiliary functions
$f_{7}$ and $f_{8}$.

\section{Physical Observables for $B_{c}\rightarrow D_{s}^{\ast} \ell^{+}\ell^{-}$}\label{po}

In this section we will present the calculations of the physical
observables such as the branching ratios ${\cal BR}$, the
forward-backward asymmetries ${\cal A}_{FB}$, the lepton
polarization asymmetries $P_{L,N,T}$ and the helicity fractions $
f_{L,T}$ of $D_{s}^{\ast }$ meson in $ B_{c}\rightarrow D_{s}^{\ast
}\ell^{+}\ell^{-}$ decay by incorporating both the weak annihilation
(WA) and the penguin amplitudes.

\subsection{The Differential Decay Rate of $B_{c}\rightarrow D_{s}^{\ast }\ell
^{+}\ell ^{-}$}

In the rest frame of $B_{c}$ meson the differential decay width of $
B_{c}\rightarrow D_{s}^{\ast }\ell^{+}\ell^{-}$ can be written as
\begin{equation}
\frac{d\Gamma (B_{c}\rightarrow D_{s}^{\ast }\ell^{+}\ell^{-})}{dq^{2}}=\frac{1%
}{\left( 2\pi \right) ^{3}}\frac{1}{32M_{B_{c}}^{3}}%
\int_{-u(q^{2})}^{+u(q^{2})}du\left\vert \mathcal{M}\right\vert ^{2}
\label{62a}
\end{equation}%
where
\begin{eqnarray}
\mathcal{M} &=&\mathcal{M}^{\text{WA}}+\mathcal{M}^{\text{PENG}} \\
q^{2} &=&(p_{l^{+}}+p_{l^{-}})^{2} \\
u &=&\left( p-p_{l^{-}}\right) ^{2}-\left( p-p_{l^{+}}\right) ^{2}
\end{eqnarray}%
Now the limits on $q^{2}$ and $u$ are
\begin{eqnarray}
4m^{2} &\leq &q^{2}\leq (M_{B_{c}}-M_{D_{s}^{\ast }})^{2}  \label{62d} \\
-u(q^{2}) &\leq &u\leq u(q^{2})  \label{62e}
\end{eqnarray}%
with%
\begin{equation}
u(q^{2})=\sqrt{\lambda \left( 1-\frac{4m^{2}}{q^{2}}\right) }
\label{62f}
\end{equation}%
and%
\begin{equation*}
\lambda \equiv \lambda (M_{B_{c}}^{2},M_{D_{s}^{\ast
}}^{2},q^{2})=M_{B_{c}}^{4}+M_{D_{s}^{\ast
}}^{4}+q^{4}-2M_{B_{c}}^{2}M_{D_{s}^{\ast }}^{2}-2M_{D_{s}^{\ast
}}^{2}q^{2}-2q^{2}M_{B_{c}}^{2}
\end{equation*}%
Here $m$ corresponds to the mass of the lepton which for our case
are the $\mu$ and $\tau$. The total decay rate for the decay
$B_{c}\rightarrow D_{s}^{\ast }\ell^{+}\ell^{-}$ can be expressed in
terms of WA, penguin amplitude and the interference of these two,
which takes the form
\begin{equation}
\frac{d\Gamma }{dq^{2}}=\frac{d\Gamma ^{\text{WA}}}{dq^{2}}+\frac{d\Gamma
^{\,\text{PENG}}}{dq^{2}}+\frac{d\Gamma ^{\text{WA-PENG}}}{dq^{2}}\label{62g}
\end{equation}%
with
\begin{eqnarray}
\frac{d\Gamma ^{\text{WA}}}{dq^{2}} &=&\frac{G_{F}^{2}\left\vert
V_{tb}V_{ts}^{\ast }\right\vert ^{2}\alpha ^{2}}{2^{11}\pi
^{5}3M_{B_{c}}^{3}M_{D_{s}^{\ast }}^{2}q^{2}}u(q^{2})\times
\mathcal{A}\left(
q^{2}\right)  \label{62h} \\
\frac{d\Gamma ^{\,\text{PENG}}}{dq^{2}} &=&\frac{G_{F}^{2}\left\vert
V_{tb}V_{ts}^{\ast }\right\vert ^{2}\alpha ^{2}}{2^{11}\pi
^{5}3M_{B_{c}}^{3}M_{D_{s}^{\ast }}^{2}q^{2}}u(q^{2})\times \mathcal{B}\left(
q^{2}\right)  \label{62i} \\
\frac{d\Gamma ^{\text{WA-PENG}}}{dq^{2}} &=&\frac{G_{F}^{2}\left\vert
V_{cb}V_{cs}^{\ast }\right\vert \left\vert V_{tb}V_{ts}^{\ast }\right\vert
\alpha ^{2}}{2^{11}\pi ^{5}3M_{B_{c}}^{3}M_{D_{s}^{\ast }}^{2}q^{2}}%
u(q^{2})\times \mathcal{C}\left( q^{2}\right) \label{62j}
\end{eqnarray}%
The function $u(q^2)$ is defined Eq. (\ref{62f}) and $\mathcal{A}(q^2)$, $\mathcal{B}(q^2)$ and $%
\mathcal{C}(q^2)$ are
\begin{align}
\mathcal{A}\left( q^{2}\right) &=\frac{1}{2}\left( 2m^{2}+q^{2}\right)\kappa^{2}\bigg[%
8\lambda M_{D_{s}^{\ast }}^{2}q^{2}\left({\cal
K}_{1}^{ann}\right)^{2}+\left({\cal
K}_{2}^{ann}\right)^{2}\bigg\{12M_{D_{s}^{\ast
}}^{2}q^{2}(\lambda +4M_{B_{c}}^{2}q^{2}) \notag \\
&+\lambda ^{2}+\lambda\left(\lambda
+4q^{2}M_{D_{s}^{\ast }}^{2}+4q^{4}\right)\bigg\}\bigg]  \label{63a}\\
\mathcal{B}(q^{2}) &=8M_{D_{s}^{\ast }}^{2}q^{2}\lambda
\bigg\{(2m^{2}+q^{2})\left\vert f_{1}(q^{2})\right\vert ^{2}
-(4m^{2}-q^{2})\left\vert f_{4}(q^{2})\right\vert ^{2}\bigg\}  +4M_{D_{s}^{\ast }}^{2}q^{2}\bigg\{(2m^{2}+q^{2})\notag \\
&\times\left( 3\left\vert f_{2}(q^{2})\right\vert ^{2}-\lambda \left\vert f_{3}(q^{2})\right\vert
^{2}\right) -(4m^{2}-q^{2})\left( 3\left\vert f_{5}(q^{2})\right\vert ^{2}-\lambda
\left\vert f_{6}(q^{2})\right\vert ^{2}\right) \bigg\}\notag \\
&+\lambda (2m^{2}+q^{2})\left\vert
f_{2}(q^{2})+\left(M_{B_{c}}^{2}-M_{D_{s}^{\ast
}}^{2}-q^{2}\right)f_{3}(q^{2})\right\vert ^{2}+ 24m^{2}M_{D_{s}^{\ast
}}^{2}\lambda\left\vert f_{7}(q^{2})\right\vert^{2} \notag \\
&-(4m^{2}-q^{2})\left\vert f_{5}(q^{2})+\left(M_{B_{c}}^{2}-M_{D_{s}^{\ast
}}^{2}-q^{2}\right)f_{6}(q^{2})\right\vert ^{2}  +\left(q^{2}-4 m^2\right) \lambda\left\vert f_{8}(q^{2})\right\vert ^{2}  \notag \\
& -12 m^{2}q^{2}\left[\Re(f_{5}f_{7}^{\ast })-\Re(f_{6}f_{7}^{\ast })\right]\label{63b}\\
\mathcal{C}(q^{2})&= 2\kappa f_{2}(q^{2}) {\cal
K}_{2}^{ann}q^{2}(2m^{2}+q^{2})\bigg\{\lambda +
6M_{D_{s}^{\ast }}^{2}\left(M_{B_{c}}^{2}-M_{D_{s}^{\ast }}^{2}+q^{2}\right)\bigg\}  \notag \\
&-\kappa\lambda\bigg[2f_{1}(q^{2}){\cal
K}_{1}^{ann}M_{D_{s}^{\ast}}^{2}q^{4} +f_{3}(q^{2}){\cal
K}_{2}^{ann}(2m^{2}+q^{2}) \left(\lambda
+q^{4}+4M_{B_{c}}M_{D_{s}^{\ast }}\right)\bigg].  \label{63c}
\end{align}
where
\begin{equation}
\kappa = \frac{8\pi^{2}M_{D_{s}^{\ast }}f_{B_{c}}f_{D_{s}^{\ast
}}}{(m_{c}^{2}-m_{s}^{2})q^{2}}. \label{kappa}
\end{equation}

\subsection{Forward-Backward Asymmetries}

The differential forward-backward asymmetry $\mathcal{A}_{FB}$ of final state lepton for the said decay can be
written as
\begin{equation}
{\frac{d\mathcal{A}_{FB}(s)}{dq^{2}}}=\int_{0}^{1}\frac{d^{2}\Gamma }{dq^{2}d\cos \theta }%
d\cos \theta -\int_{-1}^{0}\frac{d^{2}\Gamma }{dq^{2}d\cos \theta
}d\cos \theta  \label{FBformula}
\end{equation}%
From experimental point of view the normalized forward-backward
asymmetry is more useful, defined as
\begin{equation*}
\mathcal{A}_{FB}=\frac{\int_{0}^{1}\frac{d^{2}\Gamma }{dq^{2}d\cos \theta }%
d\cos \theta -\int_{-1}^{0}\frac{d^{2}\Gamma }{dq^{2}d\cos \theta
}d\cos \theta  }{\int_{-1}^{1}\frac{d^{2}\Gamma }{dq^{2}d\cos \theta
} d\cos \theta }
\end{equation*}%
The normalized $\mathcal{A}_{FB}$ for $B_{c}\rightarrow D_{s}^{\ast }\ell^{+}\ell^{-}$ can be obtained from Eq. (\ref{62a}), as%
\begin{eqnarray}
\mathcal{A}_{FB} &=& \frac{1}{d\Gamma /dq^{2}}\frac{G_{F}^{2}\alpha ^{2}}{%
2^{11}\pi ^{5}m_{B}^{3}}\left\vert V_{tb}V_{ts}^{\ast }\right\vert
^{2}q^{2}u(q^{2})\bigg\{2\kappa\Re(f_{4}{\cal K}^{ann}_{2})(M_{B_{c}}^{2}-M_{D_{s}^{\ast }}^{2}+q^{2})+2\kappa\Re(f_{5}{\cal K}^{ann}_{1}) \notag\\
&&+4{Re}[f_{2}^{\ast}f_{4}+f_{1}^{\ast }f_{5}]+2\lambda \Re[f_{3}^{\ast} f_{8}]+4\Re [f_{2}f_{8}^{\ast}] \left(-M_{B_{c}}^2+M_{D_{s}^{\ast }}^2+q^2\right) \notag\\
&&+2\kappa \Re[f_{8}]{\cal K}^{ann}_{2} \left(\lambda+\left( M_{D_{s}^{\ast }}^2+q^2\right) M_{B_{c}}^2-q^{4}\right)\bigg\}
 \label{FBA}
\end{eqnarray}%
where $\kappa$ is defined in Eq. (\ref{kappa}) and $d\Gamma /dq^{2}$
is given in Eq. (\ref{62g}).

\subsection{Lepton Polarization Asymmetries}

In the rest frame of the lepton $\ell^{-}$, the unit vectors along
longitudinal, normal and transversal component of the $\ell^{-}$ can be defined
as \cite{jam,jama,Aliev}:
\begin{subequations}
\begin{eqnarray}
s_{L}^{-\mu } &=&(0,\vec{e}^{-}_{L})=\left( 0,\frac{\vec{p}_{-}}{\left| \vec{p}%
_{-}\right| }\right) , \label{p-vectorsa} \\
s_{T}^{-\mu } &=&(0,\vec{e}^{-}_{T})=\left( 0,\frac{\vec{k} \times
\vec{p}_{-}}{\left| \vec{k}\times \vec{p}_{-}\right| }\right) ,
\label{p-vectorsb} \\
s_{N}^{-\mu } &=&(0,\vec{e}^{-}_{N})=\left( 0,\vec{e}_{T}\times \vec{e}%
_{L}\right) ,  \label{p-vectorsc}
\end{eqnarray}
\end{subequations}
where $\vec{p}_{-}$ and $\vec{k}$ are the three-momenta of the
lepton $\ell^{-}$ and $D_{s}^{*}$ meson respectively in the center mass
(c.m.) frame of $\ell^{+}\ell^{-}$ system. Lorentz transformation is used to boost
the longitudinal component of the lepton polarization to the c.m. frame of the
lepton pair as
\begin{equation}
\left( s_{L}^{-\mu }\right) _{CM}=\left( \frac{|\vec{p}_{-}|}{m},\frac{%
E\vec{p}_{-}}{m\left| \vec{p}_{-}\right| }\right)
\label{bossted component}
\end{equation}
where $E$ and $m$ are the energy and mass of the lepton. The normal
and transverse components remain unchanged under the Lorentz boost. The
longitudinal ($P_{L}$), normal ($P_{N}$) and transverse ($P_{T}$)
polarizations of lepton can be defined as:
\begin{equation}
P_{i}^{(\mp )}(q^{2})=\frac{\frac{d\Gamma }{dq^{2}}(\vec{\xi}^{\mp }=\vec{e}%
^{\mp })-\frac{d\Gamma }{dq^{2}}(\vec{\xi}^{\mp }=-\vec{e}^{\mp })}{\frac{%
d\Gamma }{dq^{2}}(\vec{\xi}^{\mp }=\vec{e}^{\mp })+\frac{d\Gamma }{dq^{2}}(%
\vec{\xi}^{\mp }=-\vec{e}^{\mp })}  \label{polarization-defination}
\end{equation}%
where $i=L,\;N,\;T$ and $\vec{\xi}^{\mp }$ is the spin direction along the
leptons $\ell^{\mp }$. The differential decay rate for polarized lepton $\ell^{\mp
}$ in $B_{c}\rightarrow D_{s}^{*}\ell^{+}\ell^{-}$ decay along any spin
direction $\vec{\xi}^{\mp }$ is related to the unpolarized decay rate (\ref{62g}) with the following relation
\begin{equation}
\frac{d\Gamma (\vec{\xi}^{\mp })}{dq^{2}}=\frac{1}{2}\left( \frac{d\Gamma }{%
dq^{2}}\right) \left[1+(P_{L}^{\mp }\vec{e}_{L}^{\mp }+P_{N}^{\mp }\vec{e}%
_{N}^{\mp }+P_{T}^{\mp }\vec{e}_{T}^{\mp })\cdot \vec{\xi}^{\mp }\right].
\label{polarized-decay}
\end{equation}%
The expressions of the numerator of longitudinal, normal and
transverse lepton polarizations can be written as
\begin{align}
P_{L}(q^2) \propto &\frac{4\lambda}{3M_{D_{s}^{\ast }}^{2}}\sqrt{\frac{q^2-4m^{2}}{q^{2}}}\times  \bigg\{\Re(f_{2}f_{5}^{\ast})\left(1+\frac{12q^2M_{D_{s}^{\ast}}^{2}}{\lambda}\right)+\lambda\Re(f_{3}f_{6}^{\ast})+8q^{2}M_{D_{s}^{\ast}}^{2}\Re(f_{1}f_{4}^{\ast})\notag\\
& +\left(-M_{B_{c}}^{2}+M_{D_{s}^{\ast }}^{2}+q^{2}\right)\left[\Re(f_{3}f_{5}^{\ast})+\Re(f_{2}f_{6}^{\ast})\right]\notag\\
&+
\frac{3}{2}m\left[\Re(f_{5}f_{8}^{\ast})+\Re(f_{6}f_{8}^{\ast})\left(-M_{B_{c}}^{2}+M_{D_{s}^{\ast
}}^{2}\right)-q^{2}\Re(f_{7}f_{8}^{\ast})\right]\notag\\
&+4M_{D_{s}^{\ast }}^{2}\kappa
\Re({\cal K}^{ann}_{1}f_{4}^{\ast})+\left[\lambda+q^{2}(M_{B}^{2}+M_{D_{s}^{\ast }}^{2}-q^{2})\right]\kappa\Re({\cal K}^{ann}_{2}f_{6}^{\ast})\notag\\
&+\left[M_{D_{s}^{\ast }}^{2}-M_{B_{c}}^{2}-\frac{6q^{2}M_{D_{s}^{\ast }}^{2}}{\lambda}(q^{2}-M_{D_{s}^{\ast }}^{2}+M_{B_{c}}^{2})\right]\kappa\Re({\cal K}^{ann}_{2}f_{2}^{\ast})\bigg\}\label{long-polarization}\\
P_{T}(q^2) \propto & \frac{m\pi}{M_{D_{s}^{\ast }}^{2}}\sqrt{\frac{\lambda}{q^{2}}}\times \bigg\{\lambda q^{2}\Re(f_{3}f_{7}^{\ast})+\lambda(M_{B_{c}}^{2}-M_{D_{s}^{\ast }}^{2})\Re(f_{3}f_{6}^{\ast})-\lambda\Re(f_{3}f_{5}^{\ast})\notag\\
&+\left(-M_{B_{c}}^{2}+M_{D_{s}^{\ast }}^{2}+q^{2}\right)\left[q^{2}\Re(f_{2}f_{7}^{\ast})+(M_{B_{c}}^{2}-M_{D_{s}^{\ast }}^{2}-q^{2})\Re(f_{2}f_{5}^{\ast})+\frac{(q^{2}-4m^2)}{2m}\Re(f_{5}f_{8}^{\ast})\right]\notag\\
& + 8q^{2}M_{D_{s}^{\ast }}^{2}\Re(f_{1}f_{2}^{\ast})
+\frac{\lambda}{2m}(q^{2}-4m^2)\Re(f_{6}f_{8}^{\ast})+\left[\lambda-q^{2}\left(3M_{D_{s}^{\ast
}}^{2}+M_{B_{c}}^{2}+q^{2}\right)\right]\Re(f_{2}f_{6}^{\ast})\notag\\
&-4q^{2}M_{D_{s}^{\ast }}^{2}\kappa\Re({\cal K}^{ann}_{1}f_{2}^{\ast})-4q^{2}M_{D_{s}^{\ast }}^{2}(M_{B_{c}}^{2}-M_{D_{s}^{\ast }}^{2}+q^{2})\kappa\Re({\cal K}^{ann}_{2}f_{1}^{\ast})\notag\\
&-\left[\lambda+q^{2}\left(M_{B_{c}}^{2}+M_{D_{s}^{\ast}}^{2}-q^{2}\right)\right]\kappa\left(\Re({\cal K}^{ann}_{2}f_{5}^{\ast})-(M_{B_{c}}^{2}-M_{D_{s}^{\ast}}^{2})\Re({\cal K}^{ann}_{2}f_{6}^{\ast})+q^{2}\Re({\cal K}^{ann}_{2}f_{7}^{\ast})\right)\notag\\
&+4\kappa^{2}q^{2}M_{D_{s}^{\ast
}}^{2}\left(M_{B_{c}}^{2}-M_{D_{s}^{\ast }}^{2}+q^{2}\right){\cal
K}_{1}^{ann}{\cal K}_{2}^{ann}\bigg\}  \label{norm-polarization}\\
P_{N}\left( q^{2}\right)  \propto & i\frac{m\pi \sqrt{ \left( q^{2}-4m_{l}^{2}\right) \lambda }}{M_{D_{s}^{\ast }}^{2}}\bigg\{M_{D_{s}^{\ast}}^{2}\left[4\Im(f_{2}f_{4}^{\ast})+4\Im(f_{1}f_{5}^{\ast})+2\kappa\Im(f_{5}{\cal K}_{1}^{ann})\right]\notag\\
&+2M_{D_{s}^{\ast}}^{2}\kappa\Im(f_{4}{\cal K}_{2}^{ann})\left(M_{B_{c}}^{2}-M_{D_{s}^{\ast }}^{2}+q^{2}\right)-\lambda\Im(f_{6}f_{7}^{\ast})-\frac{\left(\lambda+q^{2}(M_{B_{c}}^{2}+M_{D_{s}^{\ast}}^{2}-q^{2})\right)}{2m}\kappa\Im({\cal K}_{2}^{ann}f_{8}^{\ast})\notag\\
&+\left(-M_{B_{c}}^{2}+M_{D_{s}^{\ast}}^{2}+q^{2}\right)\bigg[\Im(f_{7}f_{5}^{\ast})+\frac{1}{2m}\Im(f_{2}f_{8}^{\ast})\bigg]+\left(M_{B_{c}}^{2}+3M_{D_{s}^{\ast}}^{2}-q^{2}\right)\Im(f_{5}f_{6}^{\ast})+\frac{\lambda}{2m}\Im(f_{3}f_{8}^{\ast})\bigg\}
\label{Transverse-polarization}
\end{align}%
where $\kappa$ is defined in Eq. \eqref{kappa} along with auxiliary
functions $f_{1},f_{2},\cdots,f_{8}$ in Eqs.
(\ref{621a}-\ref{621h}). Here we have dropped out the constant
factors which are however understood.

\subsection{Helicity Fractions of $D_{s}^{\ast }$ in $B_{c}\rightarrow
D_{s}^{\ast }\ell^{+}\ell^{-}$}

We now discuss helicity fractions of $D_{s}^{\ast }$ in $B_{c}\rightarrow
D_{s}^{\ast }\ell^{+}\ell^{-}$ which are interesting variable and are as such
independent of the uncertainties arising due to form factors and other input
parameters. The final state meson helicity fractions were already discussed
in literature for $B\rightarrow K^{\ast }\left( K_{1}\right) \ell^{+}\ell^{-}$
decays \cite{22,23,aa,aa1}. Even for the $K^{\ast }$ vector meson, the
longitudinal helicity fraction $f_{L}$ has been measured by Babar
collaboration for the decay $B\rightarrow K^{\ast }\ell^{+}\ell^{-}(l=e,\mu )$ in
two bins of momentum transfer and the results are \cite{babar}
\begin{eqnarray}
f_{L} &=&0.77_{-0.30}^{+0.63}\pm 0.07 , \ \ \ \ \ 0.1\leq q^{2}\leq 8.41
GeV^{2}  \notag \\
&&  \label{64} \\
f_{L} &=&0.51_{-0.25}^{+0.22}\pm 0.08 , \ \ \ \ \ q^{2}\geq 10.24GeV^{2}
\notag
\end{eqnarray}%
while the average value of $f_{L}$ in full $q^{2}$ range is
\begin{equation}
f_{L}=0.63_{-0.19}^{+0.18}\pm 0.05 , \ \ q^{2}\geq 0.1 GeV^{2}  \label{65}
\end{equation}

The explicit expression of the decay rate for $B_{c}^{-}\rightarrow
D_{s}^{\ast -}\ell^{+}\ell^{-}$ decay can be written in terms of longitudinal $%
\Gamma_{L}$ and transverse components $\Gamma_{T}$ as
\begin{eqnarray}
\frac{d\Gamma _{L}(q^{2})}{dq^{2}} &=&\frac{d\Gamma _{L}^{\text{WA}}(q^{2})}{%
dq^{2}}+\frac{d\Gamma _{L}^{\text{PENG}}(q^{2})}{dq^{2}}+\frac{d\Gamma _{L}^{%
\text{WA-PENG}}(q^{2})}{{dq^{2}}} \\
\frac{d\Gamma _{\pm }(q^{2})}{dq^{2}} &=&\frac{d\Gamma _{\pm }^{\text{WA}%
}(q^{2})}{dq^{2}}+\frac{d\Gamma _{\pm }^{\text{PENG}}(q^{2})}{dq^{2}}+\frac{%
d\Gamma _{\pm }^{\text{WA-PENG}}(q^{2})}{{dq^{2}}}  \label{65b} \\
\frac{d\Gamma _{T }(q^{2})}{dq^{2}} &=&\frac{d\Gamma _{+}(q^{2})}{dq^{2}}+%
\frac{d\Gamma _{- }(q^{2})}{dq^{2}}.
\end{eqnarray}%
where
\begin{eqnarray}
\frac{d\Gamma _{L}^{\text{WA}}(q^{2})}{dq^{2}} &=&\frac{G_{F}^{2}\left\vert
V_{cb}V_{cs}^{\ast }\right\vert ^{2}\alpha ^{2}}{2^{11}\pi ^{5}}\frac{%
u(q^{2})}{M_{B_{c}}^{3}}\times \frac{1}{3}{\cal A}_{L}^{\text{WA}}  \label{65c} \\
\frac{d\Gamma _{L}^{\text{PENG}}(q^{2})}{dq^{2}} &=&\frac{%
G_{F}^{2}\left\vert V_{tb}V_{ts}^{\ast }\right\vert ^{2}\alpha ^{2}}{%
2^{11}\pi ^{5}}\frac{u(q^{2})}{M_{B_{c}}^{3}}\times \frac{1}{3}{\cal A}_{L}^{\text{%
PENG}}  \label{65d} \\
\frac{d\Gamma _{L}^{\text{WA-PENG}}(q^{2})}{dq^{2}} &=&\frac{%
G_{F}^{2}\left\vert V_{cb}V_{cs}^{\ast }\right\vert \left\vert
V_{tb}V_{ts}^{\ast }\right\vert \alpha ^{2}}{2^{11}\pi ^{5}}\frac{u(q^{2})}{%
M_{B_{c}}^{3}}\times \frac{1}{3}{\cal A}_{L}^{\text{WA-PENG}}  \label{65e} \\
\frac{d\Gamma _{\pm }^{\text{WA}}(q^{2})}{dq^{2}} &=&\frac{%
G_{F}^{2}\left\vert V_{cb}V_{cs}^{\ast }\right\vert ^{2}\alpha ^{2}}{%
2^{11}\pi ^{5}}\frac{u(q^{2})}{M_{B_{c}}^{3}}\times \frac{2}{3}{\cal A}_{\pm }^{%
\text{WA}}  \label{65f} \\
\frac{d\Gamma _{\pm }^{\text{PENG}}(q^{2})}{dq^{2}} &=&\frac{%
G_{F}^{2}\left\vert V_{tb}V_{ts}^{\ast }\right\vert ^{2}\alpha ^{2}}{%
2^{11}\pi ^{5}}\frac{u(q^{2})}{M_{B_{c}}^{3}}\times \frac{4}{3}{\cal A}_{\pm }^{%
\text{PENG}}  \label{65g} \\
\frac{d\Gamma _{\pm }^{\text{WA-PENG}}(q^{2})}{dq^{2}} &=&\frac{%
G_{F}^{2}\left\vert V_{cb}V_{cs}^{\ast }\right\vert \left\vert
V_{tb}V_{ts}^{\ast }\right\vert \alpha ^{2}}{2^{11}\pi ^{5}}\frac{u(q^{2})}{%
M_{B_{c}}^{3}}\times \frac{2}{3}{\cal A}_{\pm }^{\text{WA-EP}}.  \label{65h}
\end{eqnarray}%

The different functions appearing in Eqs. (\ref{65c}-\ref{65h}) can be expressed in
terms of auxiliary functions (cf. Eqs. (\ref{621a}-\ref{621h})) as
\begin{align}
{\cal A}_{L}^{\text{WA}} &=\frac{\kappa^{2}}{4 q^{2} M_{D_{s}^{\ast }}^{2}}%
\bigg[\left( {\cal K}^{ann}_{1}(q^{2})\right)^{2}
\bigg\{q^{2}\lambda(\lambda +4q^{2}M_{D_{s}^{\ast
}}^{2})-4M^{2}\lambda(2\lambda+8q^{2}M_{D_{s}^{\ast }}^{2})\notag\\
&-q^{2} \left(M_{B_{c}}^{2}-M_{D_{s}^{\ast
}}^{2}-q^{2}\right)^{2}\left(\lambda-2u^{2}(q^{2})\right)\bigg\}
+\left( {\cal K}^{ann}_{2}(q^{2})\right)^{2}\bigg\{-\lambda^{2}(q^{2}-4m^{2})\notag\\
&+ 12\lambda
q^{2}((M_{B_{c}}^{2}-M_{D_{s}^{\ast}}^{2})^{2}-M_{D_{s}^{\ast}}^{2})
+q^{2}(8q^{2}M_{D_{s}^{\ast }}^{2}-\lambda
)(M_{B_{c}}^{2}-M_{D_{s}^{\ast }}^{2}+q^{2})^{2} \notag\\
&-2u^{2}(q^{2})q^{2}((M_{B_{c}}^{2}
-M_{D_{s}^{\ast}}^{2})^{2}+q^{4})
+4m^{2}((M_{B_{c}}^{2}-M_{D_{s}^{\ast
}}^{2})^{2}-q^{4})^{2}\bigg\}\bigg]  \label{65i}
\end{align}
\begin{align}{\cal A}_{L}^{\text{PENG}}
&=\frac{1}{q^{2}M_{D_{s}^{\ast }}^{2}}\bigg[24\left\vert
f_{7}(q^{2})\right\vert ^{2}m^{2}M_{D_{s}^{\ast }}^{2}\lambda
+(2m^{2}+q^{2})\left\vert (M_{B_{c}}^{2}-M_{D_{s}^{\ast
}}^{2}-q^{2})f_{2}(q^{2})+\lambda f_{3}(q^{2})\right\vert ^{2}  \notag \\
&+(q^{2}-4m^{2})\left\vert (M_{B_{c}}^{2}-M_{D_{s}^{\ast
}}^{2}-q^{2})f_{5}(q^{2})+\lambda f_{6}(q^{2})\right\vert ^{2}\bigg]+\frac{1}{2}(q^2-4m^2)\lambda \left\vert f_{8}\right\vert ^{2}  \label{65j}\\
{\cal A}_{L}^{\text{WA-PENG}} &=\frac{\kappa}{ q^{2} M_{D_{s}^{\ast
}}^{2}}\bigg[Re(f_{1}\left(q^{2}\right){\cal
K}^{ann}_{1}(q^{2}))\bigg\{(\lambda +4M_{D_{s}^{\ast
}}^{2}q^{2})\left(8m^{2}\sqrt{\lambda}+q^{2}(2u(q^{2})
-\sqrt{\lambda})\right) \notag \\
&-4M_{D_{s}^{\ast }}^{2}q^{2}\lambda \bigg\}
+Re(f_{2}\left(q^{2}\right){\cal
K}^{ann}_{2}(q^{2}))\bigg\{q^{2}u^{2}(q^{2})(M_{B_{c}}^{2}-M_{D_{s}^{\ast
}}^{2}-q^{2})\notag \\
&+6q^{2}\lambda(M_{D_{s}^{\ast
}}^{2}-M_{B_{c}}^{2})+q^{2}(\lambda-8q^{2}M_{D_{s}^{\ast}}^{2})(M_{B_{c}}^{2}-M_{D_{s}^{\ast
}}^{2}+q^{2})-4m^{2}q^{2}(4q^{2}M_{D_{s}^{\ast }}^{2}+\lambda) \bigg\}\notag \\
& +Re(f_{3}\left(q^{2}\right){\cal
K}^{ann}_{2}(q^{2}))\bigg\{\lambda^{2}(4m^{2}-q^{2})+q^{4}(q^{2}u(q^{2})\sqrt{\lambda}
-6\lambda(M_{B_{c}}^{2}+M_{D_{s}^{\ast
}}^{2}))\notag \\
&+q^{2}(M_{B_{c}}^{2}-M_{D_{s}^{\ast
}}^{2})(6\lambda-u^{2}(q^{2})) \bigg\}\bigg]  \label{65k} \\
A_{\pm }^{\text{WA }} &=\kappa^{2}\bigg[\left( 2m^{2}+q^{2}\right)
\bigg[\lambda \left( {\cal
K}^{ann}_{1}(q^{2})\right)^{2}+\left({\cal
K}^{ann}_{2}(q^{2})\right) ^{2}\left(
\lambda+4M_{D^{\ast}_{s}}^{2}q^{2}\right)\bigg]\bigg] \label{65l}\\
A_{\pm}^{\text{PENG}} &=(q^{2}-4m_{l}^{2})\left\vert
f_{5}(q^{2})\mp\sqrt{\lambda }f_{4}(q^{2})\right\vert ^{2}+\left(
q^{2}+2m_{l}^{2}\right) \left\vert f_{2}(q^{2})\pm\sqrt{\lambda
}f_{1}(q^{2})\right\vert ^{2} \label{65m}\\
A_{\pm }^{\text{WA -PENG}}
&=-\kappa\bigg\{2\sqrt{\lambda}(q^{2}-4m^{2})
Re(f_{2}\left(q^{2}\right){\cal
K}^{ann}_{1}(q^{2}))+4\lambda(q^{2}+2m^{2})
Re(f_{1}\left(q^{2}\right){\cal K}^{ann}_{1}(q^{2})) \nonumber \\
& \pm2(q^{2}+2m^2)(M_{B_{c}}^{2}-M_{D_{s}^{\ast
}}^{2}+q^{2})[2Re[(f_{1}\left(q^{2}\right){\cal
K}^{ann}_{2}(q^{2}))]\sqrt{\lambda}\mp2Re{(f_{2}\left(q^{2}\right){\cal
K}^{ann}_{1}(q^{2}))}]\bigg\} \label{65r}
\end{align}

Finally the longitudinal and transverse helicity amplitude becomes
\begin{eqnarray}
f_{L}(q^{2}) &=&\frac{d\Gamma _{L}(q^{2})/dq^{2}}{d\Gamma (q^{2})/dq^{2}}
\notag \\
f_{\pm }(q^{2}) &=&\frac{d\Gamma _{\pm }(q^{2})/dq^{2}}{d\Gamma
(q^{2})/dq^{2}}  \notag \\
f_{T}(q^{2}) &=&f_{+}(q^{2})+f_{-}(q^{2})
\end{eqnarray}%
so that \ the sum of the longitudinal and transverse helicity amplitudes is
equal to one i.e. $f_{L}(q^{2})+f_{T}(q^{2})=1$ for each value of $q^{2}$%
\cite{22}.

As we have disscussed in the Introduction that the WA contribution plays a cruicial role in the $B_c\rightarrow D^{\ast}_s \ell^{+} \ell^{-}$. This
feature can also be seen via the expressions of different
observables for example branching ratio which is given in eq. (29), there are three terms
$\mathcal{A}(q^{2})$, $\mathcal{B}(q^{2})$ and $\mathcal{C}(q^{2})$ correspond to WA, Penguin and
the cross term of Penguin and WA, although the WA contribution
$\mathcal{A}(q^{2})$ trying to suppress the effects of NP but the term
$\mathcal{C}(q^{2})$ some how compensate and enhanced the NP effects. Same is
the case for the other calculated observables. Therefore, the
interference term between the NP and WA play a cruicial role and
make the NP effects distinct form the SM.

\section{Numerical Results and Discussion}\label{num}

We present here our numerical results of the branching ratio
($\mathcal{BR}$), the forward-backward asymmetry
$(\mathcal{A}_{FB})$, lepton polarizations asymmetries $P_{L,N,T}$
and the helicity fractions ($f_{L,T})$ of $D_{s}^{\ast }$ for the
$B_{c}\rightarrow D_{s}^{\ast }\ell^{+}\ell^{-}$ decays with
$\ell=\mu,\tau$. The numerical values of the input parameters which
are used in the subsequent analysis are summarized in Table
\ref{input}:
\begin{table}[ht]
\centering
\begin{tabular}{l}
\hline\hline
$m_{B_{c}}=6.277$ GeV, $m_{D_{s}^{\ast}}=2.1123$ GeV, $m_{b}=4.28$ GeV, $m_{\mu}=0.105$ GeV\\
$m_{\tau}=1.77$ GeV, $|V_{cb}V_{cs}^{\ast}|=4.15\times
10^{-2}$, $|V_{tb}V_{ts}^{\ast}|=4.1\times
10^{-2}$, $\tau_{B}=0.453\times 10^{-12}$ sec\\
$G_{F}=1.17\times 10^{-5}$ GeV$^{-2}$, $\alpha^{-1}=137$, $f_B$ = 0.35 GeV, $f_D$ = 0.30 GeV\\
\hline\hline
\end{tabular}
\caption{Values of input parameters used in our numerical analysis \cite{pdg}.}\label{input}
\end{table}
\begin{table}[ht]
\centering
\begin{tabular}{lccc}
\hline\hline
Wilson coefficients & $C_{7}^{eff}$& $C_{9}^{eff}$& $C_{10}$\\
\hline
SM & $-0.313$& $4.334$&  $-4.669$\\
SUSY I  & $+0.376$&  $4.767$&  $-3.735$\\
SUSY II & $+0.376$&  $4.767$&  $-3.735$\\
SUSY III & $-0.376$&  $4.767$&  $-3.735$\\
SUSY SO(10) ($A_{0}=-1000$) & $-0.219$& $4.275$&  $-4.732$\\
\hline\hline
\end{tabular}
\caption{Wilson Coefficients in SM and different SUSY models but
without NHBs contributions \cite{j11,j14,jam,jama}.}\label{susy1}
\end{table}
\begin{table}[ht]
\centering
\begin{tabular}{lccc}
\hline\hline
Wilson coefficients & $C_{7}^{\prime eff}$&  $C_{9}^{\prime eff}$& $C_{10}^{\prime}$\\
\hline
SM or SUSY I,II,III &  $0$& $0$&  $0$\\
SUSY SO(10) ($A_{0}=-1000$) & $0.039+0.038i$&  $0.011+0.072i$& $-0.075-0.67i$\\
\hline\hline
\end{tabular}
\caption{Primed Wilson Coefficients in SM and different SUSY models but without NHBs contributions. Where the primed Wilson coefficients corresponds to the operators which are opposite in helicities from those of the SM operators.}\label{susy2}
\end{table}
\begin{table}[ht]
\centering
\begin{tabular}{lcccc}
\hline\hline
Wilson coeff. & $C_{Q_{1}}$&  $C_{Q_{1}}^{\prime}$& $C_{Q_{2}}$& $C_{Q_{2}}^{\prime}$\\
\hline
SM &  $0$& $0$&  $0$& $0$\\
SUSY I &  $0$& $0$&  $0$& $0$\\
SUSY II & $6.5(16.5)$&  $0$&  $-6.5(-16.5)$&0\\
SUSY III & $1.2(4.5)$&  $0$&  $-1.2(-4.5)$&$0$\\
SUSY SO(10) & $0.106+0i$& $-0.247+0.242i$&  $-0.107+0i$&$-0.25+0.246i$\\
($A_{0}=-1000$) & $(1.775+0.002i)$& $(-4.148+4.074i)$&  $(-1.797-0.002i)$&$(-4.202+4.128i)$\\
\hline\hline
\end{tabular}
\caption{Wilson coefficient corresponding to NHBs contributions in different SUSY scenarios. Where the primed Wilson coefficients are for the primed operators from NHBs contribution in SUSY SO(10) GUT model. The values in the parentheses are for the $\tau$ lepton case.}\label{susy3}
\end{table}

The values of Wilson coefficients used in our numerical analysis are
taken from Refs. \cite{j11,j14,jam,jama} and summarized in Tables
\ref{susy1}, \ref{susy2} and \ref{susy3}. In the following analysis,
we will focus on the parameter space of large $\tan\beta$, where the
NHBs effects are important owing to the fact that the Wilson
coefficients corresponding to NHBs are proportional to
$(m_{b}m/m^{2}_{h})\tan^{3}\beta$, with $h=h^{0},A^{0}$. Here, the
$\tan\beta$ contributes from the chargino-up-type squark loop and
the $\tan^{2}\beta$ appears from the exchange of the NHBs. In the
Ref. \cite{nhb,nhba} it is pointed out that at large value of
$\tan\beta$ the $C_{Q_{i}}^{(\prime)}$ compete with
$C_{i}^{(\prime)}$ and can outpace $C_{i}^{(\prime)}$ in some
region. Depending on the magnitude and sign of the SUSY parameters
one can think of many options in the parameter space, but
experimental results i.e., the decay rate of $b \to s\gamma$ and $b
\to s \ell^+\ell^-$ restrict us to consider the following scenarios
for MSSM:
\begin{itemize}
\item SUSY I: refers to the regions where SUSY destructively contributes and changes the sign of $C_{7}^{eff}$, which will have drastic effects on the observables, but without contribution of
NHBs.
\item SUSY II: corresponds to the region where $\tan\beta$ is large and the masses of the sparticles are relative small.
\item SUSY III: points to the regions where $\tan\beta$ is large and the masses of superpartners are also relatively large, i.e. $\geq$ 450 GeV$^2$.
\end{itemize}
In SUSY I and SUSY II scenarios one can accommodate the non-zero
crossing of the forward-backward asymmetry in $B\to
K^{\ast}\mu^{+}\mu^{-}$ decay \cite{L9,L10,L11,L12}. Since the
primed Wilson coefficients are for the primed operators in Eq. (2.2)
which appear in the SUSY SO(10) GUT model. In Table \ref{susy3} it
is mentioned that the values of $C_{Q_{i}}^{(\prime)}$ is different
for the different choice of final state lepton, which is due to the
fact that contributions from the NHBs are proportional to the lepton
mass.

The other two conditions, apart from the large $\tan\beta$ limit,
which are responsible for the dominant contributions from NHBs are:
(i) the mass values of the lightest chargino and lightest stop
should not be too large and (ii) the mass splitting of charginos and
stops should be relatively large, which also indicate large mixing
between stop sector and chargino sector as pointed out in Ref.
\cite{j11}. Since the SUSY effects are very sensitive to the sign of
the Higgs mass term and SUSY contributes destructively appears when
the sign of this term becomes minus. In Ref. \cite{j11} it is shown
that there exist some regions of SUSY parameter space where NHBs can
dominantly contribute to the FCNC process $b \to s \ell^+\ell^-$ due
to flip of the sign of $C_7$ from positive to negative, within the
constraint on $b \to s\gamma$. Also it is noticed that when the
masses of sparticles are relatively large, say about 450 GeV, there
exits considerable regions in the SUSY parameter space where NHBs
could contribute dominantly as in the case of SUSY III and the SUSY
SO(10) models. However, in these scenarios the sign of $C_7$ remains
unaltered with respect to the SM sign because of the cancellation of
the contributions of charged Higgs and charginos with each other.

It is worth mentioning here that the decay $B_q \to \ell^+\ell^-$ is
a clean channel to probe for the NHBs effects in SUSY models at
large $\tan\beta$. Since its branching ratio in the SUSY models can
be written as \cite{j57}
\begin{eqnarray}
{\cal BR}(B_s \to \mu^+\mu^-)&=& \frac{G_{F}^{3}\alpha^{2}}{64\pi^{3}}m^{3}_{B_{s}}\tau_{B_{s}}f^{2}_{B_{s}}\left\vert V_{tb}^{\ast}V_{ts}\right\vert^{2}\sqrt{1-\frac{4m^{2}}{m_{B_{s}}^{2}}}\notag\\
&&\times\left[\left(1-\frac{4m^{2}}{m_{B_{s}}^{2}}\right)C_{Q_{1}}^{2}+\left(C_{Q_2}+\frac{2m}{m_{B_{s}}}C_{10}\right)^{2}\right],\label{brmm}
\end{eqnarray}
where $C_{Q_1}$ and $C_{Q_2}$ are referring to the NHBs contributions which are absent in the SM but in MSSM they are proportional to $(m_{b}m/m^{2}_{h})\tan^{3}\beta$, with $h=h^{0},A^{0}$. The value of $C_{10}$ is large in the SM but it is suppressed by the factor $2m/m_{B_s}$, therefore, the corresponding SM value of branching ratio is \cite{recentcdf}
\begin{equation}
{\cal BR}_{SM}(B_s \to \mu^+\mu^-)=(3.19\pm0.19)\times10^{-9}.
\end{equation}
The branching ratio corresponding to different values of the NHBs parameters can be enhanced by a factor ranging from $10^1-10^3$, but the recent upper bound by CDF collaborations, $3.9 \times 10^{-8}$ at $95\%$ confidence level \cite{recentcdf}, is about 12 times larger than that of SM prediction. It can be noticed from Eq. \eqref{brmm} that the branching ratio for $B_s \to \mu^+\mu^-$ is directly proportional to the NHBs contributions. Therefore, this stringently constraints the parameter space of the SUSY models and especially the large value of the Wilson Coefficients $C_{Q_1}$ and $C_{Q_2}$ are severely constrained. Therefore, the precise measurement of this decay at the future experiments will help us to get useful constraints on the SUSY parameters and if considerably large deviation from SM predication is measured, then along with the signal of supersymmetry, this would have important implications on the Higgs searches at LHC.

The purpose of the present study of the SUSY effects in $B_{c}\to
D_{s}^{\ast}\ell^{+}\ell^{-}$, with $\ell=\mu,\tau$, is to
incorporate the constraints provided by $B_s \to \mu^+\mu^-$ as well
as to see the effects of NHBs at the larger extent in these FCNC
processes. Since this decay mode is unique in a sense that it
contains WA contributions along with the penguin and so it is
interesting to see how SUSY affects different physical observables
in this process. For our numerical analysis we have set the values
of $C_{Q_1}$ and $C_{Q_2}$ between 0 to 6.5(16.5) for muons and
(tauons) to check the dependency of different observables on these
NHBs contributions. We hope that in the future experiments such as
the Tevaran and the LHC, with the more data on these decays, will
help us to test more precisely the constraints obtained from the
decay channel $B_s \to\mu^{+}\mu^{-}$.

\begin{figure}[ht]
\begin{tabular}{cc}
\hspace{0.5cm}($\mathbf{a}$)&\hspace{1.2cm}($\mathbf{b}$)\\
\includegraphics[scale=0.51]{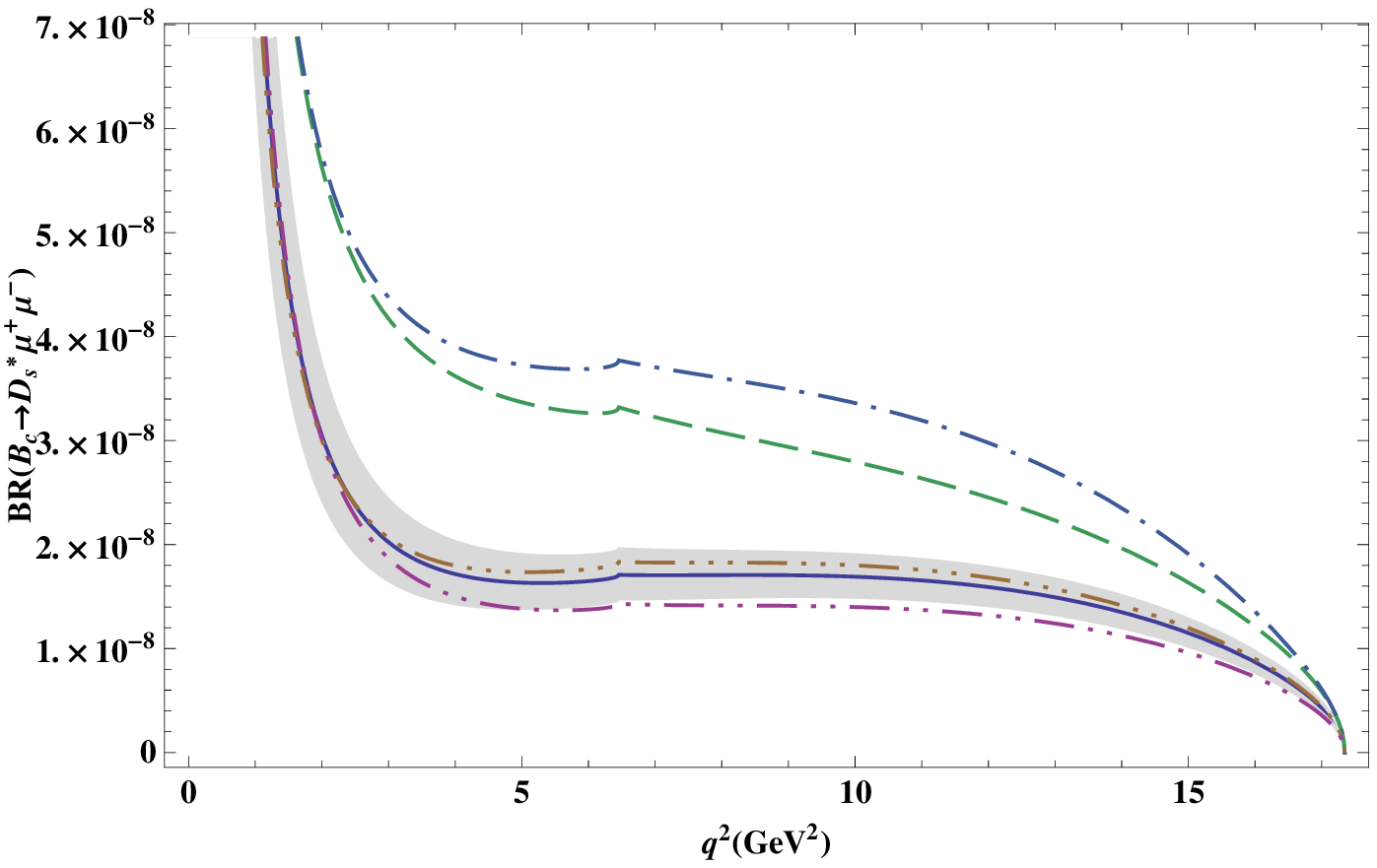}&\includegraphics[scale=0.51]{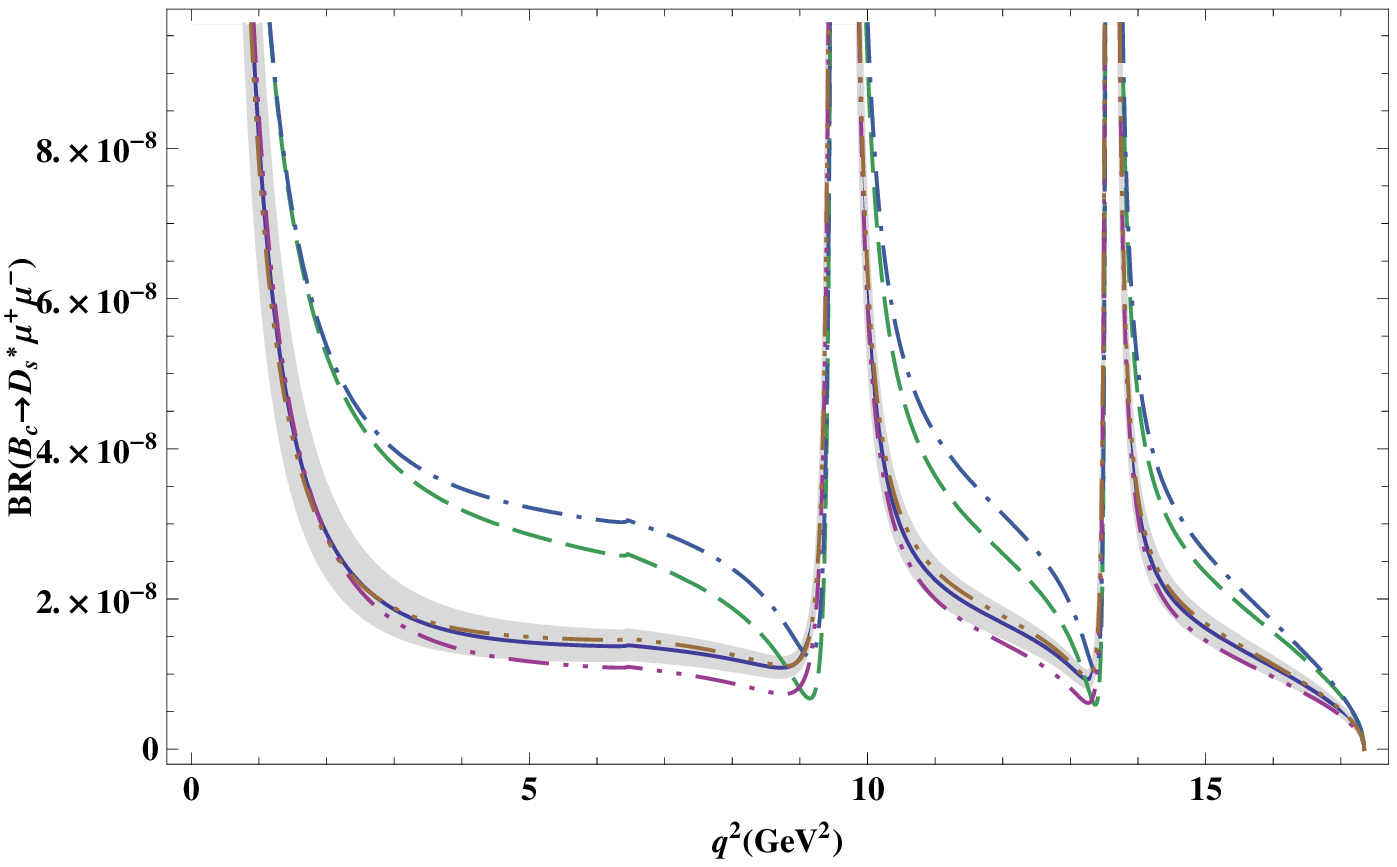}\\
\hspace{0.5cm}($\mathbf{c}$)&\hspace{1.2cm}($\mathbf{d}$)\\
\includegraphics[scale=0.51]{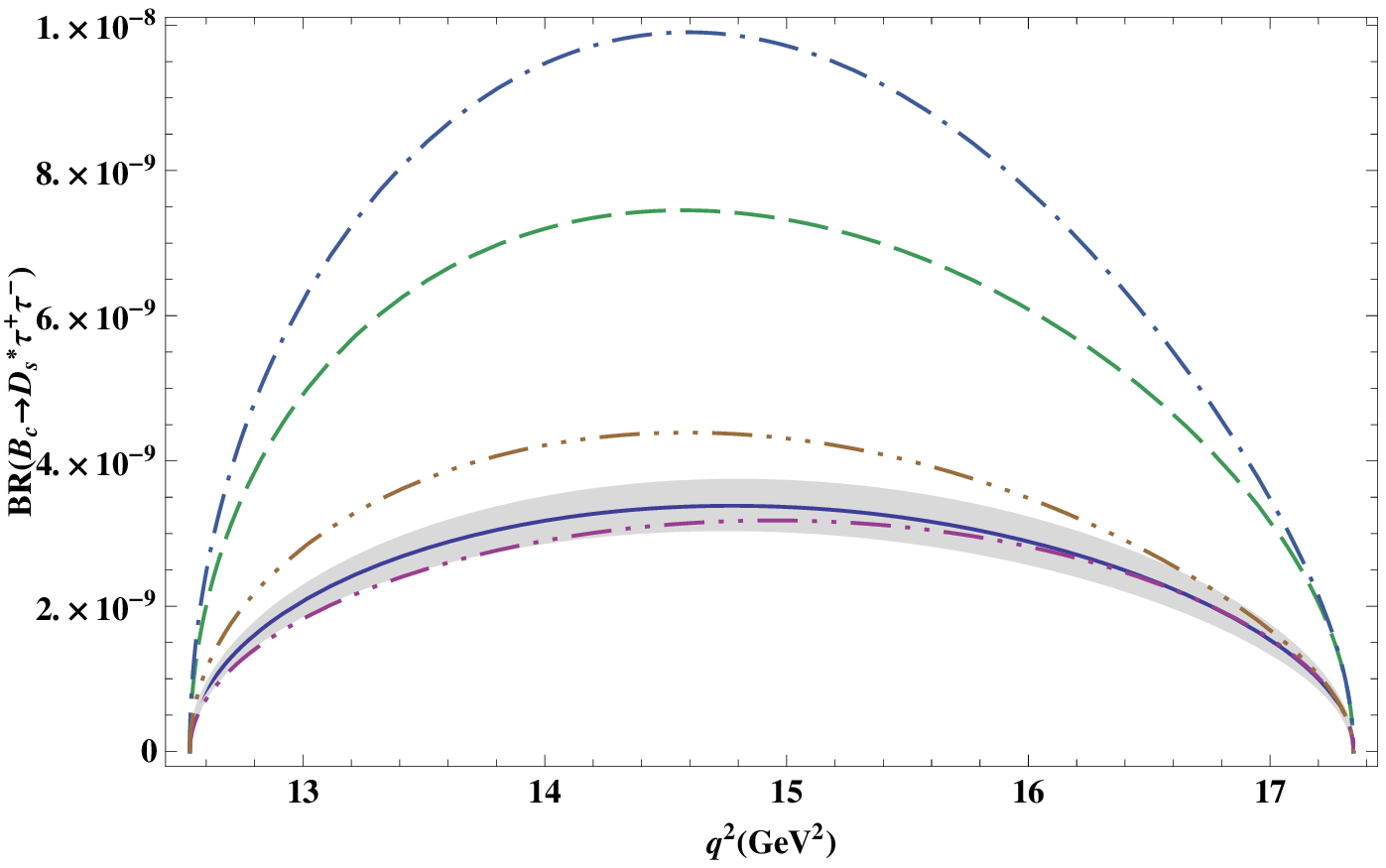}&\includegraphics[scale=0.51]{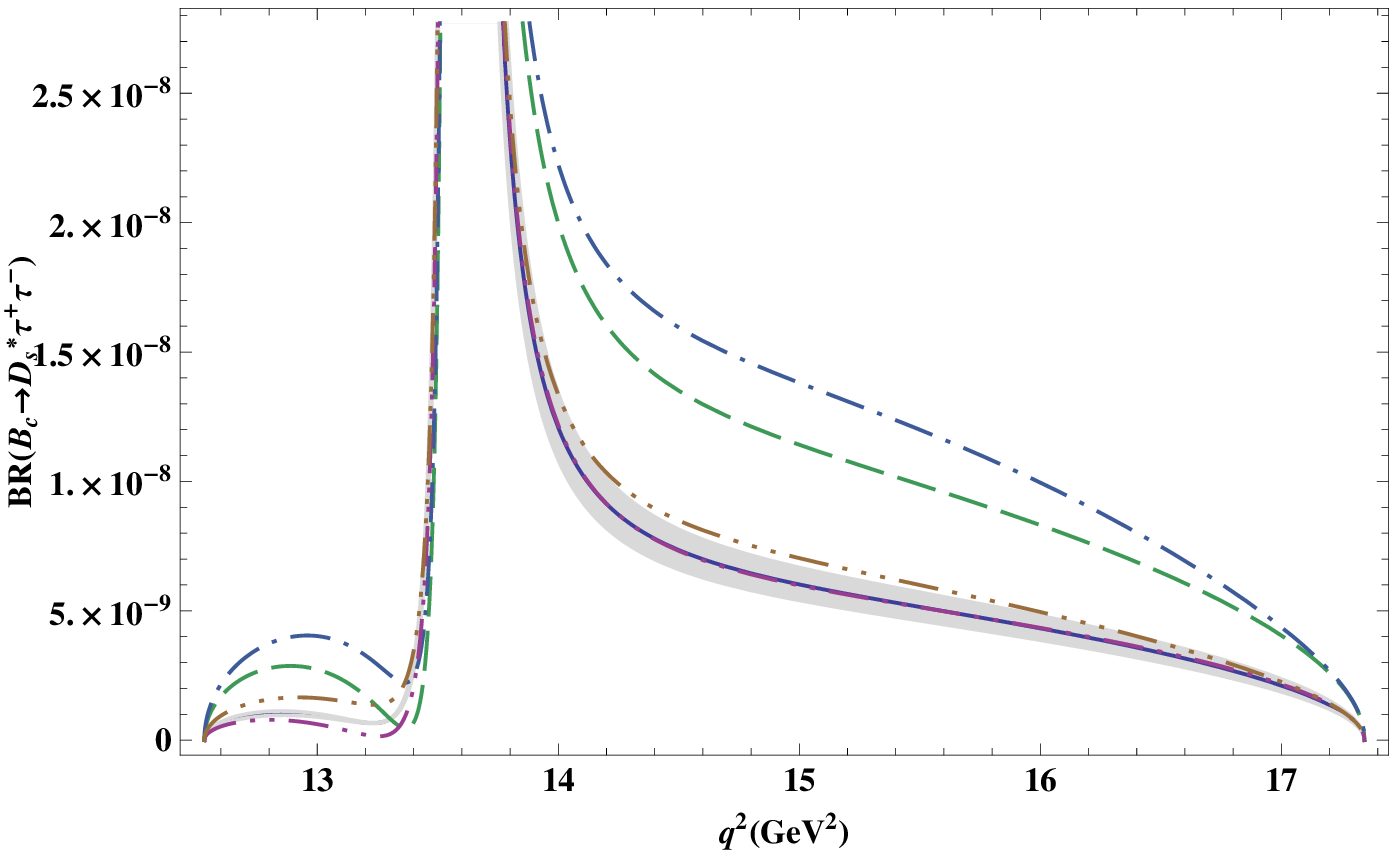}\end{tabular}
\caption{The dependence of branching ratio of $B_{c}\to D_{s}^{\ast}\ell^{+}\ell^{-}$ $(\ell=\mu,\tau)$ on $q^{2}$ without long-distance contributions (a,c) and with long-distance contributions (b,d) for different scenarios of MSSM and SUSY SO(10) GUT model. In all the graphs, the solid, dashed, dashed-dot, dashed-double dot and dashed-triple dot curves correspond to the SM, SUSY I, SUSY II, SUSY III and SUSY SO(10) GUT model, respectively.} \label{br}
\end{figure}

First, we discuss the branching ratios (${\cal BR}$) for the decays
$B_{c}\to D_{s}^{\ast}\ell^{+}\ell^{-}$, with $\ell=\mu,\tau$, which
we have plotted as a function of $q^2$ (GeV$^{2})$ in Fig. \ref{br},
without and with long-distance contributions in the Wilson
coefficients, both in the SM and in the SUSY scenarios. This figure
depict that the values of $\mathcal{BR}$s, both for the case of
muons and tauons as final state leptons, get sizeably influenced due
to the SUSY effects which come through the new parameters, i.e the
modified and new Wilson coefficients corresponding to the operators
described in Eq. (2.2). One can see clearly from these graphs that
the increment in the values of ${\cal BR}$s in the SUSY I model is
due to the relative change of the sign of $C_{7}^{eff}$ with respect
to that of the SM; while the large deviation in SUSY II model from
the SM values is mainly due to the contributions of NHBs and due to
the relative change of the sign of $C_{7}^{eff}$. Since the values
of Wilson Coefficients corresponding to NHBs is small for the cases
of SUSY III and SUSY SO(10) GUT models, so one should expect small
deviation in the ${\cal BR}$s from the SM values and this signature
is clear in Fig. \ref{br}. Moreover, the NP effects due to the SUSY,
manifest in the $\mathcal{BR}$s throughout the $q^{2}$ region
irrespective of the mass of the final state leptons. In addition,
one can also extract the constructive behavior of SUSY I and SUSY II
to the $\mathcal{BR}$ from Table \ref{brtable}. Furthermore, we have
also plotted the ${\cal BR}$s with the long-distance contribution
corresponding to the SM and SUSY models in Fig. \ref{br}(b,d).

\begin{table}[ht]
\centering
\begin{tabular}{lcccc}
\hline\hline
Branching & $B_{c}\to D_{s}^{\ast}\mu^{+}\mu^{-}$&  $B_{c}\to D_{s}^{\ast}\mu^{+}\mu^{-}$& $B_{c}\to D_{s}^{\ast}\tau^{+}\tau^{-}$& $B_{c}\to D_{s}^{\ast}\mu^{+}\mu^{-}$\\
ratios & Without LD & With LD & Without LD & With LD \\
\hline
SM &  $1.60\times10^{-6}$& $2.81\times10^{-5}$&  $1.33\times10^{-8}$& $1.19\times10^{-6}$\\
SUSY I &  $1.91\times10^{-6}$& $2.85\times10^{-5}$&  $3.06\times10^{-8}$& $1.21\times10^{-6}$\\
SUSY II & $2.29\times10^{-6}$&  $2.89\times10^{-5}$&  $2.89\times10^{-8}$& $1.22\times10^{-6}$\\
SUSY III & $1.58\times10^{-6}$&  $2.81\times10^{-5}$&  $1.35\times10^{-8}$&$1.19\times10^{-6}$\\
SUSY SO(10) & $1.60\times10^{-6}$& $2.81\times10^{-5}$&  $1.52\times10^{-8}$&$1.19\times10^{-6}$\\
\hline\hline
\end{tabular}
\caption{Branching ratio for $B_{c}\to D_{s}^{\ast}\ell^{+}\ell^{-}$ ($\ell=\mu,\tau$) in the SM and different SUSY models.}\label{brtable}
\end{table}

It is important to mention that as an exclusive decay, there are
different sources of uncertainties involved in the analysis of the
above mentioned decay. The major source of uncertainties in the
numerical analysis of $B_{c}\to D_{s}^{\ast}\ell^{+}\ell^{-}$
($\ell=\mu,\tau$) decays originate from the $B_{c}\to D_{s}^{\ast}$
transition form factors calculated in the QCD sum rule approach
\cite{53a} as summarized in Table \ref{formfactor}. But it is also
important to stress that these hadronic uncertainties have almost no
influence on the various asymmetries including the forward-backward
asymmetries, lepton polarization asymmetries and helicity fractions
of $D_{s}^{\ast}$ in the decays $B_{c}\to
D_{s}^{\ast}\ell^{+}\ell^{-}$ because of the cancellation among
different polarization states and this make them a good tool to
probe beyond the SM.

\begin{figure}[ht]
\begin{tabular}{cc}
\hspace{0.5cm}($\mathbf{a}$)&\hspace{1.2cm}($\mathbf{b}$)\\
\includegraphics[scale=0.51]{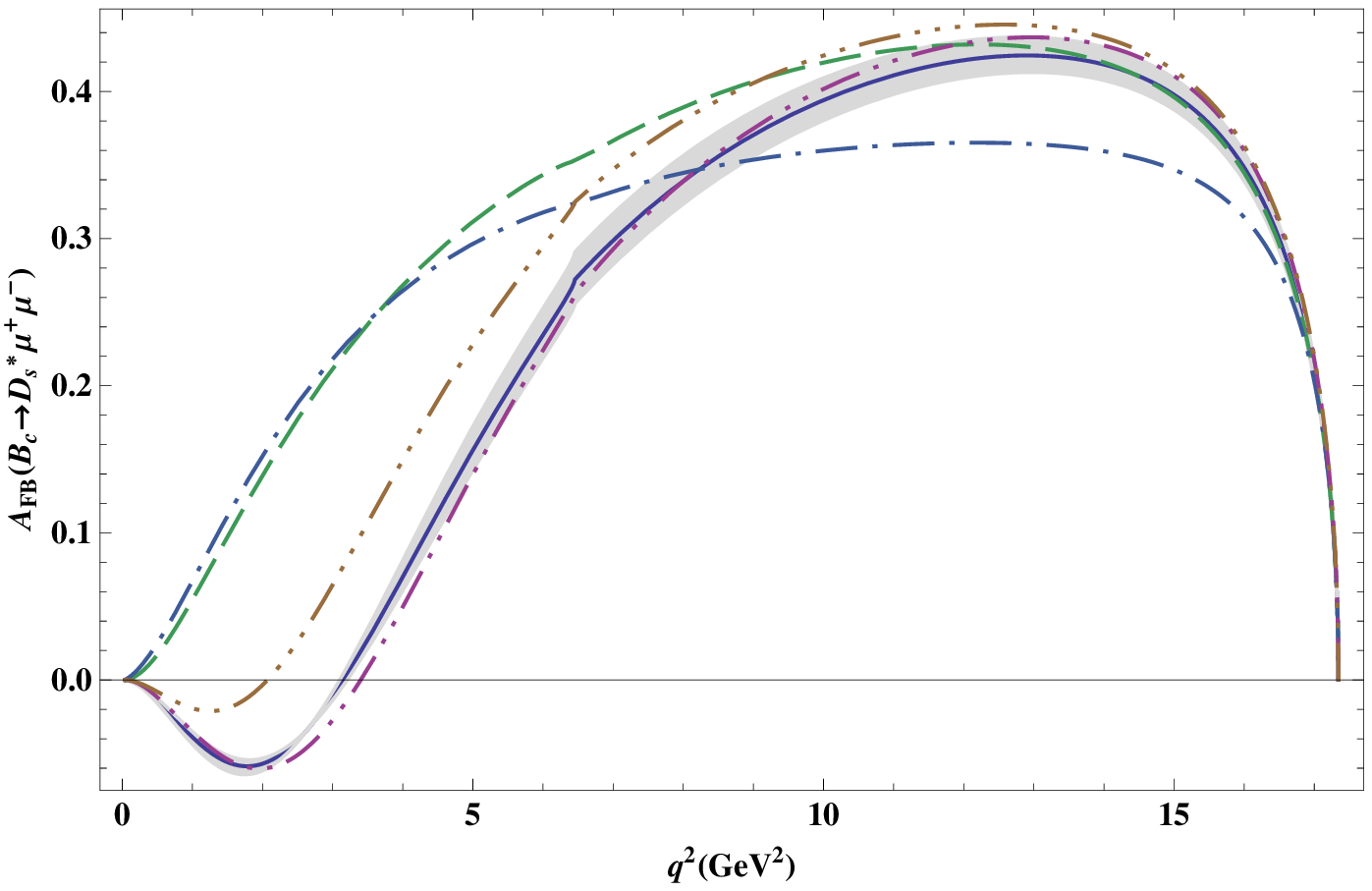}&\includegraphics[scale=0.51]{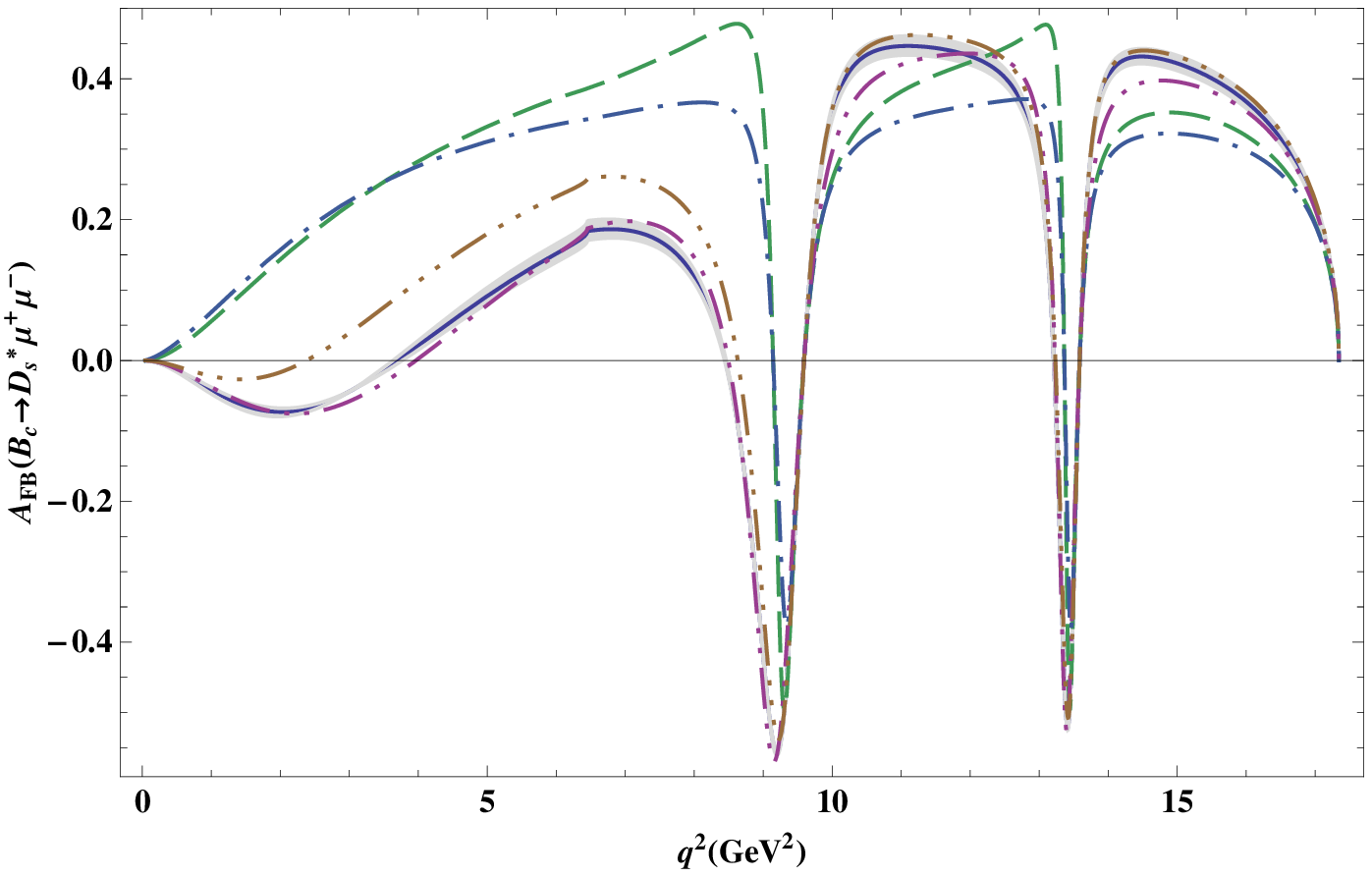}\\
\hspace{0.5cm}($\mathbf{c}$)&\hspace{1.2cm}($\mathbf{d}$)\\
\includegraphics[scale=0.51]{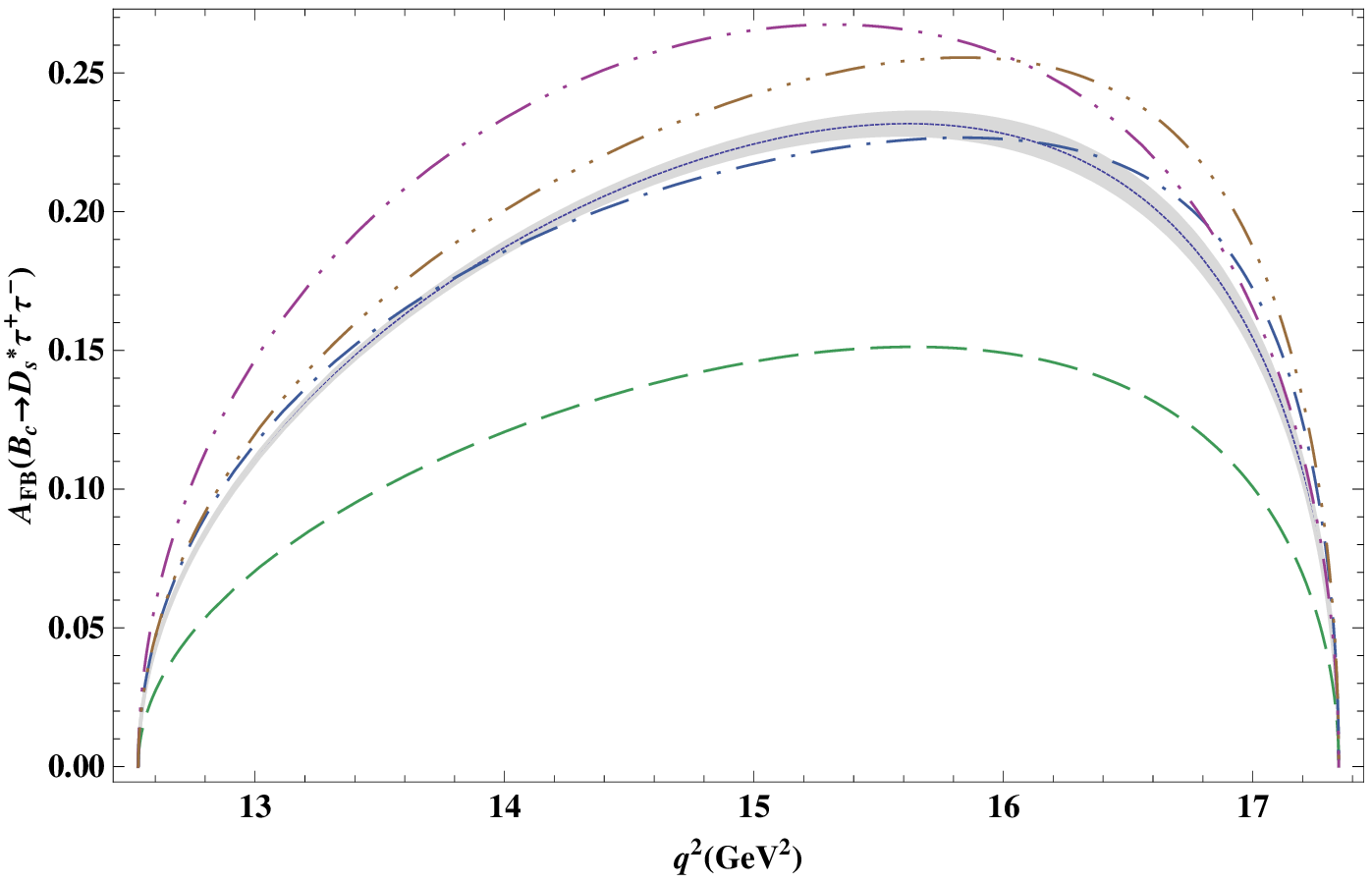}&\includegraphics[scale=0.51]{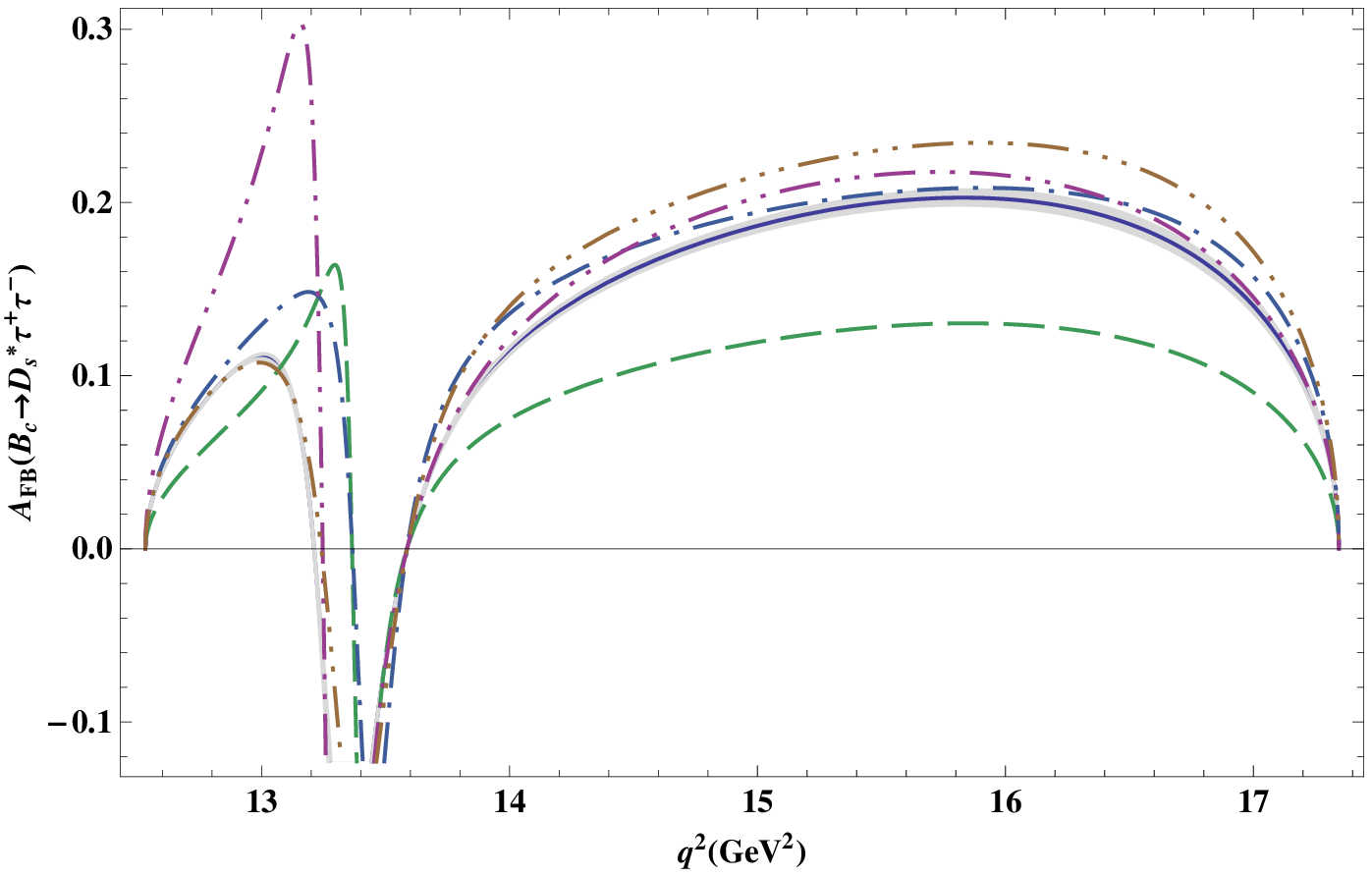}\end{tabular}
\caption{The dependence of forward-backward asymmetry for the decays
$B_{c}\to D_{s}^{\ast}\ell^{+}\ell^{-}$ $(\ell=\mu,\tau)$ on $q^{2}$
without long-distance contributions (a,c) and with long-distance
contributions (b,d) for different scenarios of MSSM and SUSY SO(10)
GUT model.} \label{fbamt}
\end{figure}

To illustrate the generic effects due to the SUSY models on the
forward-backward asymmetry $\mathcal{A}_{FB}$, we plot
$\frac{d(\mathcal{A}_{FB})}{dq^{2}}$ as a function of $q^{2}$ in
Fig. \ref{fbamt}.
For the zero position of $\mathcal{A}_{FB}$ it is
also argued that the uncertainty in the zero position of the
$\mathcal{A}_{FB}$ due to the hadronic uncertainties is negligible
\cite{aali} to leading order in $\alpha_{s}$. Therefore, the zero
position of the $\mathcal{A}_{FB}$ can serve as a stringent test for
the NP effects including the SUSY. Figures \ref{fbamt}(a) and
\ref{fbamt}(b) describe the ${\cal A}_{FB}$ for $B_{c}\to
D_{s}^{\ast}\mu^{+}\mu^{-}$ with and without long-distance
contributions in the Wilson coefficients respectively, where the
different SUSY models show clear deviation from the SM curve. Figure
\ref{fbamt}(a) clearly shows that SUSY I and SUSY II do not cross
the zero of the ${\cal A}_{FB}$ because of the fact that the zero
crossing of ${\cal A}_{FB}$ of SM is solely due to the opposite
signs of $C_{7}^{eff}$ and $C_{9}^{eff}$ but in SUSY I and SUSY II
the signs of these coefficients are same and hence the zero point of
${\cal A}_{FB}$ disappears in both cases. Furthermore, in the SUSY
III model, due to the opposite sign of $C^{eff}_7$ and $C^{eff}_9$,
the ${\cal A}_{FB}$ passes from the zero but the zero position
shifts mildly to the right (about $0.2$ GeV$^2$) from that of the SM
value, i.e 3.3 GeV$^2$, due to the contribution from the NHBs.
Moreover, the SUSY SO(10) GUT model also shows the similar behavior
except that the zero position of the ${\cal A}_{FB}$ shifts
considerably to the left (about $1$ GeV$^2$) from that of the SM
value. Hence, the precise measurement of the zero position of
$\mathcal{A}_{FB}$ for the decay $B_{c}\to
D_{s}^{\ast}\mu^{+}\mu^{-}$ will be a very good observable to yield
any indirect imprints of NP due to SUSY models and can serve as a
good tool to distinguish among the different variants of SUSY
models.

Besides the zero position of $\mathcal{A}_{FB}$, the magnitude of
$\mathcal{A}_{FB}$ is also an important tool to establish NP.
Particularly, this con be more exciting when the tauons are the
final state leptons, where the zero of the $\mathcal{A}_{FB}$ is
absent. In the Figs. \ref{fbamt}(c) and \ref{fbamt}(d) the ${\cal
A}_{FB}$ of $B_{c}\to D_{s}^{\ast}\tau^{+}\tau^{-}$ is plotted with
and without long-distance contributions, respectively, where the
different SUSY models present a distinct variations from that of the
SM magnitude. A closer look on the pattern of Fig. \ref{fbamt}(c)
indicates that the SUSY I and SUSY II models decrease the magnitude
of $\mathcal{A}_{FB}$ from its SM value. Whereas, SUSY III has very
mild deviation from that of the SM behavior but SUSY SO(10) GUT
model has an increment in the magnitude of $\mathcal{A}_{FB}$
compared to that of the SM value for the case where tauons are in
the final state. It is valuable to comment here that just like the
zero position of the $\mathcal{A}_{FB}$, the magnitude of
$\mathcal{A}_{FB}$ depends on the values of the Wilson coefficients
$C_{7}^{eff}, C_{9}^{eff}, C_{10}$ and $C_{Q_i}^{(\prime)}$ and the
effects due to the hadronic uncertainties on the magnitude of
$\mathcal{A}_{FB}$ are almost insensitive. Hence, the deviation in
the magnitude of $\mathcal{A}_{FB}$ due to the SUSY parameters is
prominent and can be measured at the experiments which indeed help
us to understand the constraints on the parameter space of the SUSY.

\begin{figure}[ht]
\begin{tabular}{cc}
\hspace{.5cm}($\mathbf{a}$)&\hspace{1.2cm}($\mathbf{b}$)\\
\includegraphics[scale=0.51]{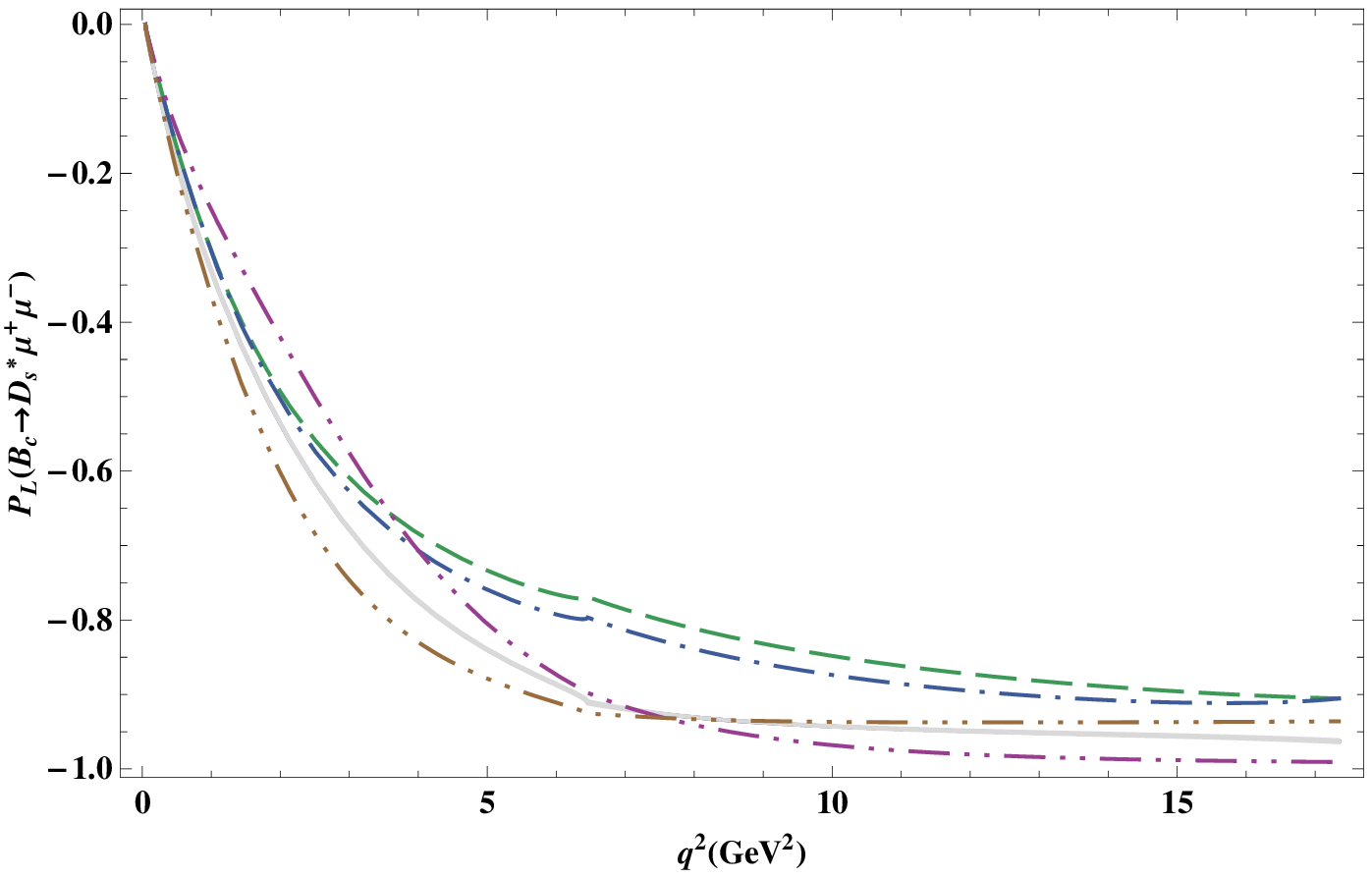}&\includegraphics[scale=0.51]{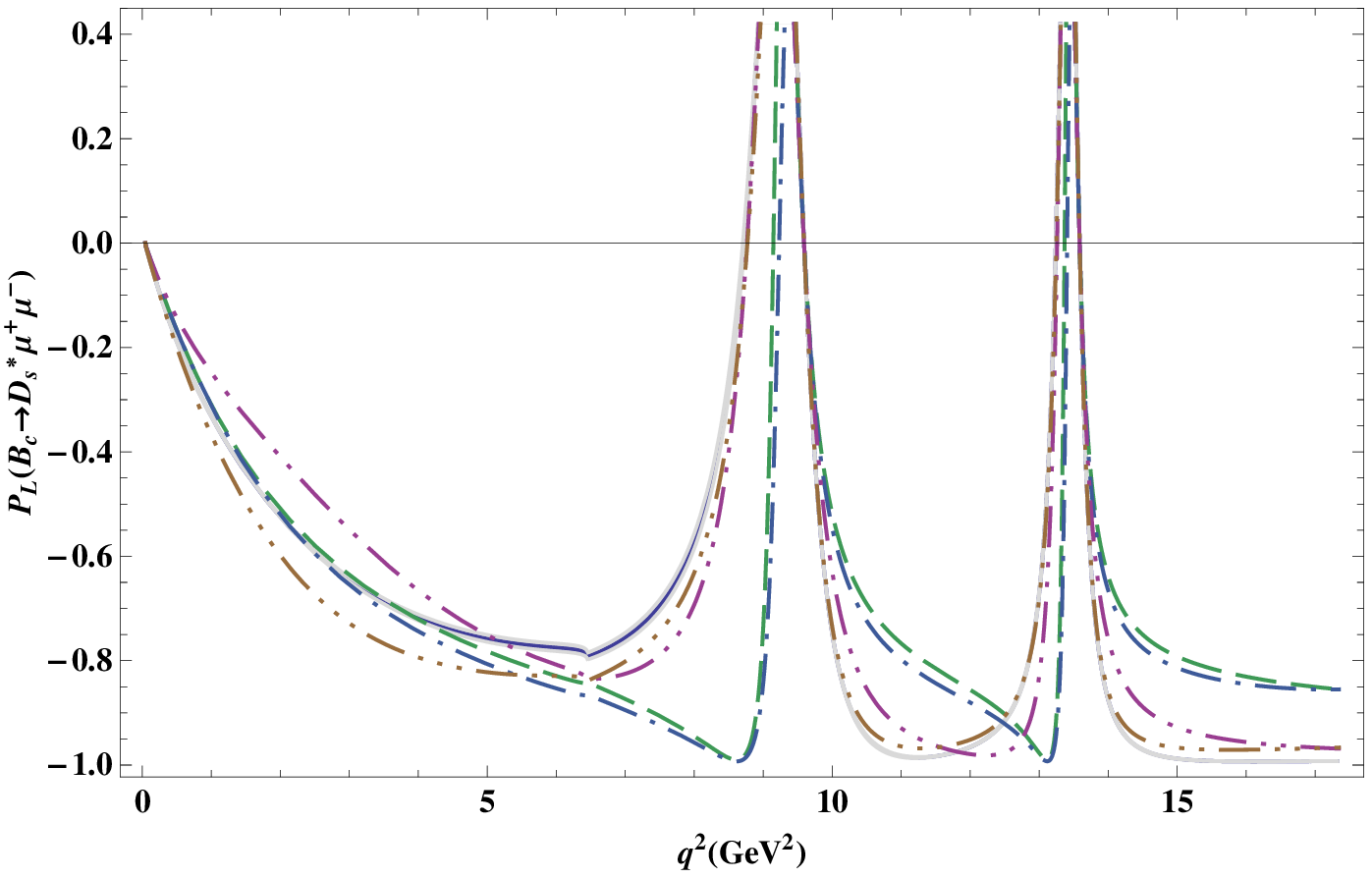}\\
\hspace{.5cm}($\mathbf{c}$)&\hspace{1.2cm}($\mathbf{d}$)\\
\includegraphics[scale=0.51]{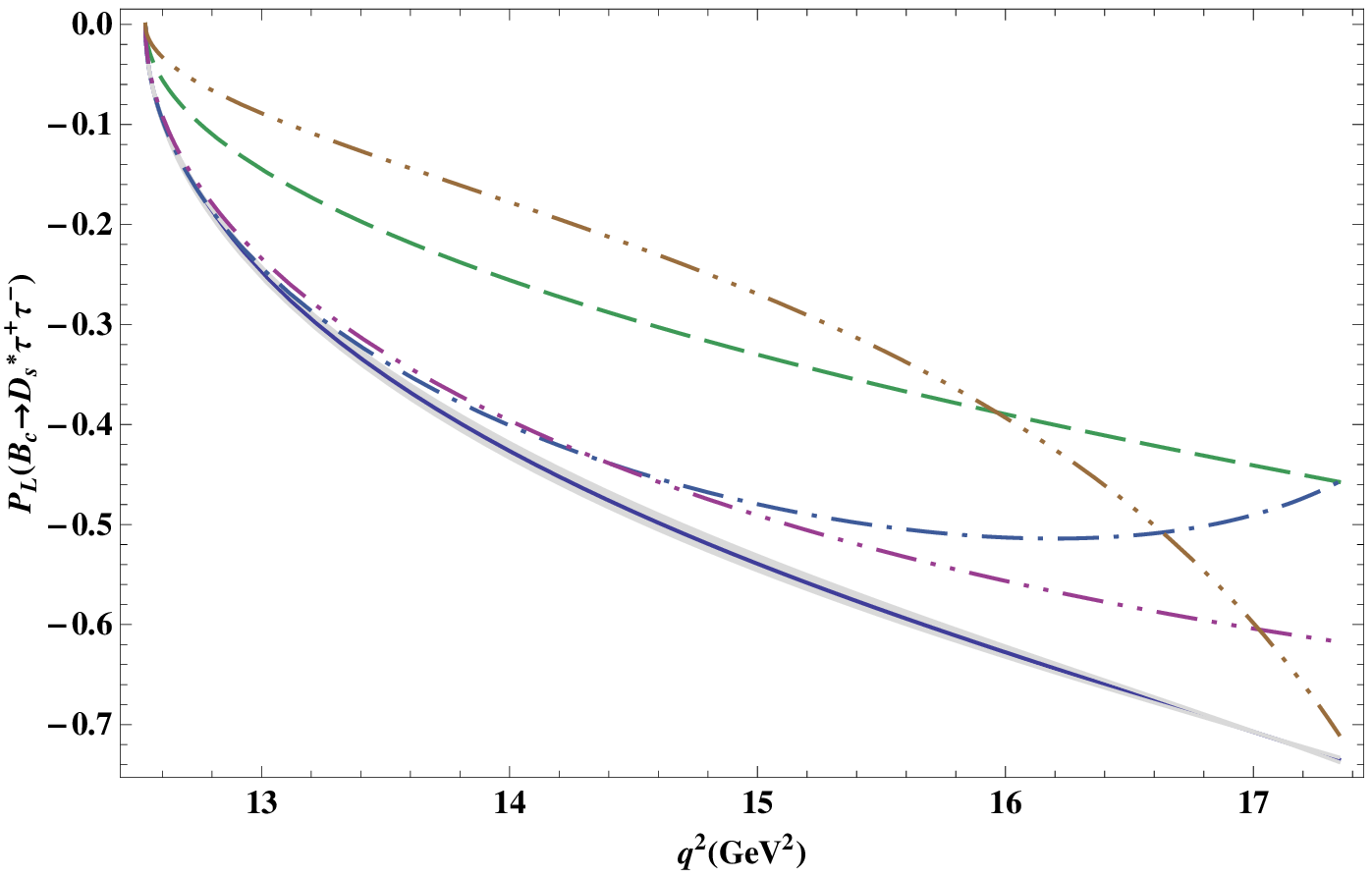}&\includegraphics[scale=0.51]{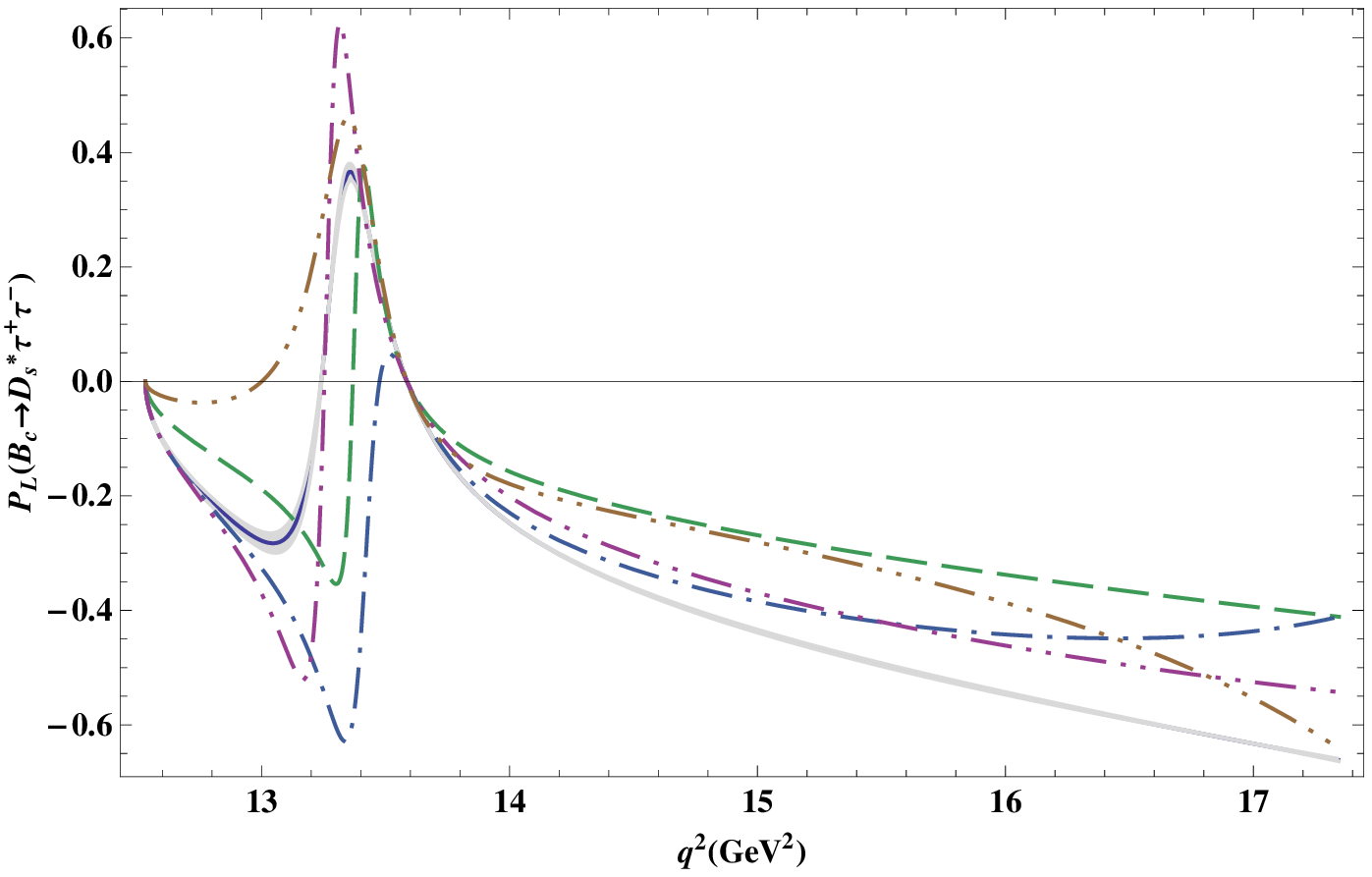}
\end{tabular}
\caption{The dependence the probabilities of the longitudinal lepton
polarization asymmetries, $P_{L}$, for the decays $B_{c}\to
D_{s}^{\ast}\ell^{+}\ell^{-}$ $(\ell=\mu,\tau)$ on $q^{2}$ without
long-distance contributions (a,c) and with long-distance
contributions (b,d), respectively, for different scenarios of MSSM
and SUSY SO(10) GUT model.} \label{lp}
\end{figure}

\begin{figure}[ht]
\begin{tabular}{cc}
\hspace{.5cm}($\mathbf{a}$)&\hspace{1.2cm}($\mathbf{b}$)\\
\includegraphics[scale=0.51]{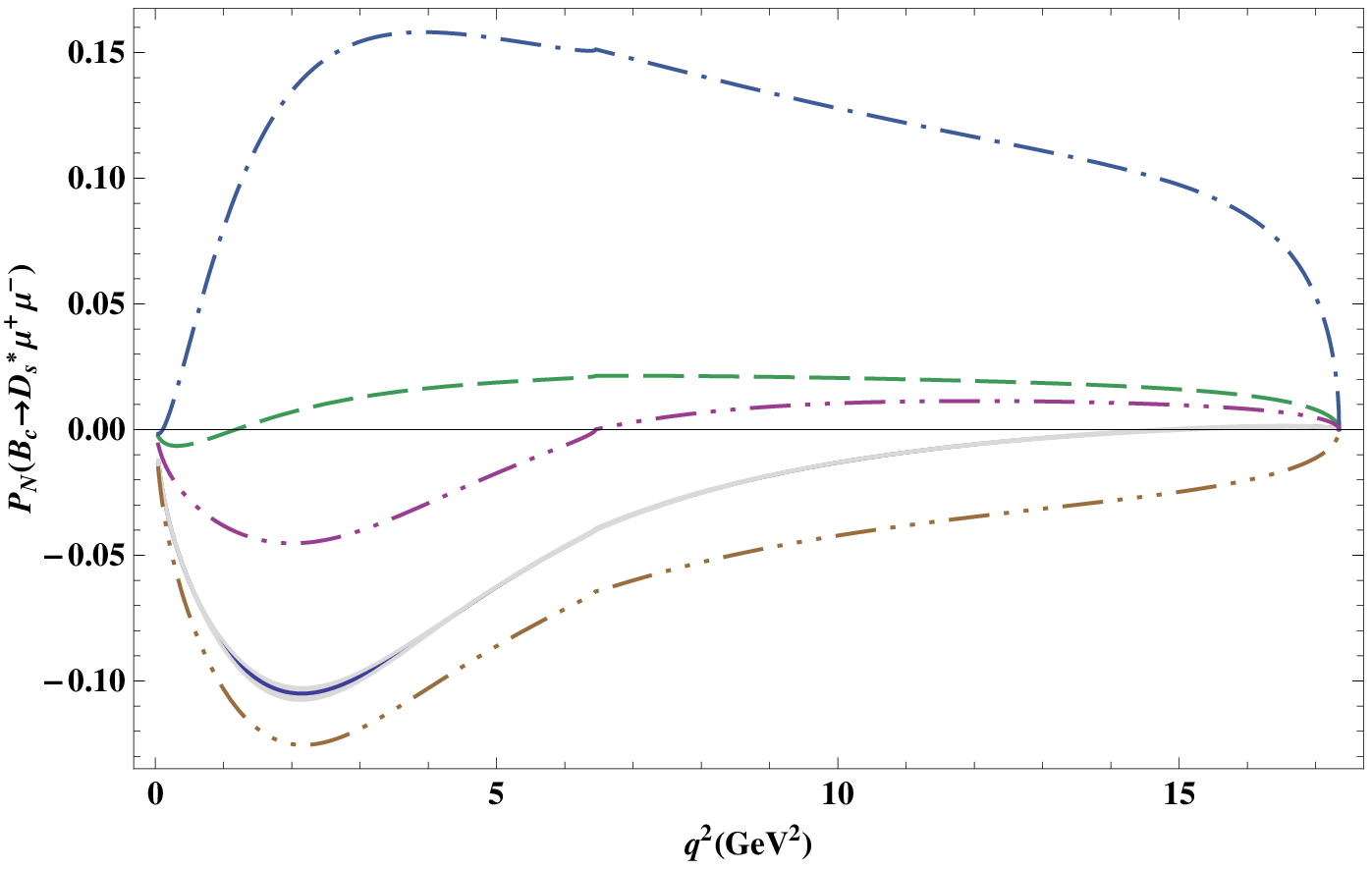}&\includegraphics[scale=0.51]{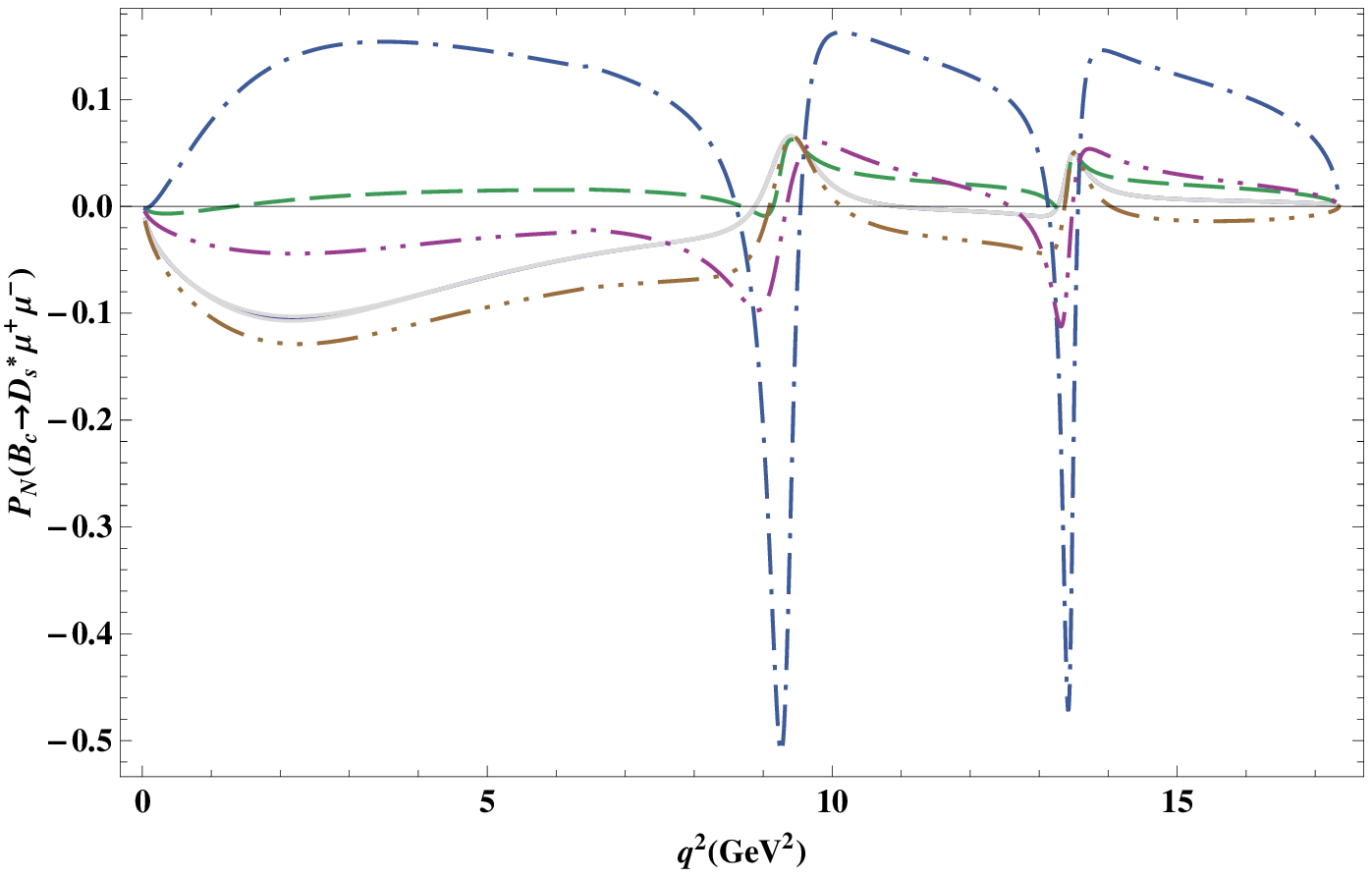}\\
\hspace{.5cm}($\mathbf{c}$)&\hspace{1.2cm}($\mathbf{d}$)\\
\includegraphics[scale=0.51]{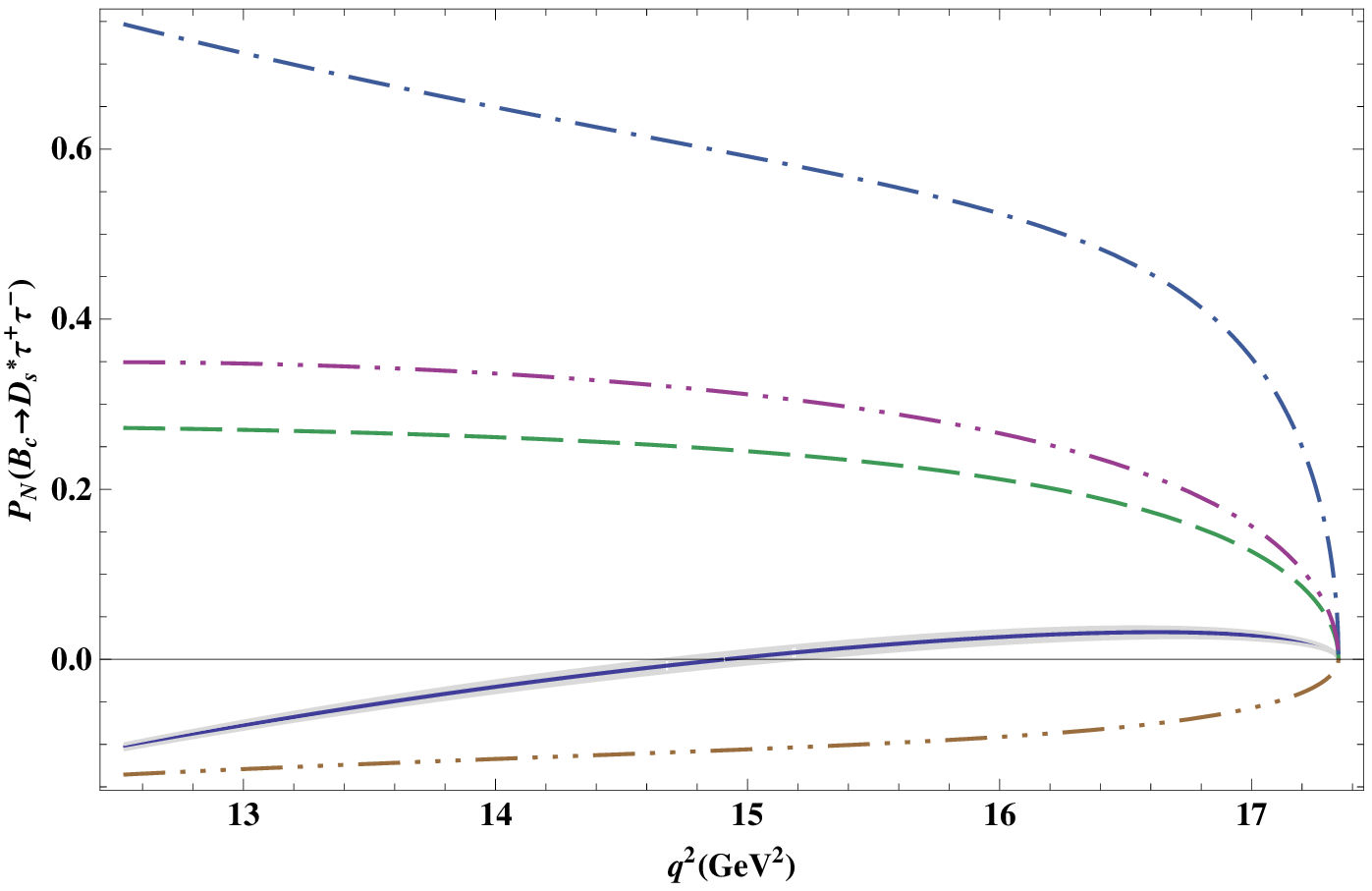}&\includegraphics[scale=0.51]{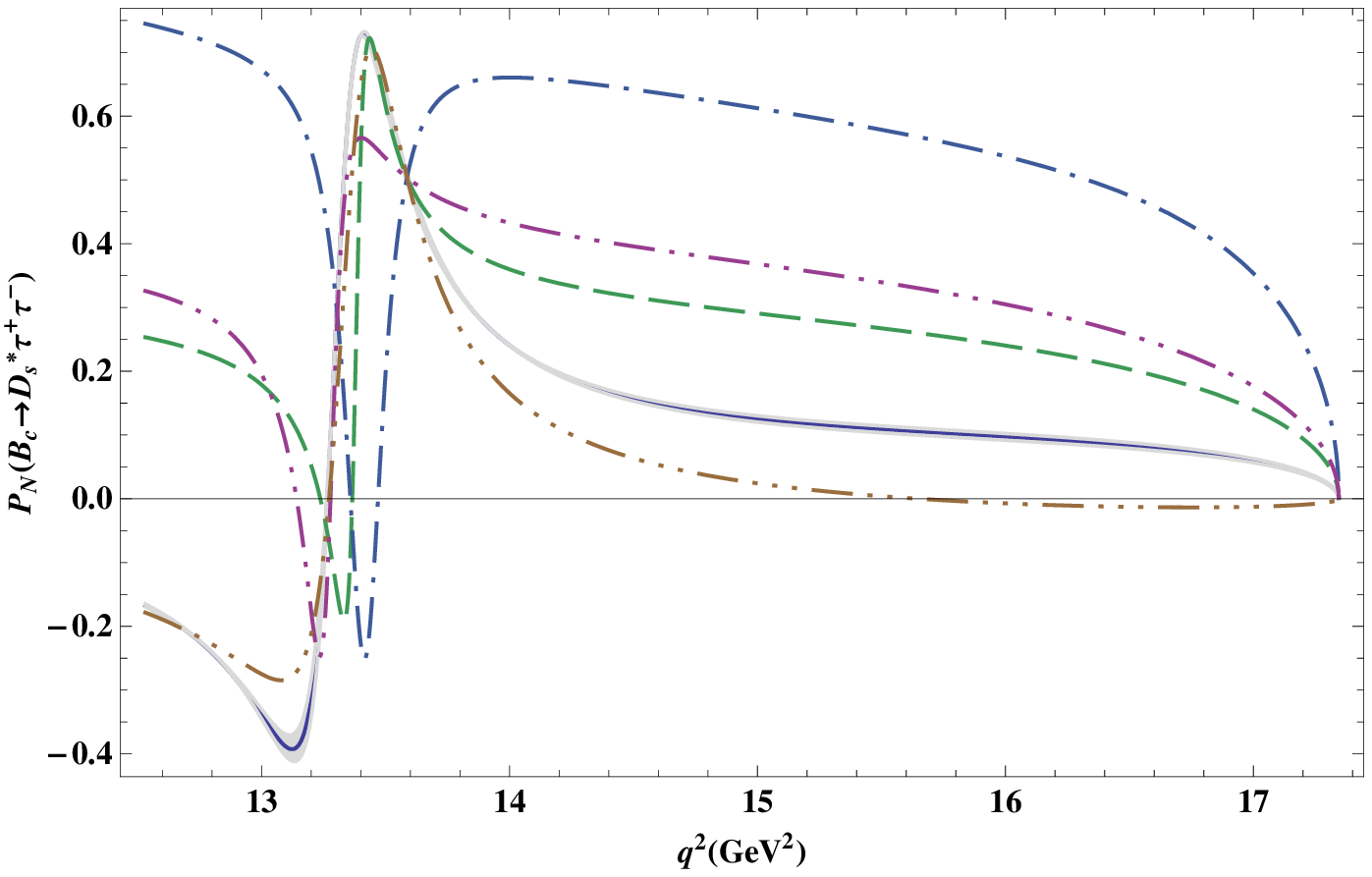}
\end{tabular}
\caption{The dependence the probabilities of the normal lepton
polarization asymmetries, $P_{L}$, for the decays $B_{c}\to
D_{s}^{\ast}\ell^{+}\ell^{-}$ $(\ell=\mu,\tau)$ on $q^{2}$ without
long-distance contributions (a,c) and with long-distance
contributions (b,d), respectively, for different scenarios of MSSM
and SUSY SO(10) GUT model.} \label{np}
\end{figure}

\begin{figure}[ht]
\begin{tabular}{cc}
\hspace{.5cm}($\mathbf{a}$)&\hspace{1.2cm}($\mathbf{b}$)\\
\includegraphics[scale=0.51]{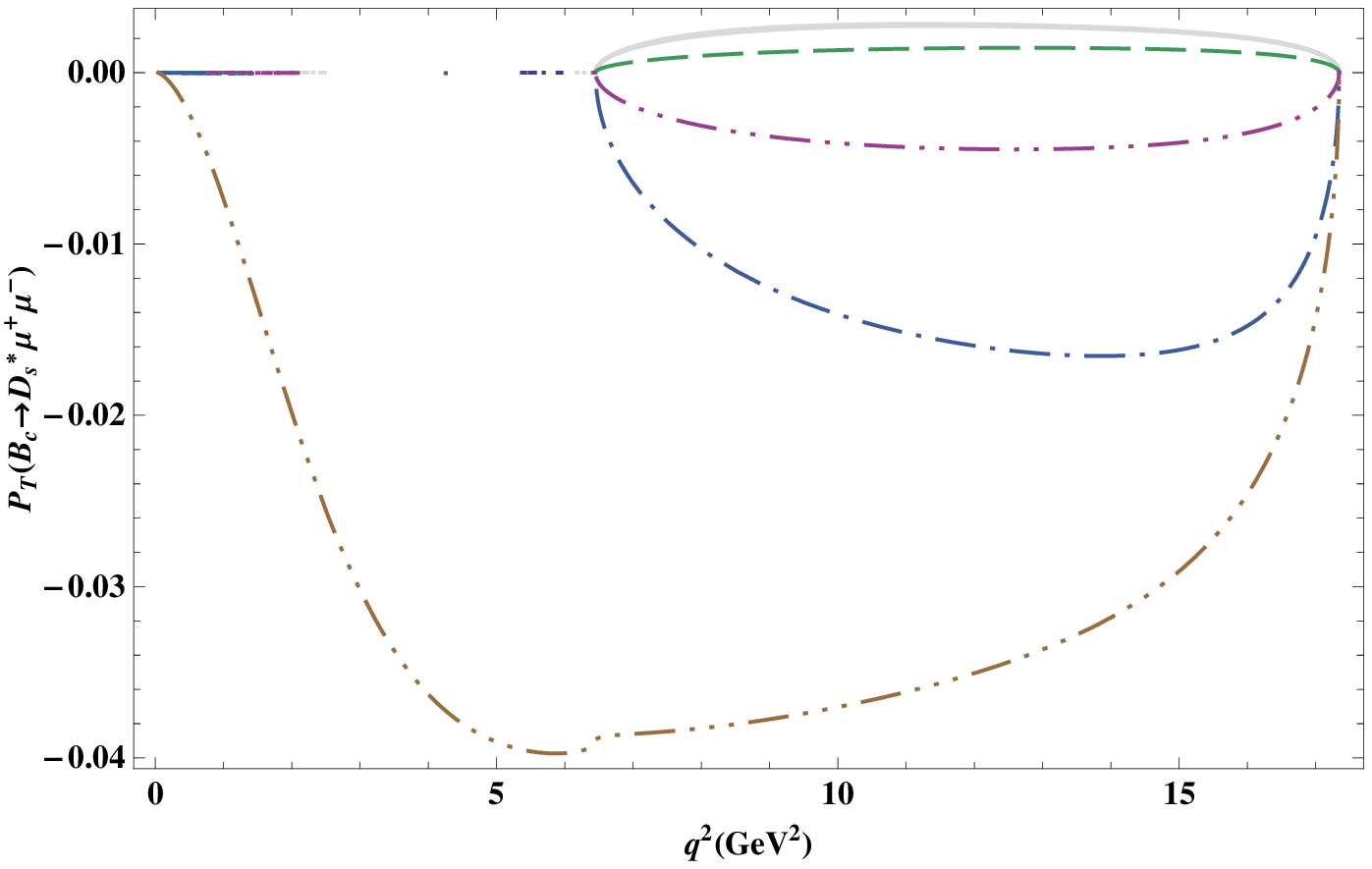}&\includegraphics[scale=0.51]{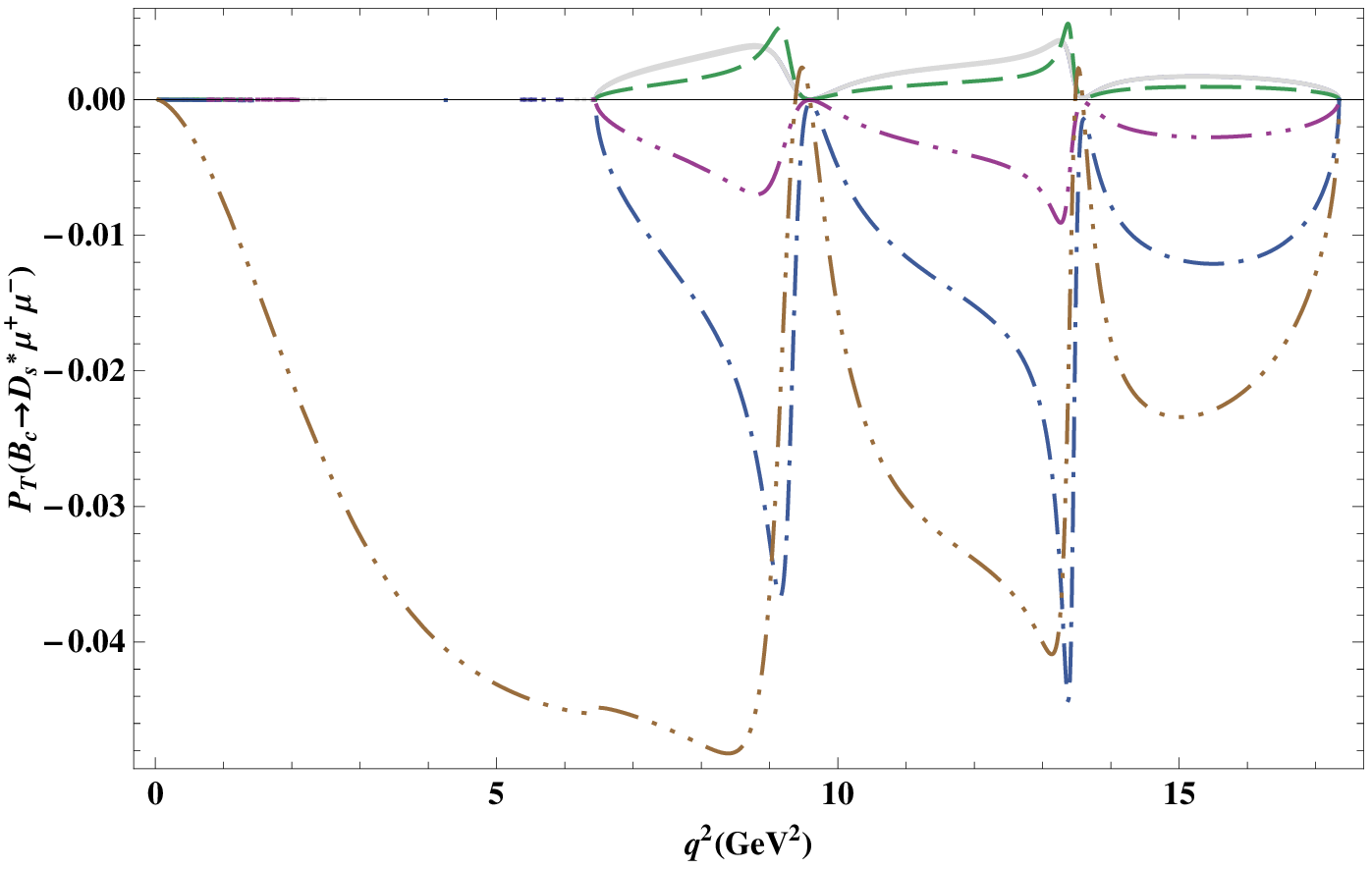}\\
\hspace{.5cm}($\mathbf{c}$)&\hspace{1.2cm}($\mathbf{d}$)\\
\includegraphics[scale=0.51]{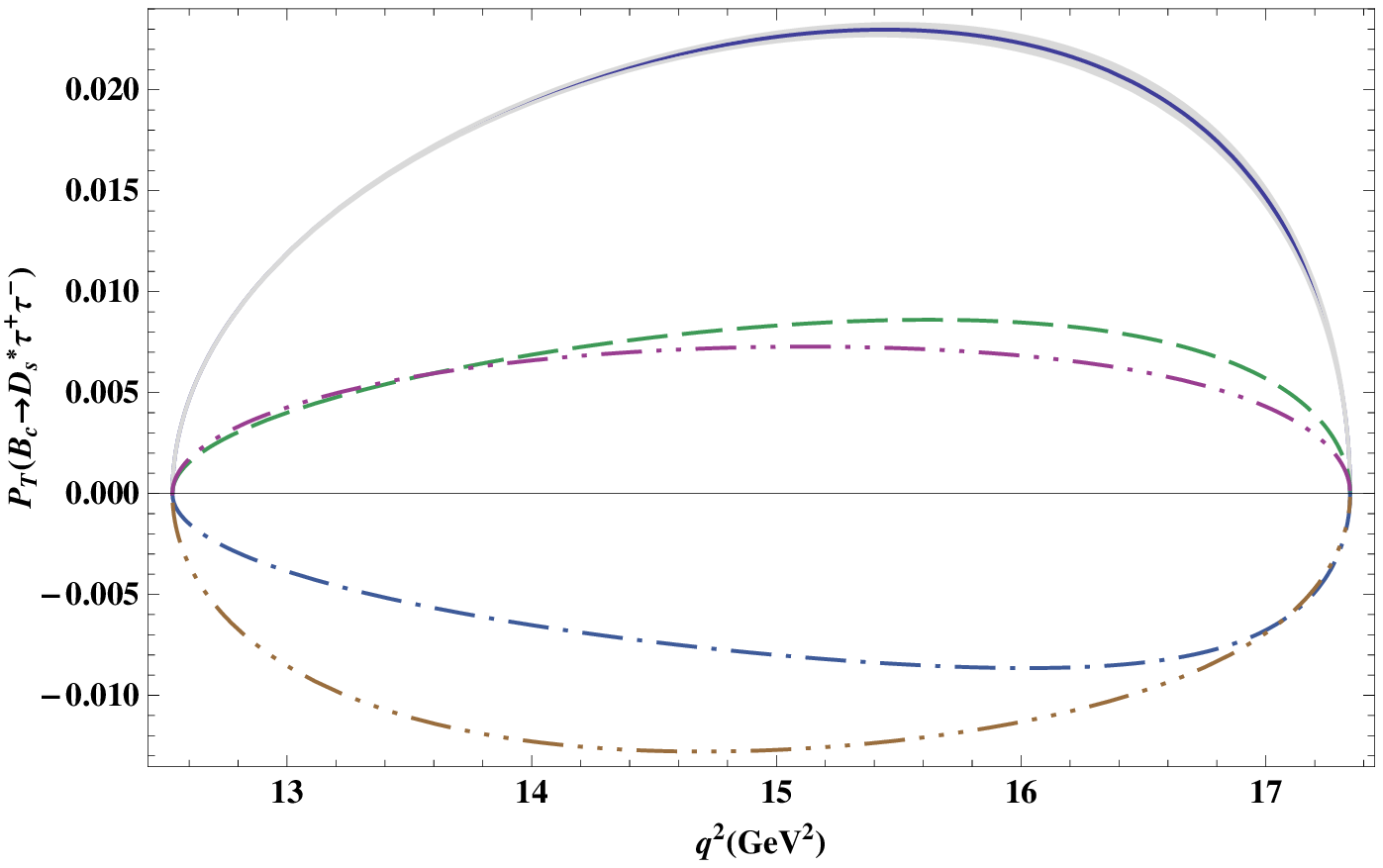}&\includegraphics[scale=0.51]{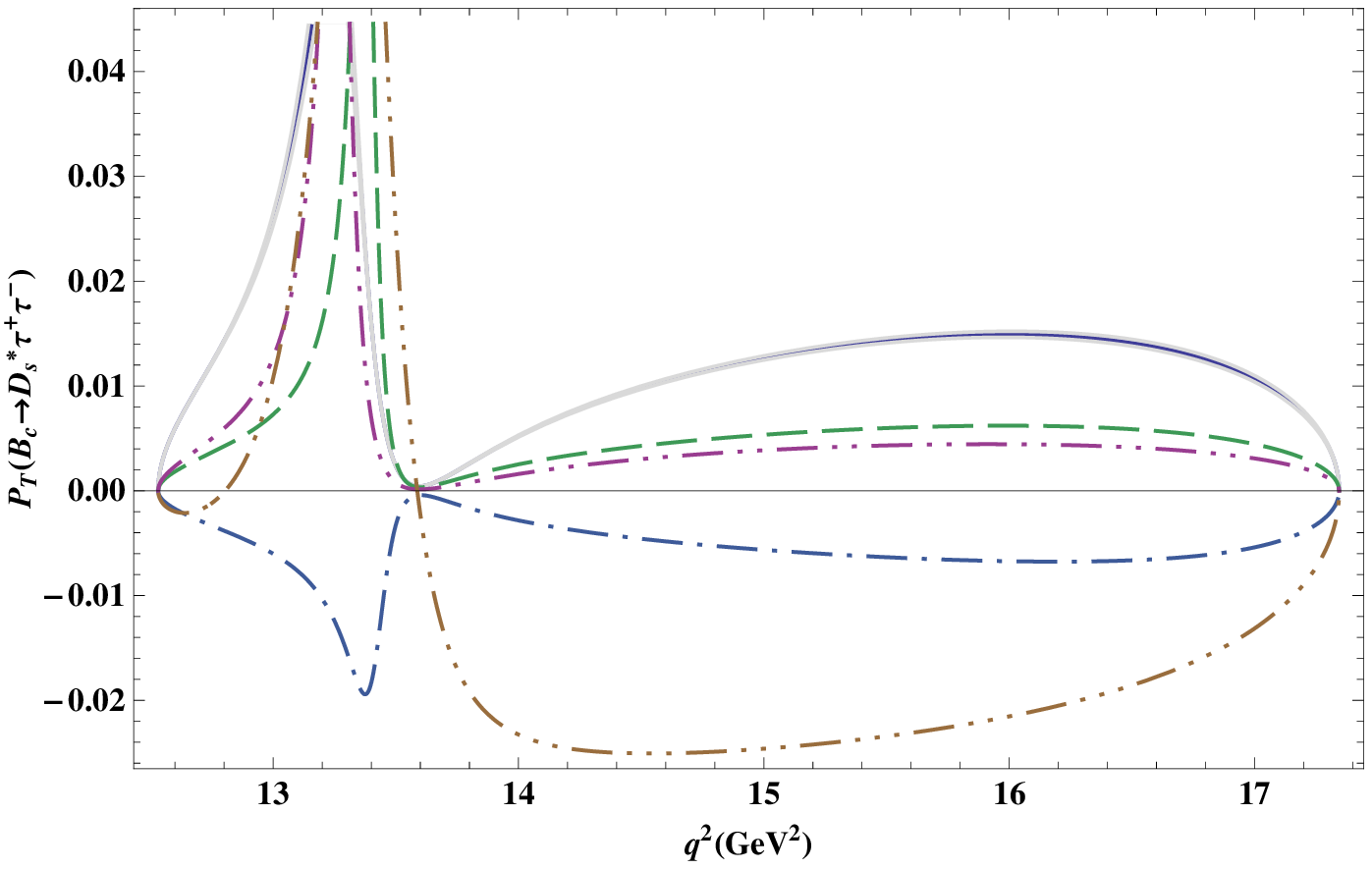}
\end{tabular}
\caption{The dependence the probabilities of the transverse lepton
polarization asymmetries, $P_{L}$, for the decays $B_{c}\to
D_{s}^{\ast}\ell^{+}\ell^{-}$ $(\ell=\mu,\tau)$ on $q^{2}$ without
long-distance contributions (a,c) and with long-distance
contributions (b,d), respectively, for different scenarios of MSSM
and SUSY SO(10) GUT model.} \label{tp}
\end{figure}

We now consider another interesting observable to get the
complementary information about NP in $B_{c}\to
D_{s}^{\ast}\ell^{+}\ell^{-}$ $(\ell=\mu,\tau)$ decays, i.e. the
lepton polarization asymmetries which are shown in Figs \ref{lp},
\ref{np} and \ref{tp}.
Since different polarization asymmetries depend on the Wilson
Coefficients so one can expect large dependency of these asymmetries
on different SUSY variants and hence making these observables
fertile to extract the NP. Therefore, we expect that even in the
SUSY I model where the value of the Wilson coefficient corresponding
to NHBs is zero, the values of these polarization asymmetries would
be mainly modified from those of the SM value, because of the change
in the sign of the terms proportional to $C_{7}^{eff}C_{10}$. Now we
focus on the longitudinal polarization asymmetry for the decays
$B_{c}\to D_{s}^{\ast}\ell^{+}\ell^{-}$ $(\ell=\mu,\tau)$ without
and with long-distance contributions plotted in Figs. \ref{lp}(a,c)
and \ref{lp}(b,d), respectively. Due to this sign alteration of
Wilson coefficients in different SUSY models, Fig. \ref{lp} evince
the variations in the magnitude of the longitudinal polarization
asymmetries from those of the SM. This value is expected to increase
in SUSY I and SUSY II model compared to that of the SM value due to
the opposite sign $C_{7}^{eff}C_{10}$ and because of the NHBs
contribution in the later case. By looking at the Eq.
(\ref{long-polarization}) we can see that the contribution of NHBs,
coming in the auxiliary functions $f_7$ and $f_8$, is compensated by
the mass of the final state lepton. Therefore the large deviation is
expected for heavy mass of the final state lepton and large value of
NHBs contribution i.e. SUSY I and SUSY II. Whereas, the SUSY III and
SUSY SO(10) GUT models show mild effect in the case of muons as the
final state lepton while their effects are quite distinguishable in
the case of tauons as the final state leptons.

Figures \ref{np}(a,c) (without long-distance contribution) and
\ref{np}(b,d) (with long-distance contribution) show the dependence
of normal lepton polarization asymmetries on the square of momentum
transfer for the said decays. One can notice that the values of said
asymmetries are quite sensitive to the large contribution of NHBs in
SUSY II and SUSY III models which is also clear from Eq.
(\ref{norm-polarization}) whereas, it is mildly effected for the
case of SUSY SO(10) GUT model due to small contributions of NHBs and
due to the complex part of the Wilson coefficients.
It is important to note that when there is a large contribution from
the NHBs in $P_{N}$ it will changes its sign as indicated in Fig.
\ref{np} for the case of SUSY II. As the hadronic uncertainties are
insignificant in these asymmetries so the contribution from
different SUSY variants are quite distinguishable for the case when
the final state leptons are muons and even more prominent when these
leptons are tauons. This can be established from Eq.
(\ref{norm-polarization}) when $P_N$ is proportional to the final
state leptonic mass. Furthermore, as the normal polarization is
proportional to the $\lambda $ which approaches to zero at high
$q^{2}$ region and hence the normal polarization asymmetries are
suppressed by $\lambda $ in this region which is depicted in Fig.
\ref{np}.

We now discuss the dependence of transverse polarization asymmetries
on square of momentum transfer for the decays $B_{c}\to
D_{s}^{\ast}\ell^{+}\ell^{-}$, without and with long-distance
effects, plotted in Figs. \ref{tp}(a,b) and \ref{tp}(c,d) for
$\ell=\mu$ and $\tau$, respectively.
One can see from Eq. (\ref{Transverse-polarization}), that it is
proportional to the imaginary parts of the Wilson coefficients which
are absent in the SM and in SUSY I, SUSY II and SUSY III model. But
in the SUSY SO(10) GUT model one would expect the non-zero
transverse polarization in for the $B_{c}\to
D_{s}^{\ast}\mu^{+}\mu^{-}$ ($\tau^{+}\tau^{-}$) decays due to
complex flavor non-diagonal down-type squark mass matrix of 2nd and
3rd generations of order one at GUT scale, which can induce the
complex couplings and consequently lead to complex Wilson
coefficients. But these effects are very small in the said decay
channels. These signatures of the transverse polarization
asymmetries are depicted in Figs. \ref{tp}(a,c) and Figs.
\ref{tp}(b,d) without and with long-distributions, respectively.
Here we can see that its value is for small to measure
experimentally both for the case of muons and tauons as the final
state leptons.

\begin{figure}[ht]
\begin{tabular}{cc}
\hspace{.5cm}($\mathbf{a}$)&\hspace{1.2cm}($\mathbf{b}$)\\
\includegraphics[scale=0.51]{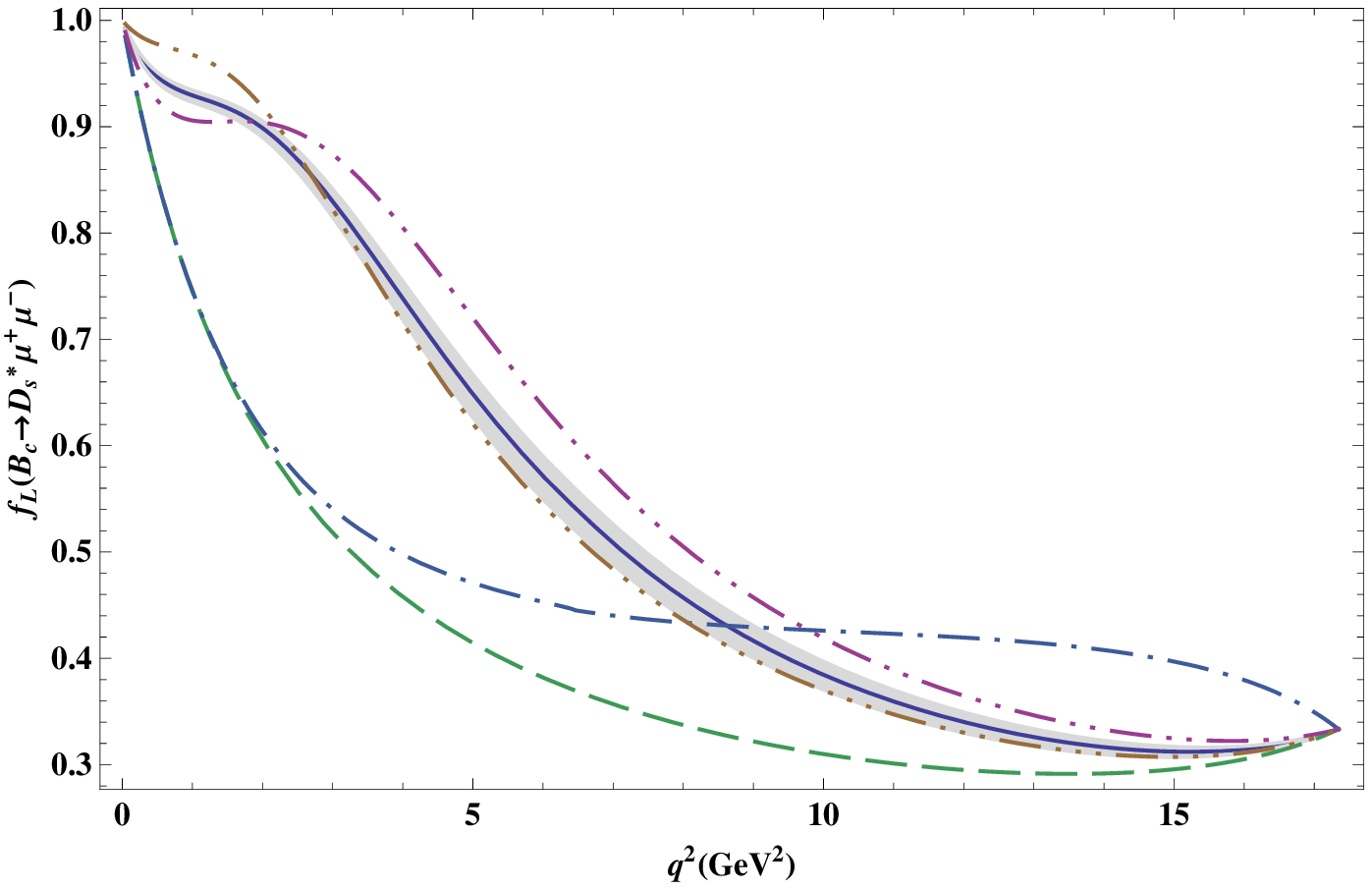}&\includegraphics[scale=0.51]{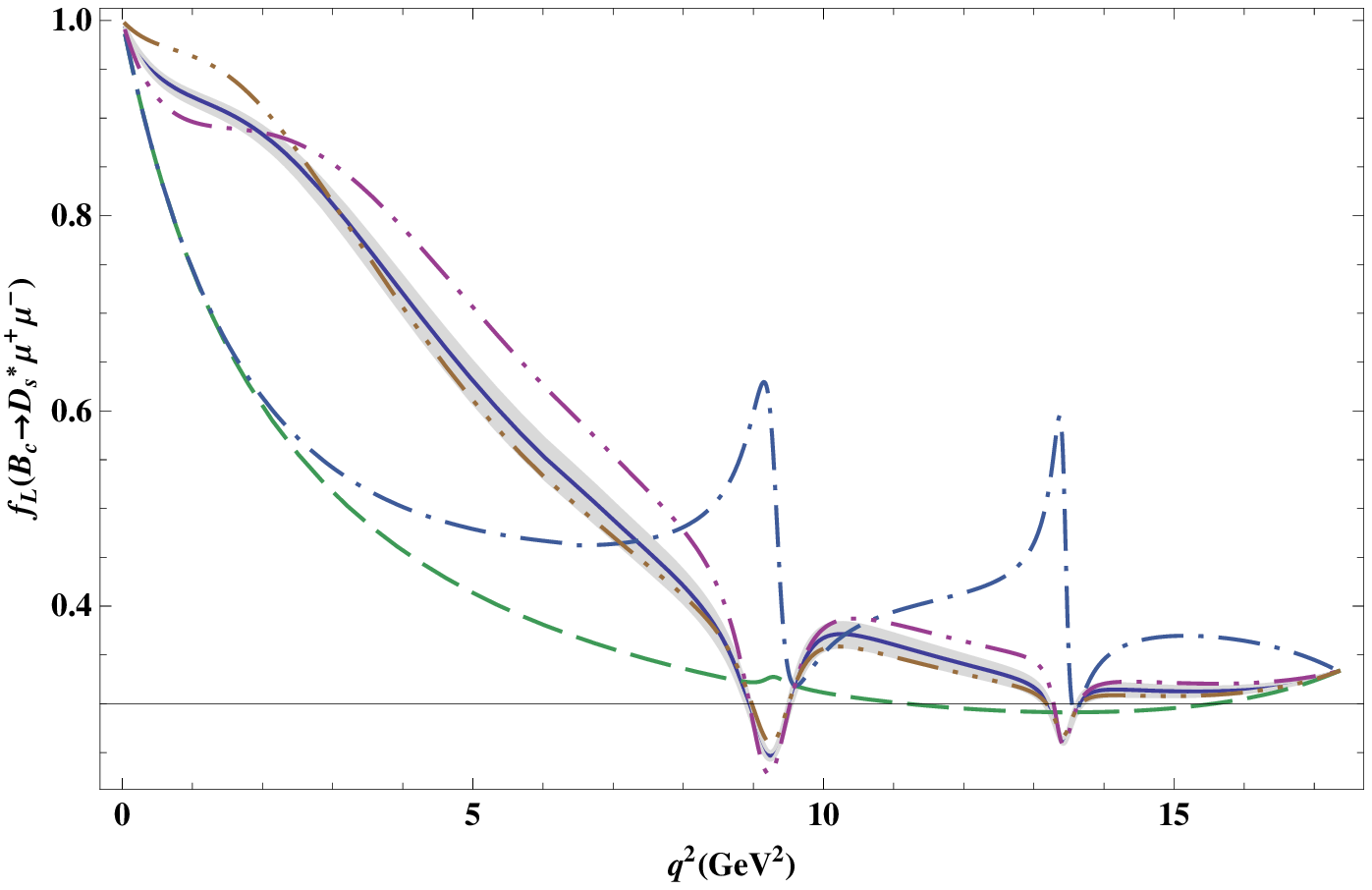}\\
\hspace{.5cm}($\mathbf{c}$)&\hspace{1.2cm}($\mathbf{d}$)\\
\includegraphics[scale=0.51]{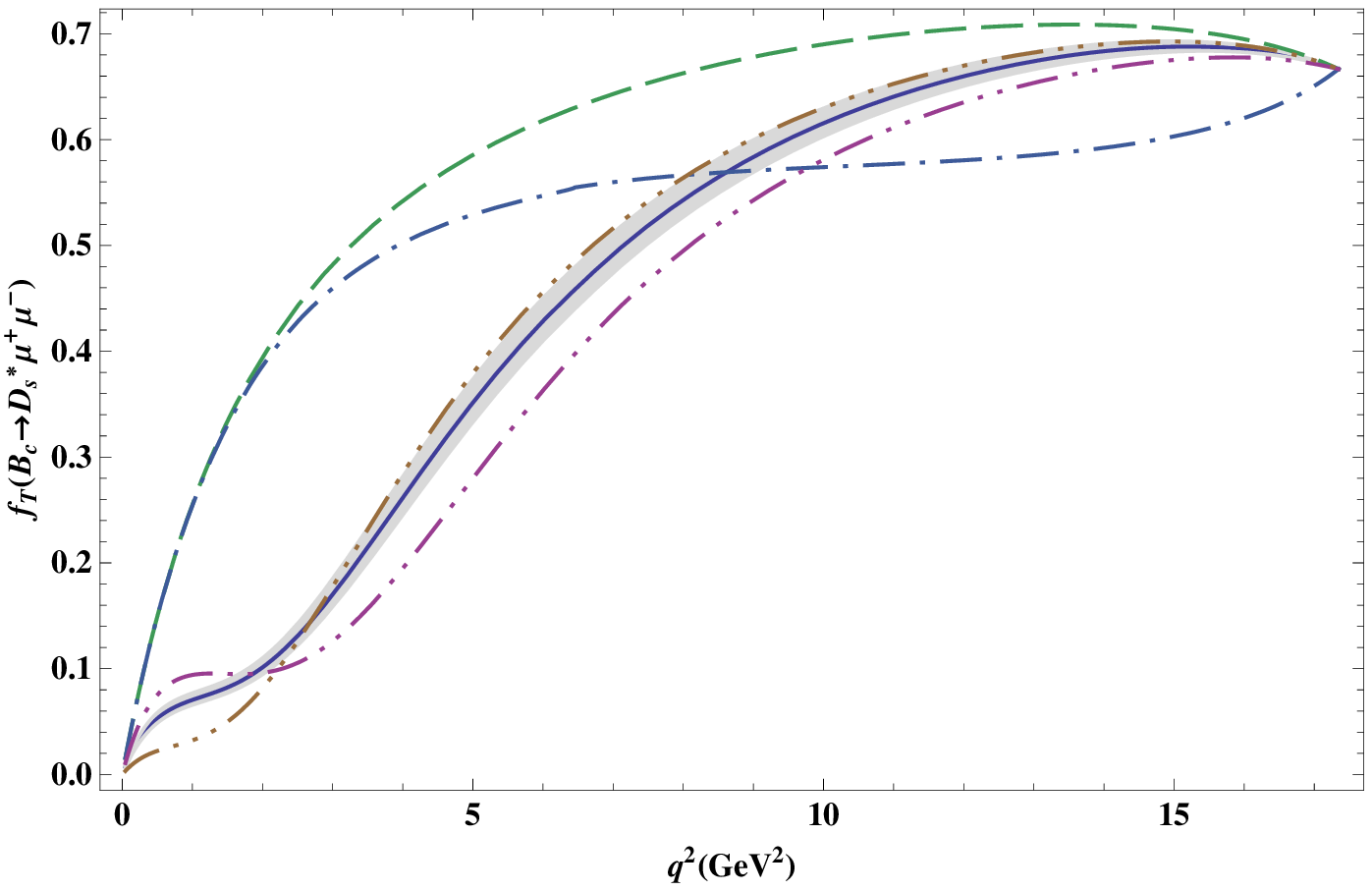}&\includegraphics[scale=0.51]{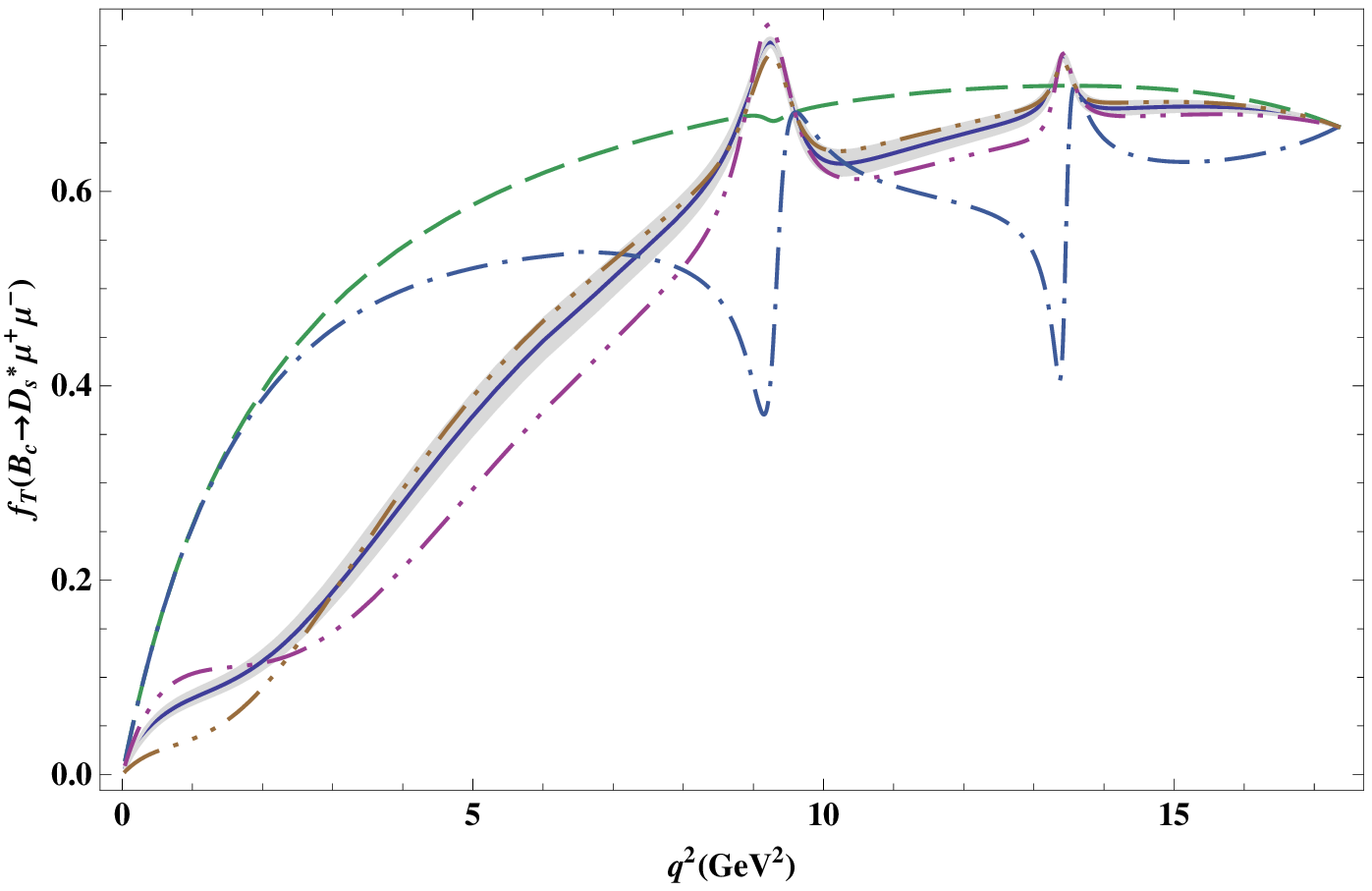}
\end{tabular}
\caption{The dependence the probabilities of the longitudinal and
transverse helicity fractions, $f_{L,T}$, of $D_{s}^{\ast}$ in
$B_{c}\to D_{s}^{\ast}\mu^{+}\mu^{-}$ decays on $q^{2}$ without
long-distance contributions (a,c) and with long-distance
contributions (b,d), respectively, for different scenarios of MSSM
and SUSY SO(10) GUT model.} \label{hfm}
\end{figure}

\begin{figure}[ht]
\begin{tabular}{cc}
\hspace{.5cm}($\mathbf{a}$)&\hspace{1.2cm}($\mathbf{b}$)\\
\includegraphics[scale=0.51]{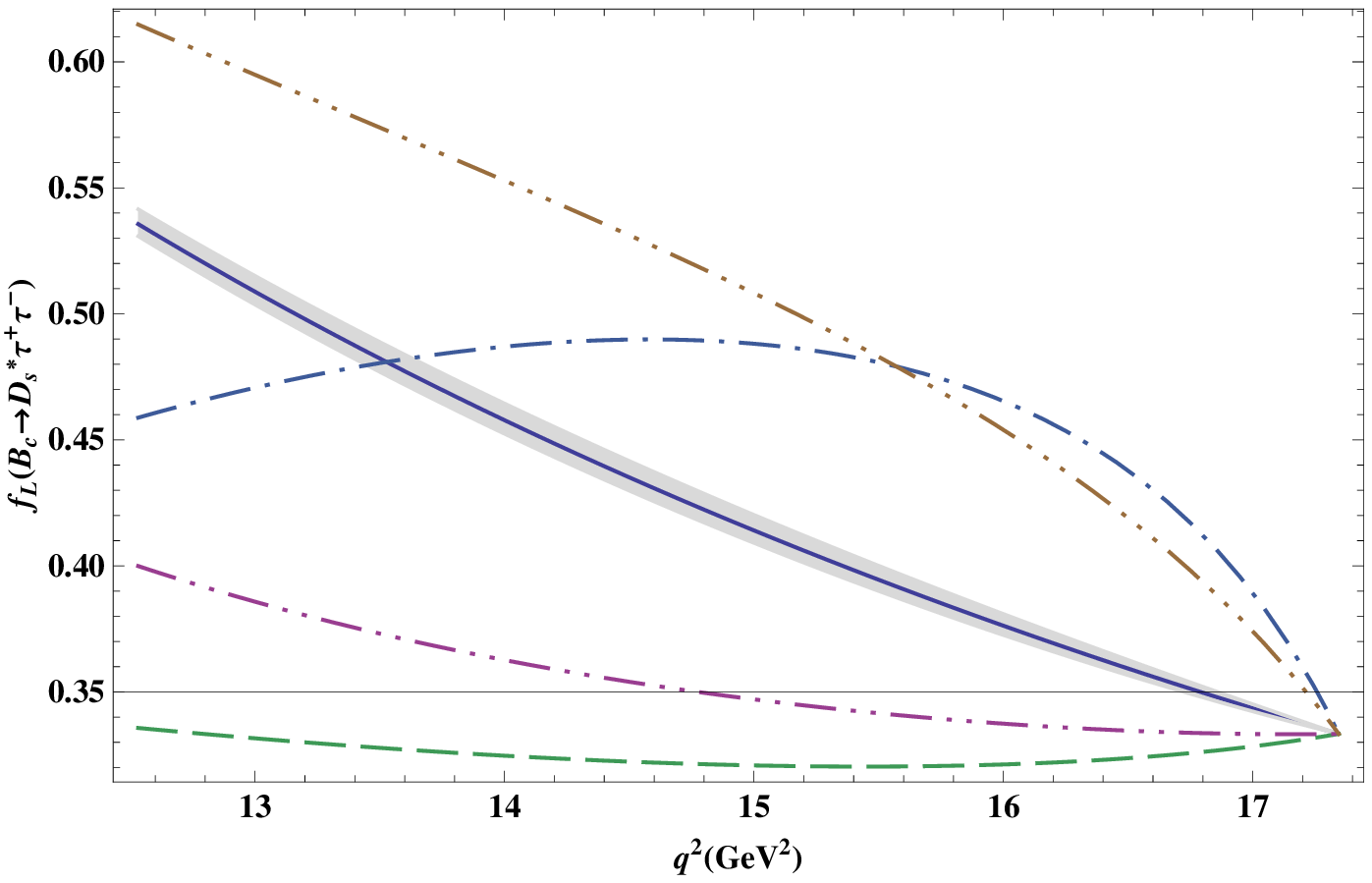}&\includegraphics[scale=0.51]{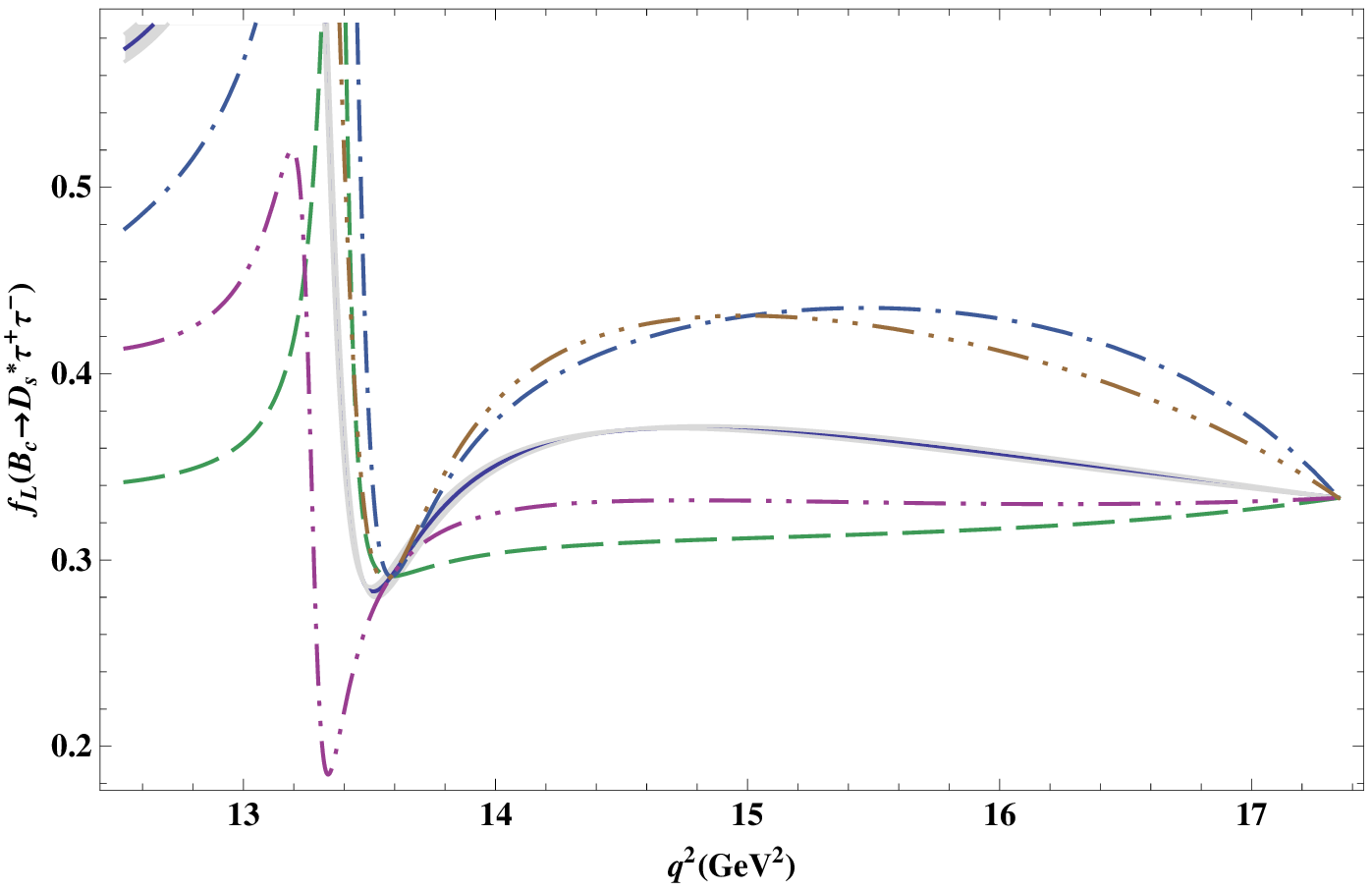}\\
\hspace{.5cm}($\mathbf{c}$)&\hspace{1.2cm}($\mathbf{d}$)\\
\includegraphics[scale=0.51]{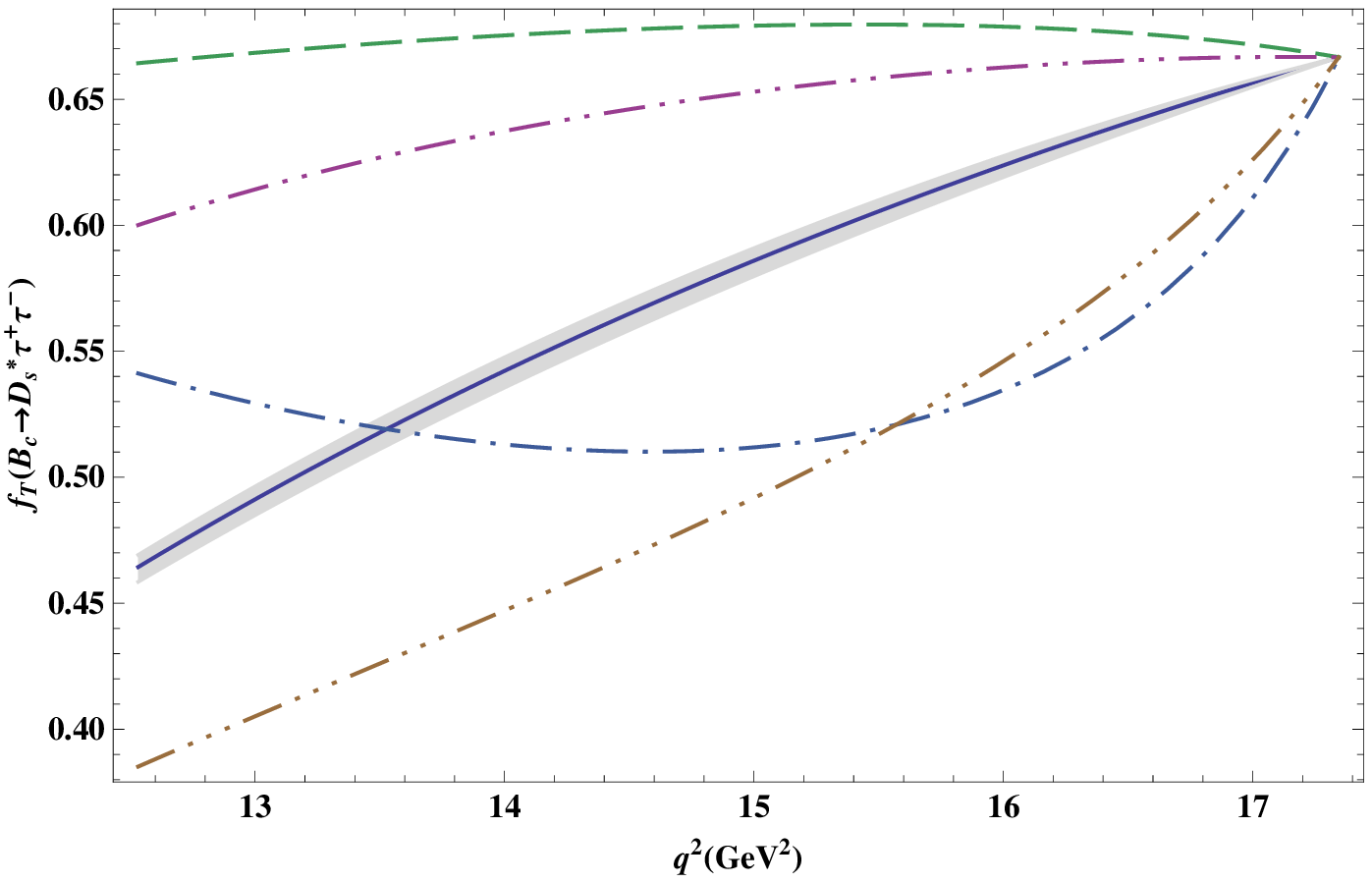}& \includegraphics[scale=0.51]{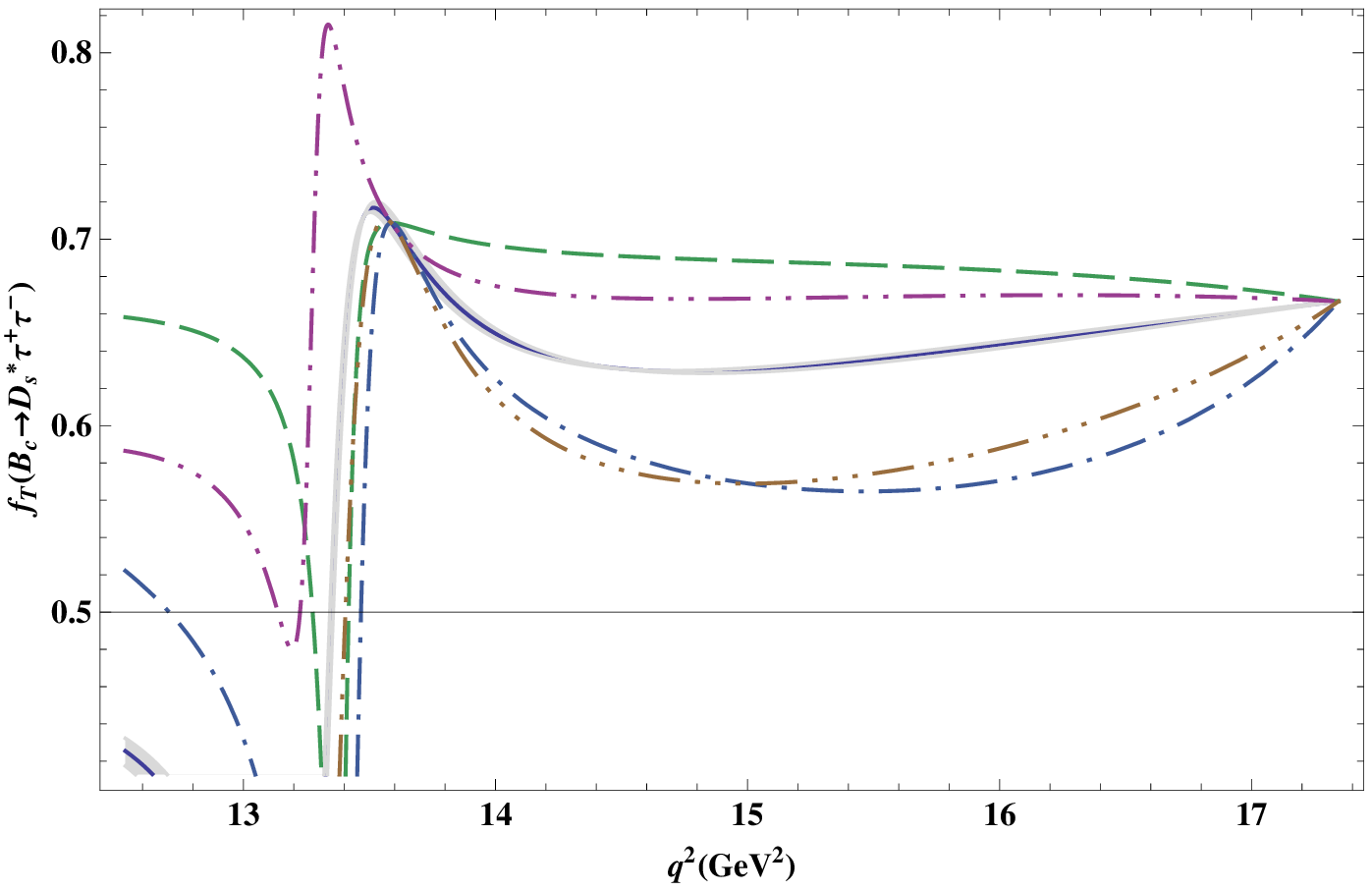}
\end{tabular}
\caption{The dependence the probabilities of the longitudinal and
transverse helicity fractions, $f_{L,T}$, of $D_{s}^{\ast}$ in
$B_{c}\to D_{s}^{\ast}\tau^{+}\tau^{-}$ decays on $q^{2}$ without
long-distance contributions (a,c) and with long-distance
contributions (b,d), respectively, for different scenarios of MSSM
and SUSY SO(10) GUT model.} \label{hft}
\end{figure}

Apart from the above mentioned observables there is another physical
observable sensitive the NP in $B_{c}\rightarrow
D_{s}^{\ast}\ell^{+}\ell^{-}$ transitions, i.e. the helicity
fractions of $D_{s}^{\ast}$ vector meson produced in the final
state. The measurement of longitudinally $K^{*}$ helicity fractions
($f_{L}$) in the decay modes $B\rightarrow K^{*}\ell^{+}\ell^{-}$ by
the BABAR Collaborations \cite{babar} put enormous interest in this
observable. Additionally, it is also shown that the helicity
fractions of final state meson, just like $\mathcal{BR}$,
$\mathcal{A}_{FB}$ and $P_{L,N,T}$, are also very good observables
to dig out the NP \cite{ali,22,23,aa,aa1}. In this respect, it is
natural to study the helicity fractions for the complementary FCNC
processes such as $B_{c}\rightarrow D_{s}^{\ast}\ell^{+}\ell^{-}$
($\ell=\mu,\tau$) in and beyond the SM. For this purpose, we have
plotted the longitudinal ($f_{L}$) and transverse ($f_{T}$) helicity
fractions of $D_{s}^{\ast}$ for the SM and different SUSY models in
Figs. \ref{hfm} and \ref{hft} for the final state leptons as muons
and tauons, respectively. In these graphs the values of the
longitudinal ($f_{L}$) and transverse ($f_{T}$) helicity fractions
of $D_{s}^{\ast}$ are plotted against $q^{2}$ and one can clearly
see that at each value of $q^{2}$ the sum of $f_{L}$ and $f_{T}$ is
equal to one.

Figure \ref{hfm} depicts the case of muons as final state leptons,
the effects of the different SUSY scenarios on the longitudinal
(transverse) helicity fractions of $D_{s}^{\ast}$ are well
distinguishable throughout the $q^{2}$ region. Here one can notice
that for the case of SUSY I, when the contributions of NHBs is
neglected, the deviation from that of the SM values is prominent.
Similarly for the SUSY II scenario, when the NHBs contributions are
large and $\tan\beta$ is also large, the NP effects are spotlighted.
It is also clear from Fig. \ref{hfm} that although the influence of
the SUSY III and SUSY SO(10) models are mild for the case of muons
as final state leptons but one can observe that for the case of
tauons in the final state these effects are quite enhanced from that
of the SM values (see Fig. \ref{hft}). Moreover, Fig. \ref{hfm} also
manifests the variations in the values of $f_{L}$ $(f_{T})$ for the
different SUSY variants with respect to that of the the SM values,
which can be a good tool to put stringent constraints on the
parameter space of different SUSY models.

Now we turn our attention to the case, where tauns are the final
state leptons, the helicity fractions of $D_{s}^{\ast}$ are shown in
Figs. \ref{hft}(a,c) and \ref{hft}(b,d) without and with the
long-distance contributions, respectively.
One can easily extract that similarly to the case of muons, there is
also prominent deviations in the values of the helicity fractions
for all the SUSY models from that of the SM values. However, the
effects for SUSY III and SUSY SO(10) models are more prominent as
compare to the previous case where the muons are the final state
leptons. These figures have also enlightened the variation in the
extremum values of helicity fractions from the SM due to the change
in the SUSY parameters. The deviation in extremum values are very
well marked up at 12.5 GeV$^2$ for all the SUSY models, for example,
the extremum value of longitudinal (transverse) helicity fraction is
changed from its SM value $0.47(0.53)$ to $0.32 (0.68)$, $0.385
(0.615)$, $0.40 (0.60)$ and $0.53 (0.47)$ for SUSY I, SUSY II, SUSY
III and SUSY SO(10), respectively, which is suitable amount of
deviation to measure. Hence, the measurement of the extremum values
of $f_{L}$ and $f_{T}$ in the case of $B_{c}\to
D_{s}^{\ast}\tau^{+}\tau^{-}$ can be used as a good tool in studying
the NP beyond the SM and to distinguish among the different SUSY
models.

\section{Conclusion}\label{con}
In our study on the rare $B_{c}\to D_{s}^{\ast}\ell^{+}\ell^{-}$
decays, with $\ell=\mu$, $\tau$, we have calculated branching ratio
($\mathcal{BR}$), the forward-backward asymmetry $\mathcal{A}_{FB}$
of the leptons, the polarization asymmetries $P_{L,N,T}$ of final
state leptons and the helicity fractions $f_{L,T}$ of the final
state vector meson $D_{s}^{\ast}$ and analyzed the implications of
different SUSY models on these observable for the said decays. The
main outcomes of our analysis can be summarized as follows:
\begin{itemize}
\item We have observed that the $\mathcal{BR}$s deviate sizeably from the SM value in different SUSY models. The study has shown that the $\mathcal{BR}$ is increased considerably for SUSY I and SUSY II mainly because of the change of the sign of Wilson coefficient $C_{7}^{eff}$. But for SUSY III and SUSY SO(10) the values of ${\cal BR}$ are mildly effected from that of the SM values because the small contributions of NHBs and due to the fact that the sign of Wilson coefficient $C_{7}^{eff}$ remains the same as the SM. Hence the accurate measurement of the $\mathcal{BR}$s for these decays would help us to say something about the physics beyond the SM including SUSY models.

\item Along with the $\mathcal{BR}$, our analysis show that $\mathcal{A}_{FB}$, especially the zero position of the $\mathcal{A}_{FB}$, is a fertile observable to extract the NP including SUSY. We have found that for SUSY I and SUSY II, $\mathcal{A}_{FB}$ does not cross the zero position unlike the SM for $B_{c}\to D_{s}^{\ast}\mu^{+}\mu^{-}$, because of the same signs of Wilson coefficient $C_{7}^{eff}$ and $C_{9}^{eff}$ for these two scenarios of SUSY. This signature is similar to that of the observed signals for $B \to K^{\ast}\mu^{+}\mu^{-}$. Moreover, the shift in the zero positions of $\mathcal{A}_{FB}$ for SUSY III and SUSY SO(10) models provide a promising signature of the NP which can be tested experimentally. Hence, the measurement of the magnitude as well as the zero crossing position of $\mathcal{A}_{FB}$ of the considered decays can provide a stringent test for MSSM models and SUSY SO(10) GUT models.

\item The longitudinal, normal and transverse polarization asymmetries of leptons are calculated in different SUSY scenarios for the rare semileptonic charmed $B_{c}$ meson which can be tested in experiments with great precision. It is found that the SUSY effects could be measured at future experiments and will shed light on the NP signal beyond the SM. The transverse polarization asymmetry is the most interesting observable to look for the SUSY SO(10) effects where its value is non-zero in almost all $q^2$ region unlike for the cases of the SM and MSSM models. It is measurable at future experiments such as the LHC and the BTeV machines where a large number of $b\bar{b}$ pairs are expected to be produced.

\item We have calculated the helicity fractions $f_{L,T}$ of the final state vector meson $D_{s}^{\ast}$ to extract the comparative study of different SUSY models and the SM. The study has shown that the deviation from the SM values of the helicity fractions are quite large with tauons in final state for the MSSM models as well as SUSY SO(10) model. It is also shown that there is a noticeable change due to SUSY parameter in the position of the extremum values of the longitudinal and transverse helicity fractions of $D_{s}^{\ast}$ meson for the case of tauons as a final state leptons. Hence, the helicity fraction of $D_{s}^{\ast}$ meson in the decays $B_{c}\to D_{s}^{\ast}\ell^{+}\ell^{-}$ $(\ell=\mu,\tau)$ can be a stringent test in finding the status of the SUSY models at the LHC.
\end{itemize}

To sum up, the more data to be available from Tevatron and LHCb will
provide a powerful testing ground for the SM and to put some
constraints on the SUSY parameter space in particular the value of
$\tan\beta$ and the masses of lightest chargino and lightest stop
squark. Our comparative analysis of the SM and the different SUSY
models on the observables for $B_{c}\to
D_{s}^{\ast}\ell^{+}\ell^{-}$ $(\ell=\mu,\tau)$ decays can be useful
for probing and its extension to various SUSY models the existence
of the supersymmetry.

\section*{Acknowledgments}
The authors would like to thank Professor Riazuddin and Professor
Fayyazuddin for their valuable guidance and helpful discussions.

\end{document}